\providecommand{\materialflag}{1}
\providecommand{\journalflag}{0}

\ifnum\journalflag=0
    \documentclass[a4paper, twoside, reqno]{amsart}
\fi

\usepackage[
    backend=biber,
    natbib=true,
    citestyle=authoryear-comp,
    bibstyle=authoryear,
    maxcitenames=2,
    maxbibnames=9,
    uniquelist=false,
    uniquename=false,
    dashed=false,
    firstinits=true,
    doi=false,
    isbn=false,
    date=year]{biblatex}

\usepackage[
    hidelinks,
    colorlinks=true,
    allcolors=blue,
    pdfproducer={Latex with hyperref},
    pdfcreator={pdflatex}]{hyperref}

\ifnum\journalflag=0
    \usepackage[foot]{amsaddr}
\fi

\usepackage{glossaries-prefix}
\usepackage{marvosym}
\usepackage{graphicx}
\usepackage{float}
\usepackage[font=footnotesize]{caption}
\usepackage{subcaption}
\usepackage{booktabs}
\usepackage{tabularx}
\usepackage{bbm} %
\usepackage{upgreek}
\usepackage{enumitem}
\usepackage{pifont}
\usepackage{textcase}

\newcommand{\papertitle}{Bayesian taut splines for estimating the number of modes}

\newcommand{\myorcid}[1]{%
    \href{https://orcid.org/#1}{%
        \includegraphics[height=2ex]{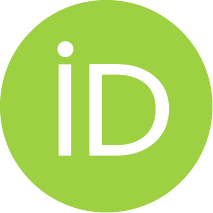}%
        \url{https://orcid.org/#1}
    }%
}

\newcommand{\joseechacon}{Jos\'e E. Chac\'on}
\newcommand{\addressjoseechaon}{Departamento de Matem\'aticas, Universidad de Extremadura, Badajoz, Spain.}
\newcommand{\emailjoseechacon}{jechacon@unex.es}
\newcommand{\orcidjoseechacon}{0000-0002-3675-1960}
\newcommand{\footjoseechacon}{$^{\dagger}$}
\newcommand{\footaddressjoseechaon}{\footjoseechacon\addressjoseechaon}
\newcommand{\orcidlinkjoseechacon}{\myorcid{\orcidjoseechacon}}
\newcommand{\footorcidjoseechacon}{\footjoseechacon\orcidlinkjoseechacon}

\newcommand{\javierfdezserrano}{Javier Fern\'andez Serrano}
\newcommand{\addressjavierfdezserrano}{Departamento de Matem\'aticas, Universidad Aut\'onoma de Madrid, Madrid, Spain.}
\newcommand{\emailjavierfdezserrano}{javier.fernandezs01@estudiante.uam.es}
\newcommand{\orcidjavierfdezserrano}{0000-0001-5270-9941}
\newcommand{\footjavierfdezserrano}{$^{\ddagger}$}
\newcommand{\footaddressjavierfdezserrano}{\footjavierfdezserrano\addressjavierfdezserrano}
\newcommand{\orcidlinkjavierfdezserrano}{\myorcid{\orcidjavierfdezserrano}}
\newcommand{\footorcidjavierfdezserrano}{\footjavierfdezserrano\orcidlinkjavierfdezserrano}

\newcommand{\mysupplement}{\gls*{sm}~\citep{Supplement}}

\let\oldparagraph\paragraph
\renewcommand{\paragraph}[1]{%
    \ifnum\journalflag=0
        \vspace*{0.5em}
    \fi
    \oldparagraph{\textit{#1}}
}

\newcommand{\keywordnumberofmodes}{number of modes}
\newcommand{\keywordbayesianinference}{Bayesian inference}
\newcommand{\keywordcompositionalspline}{compositional spline}
\newcommand{\keywordkde}{kernel density estimation}
\newcommand{\keywordmodelselection}{model selection}
\newcommand{\keywordmodetesting}{mode testing}

\newcommand{\mscnonparamestimation}{62G05}
\newcommand{\mscdensityestimation}{62G07}
\newcommand{\mscbayesianinference}{62F15}
\newcommand{\mscbayesianproblems}{62C10}
\newcommand{\mscfuzziness}{62C86}

\newcommand{\numhidalgostamps}{485}
\newcommand{\testbedid}{\fifthmodel{}}
\newcommand{\testbedsamplesize}{200}
\newcommand{\testbedmcmcsamplesize}{700}
\newcommand{\testbedsplinedimension}{22}
\newcommand{\numberofmodesred}{2}
\newcommand{\percentagered}{57\%}
\newcommand{\numberofmodesblue}{3}
\newcommand{\percentageblue}{43\%}
\newcommand{\numberofmodessquare}{2}
\newcommand{\numberofmodescircle}{3}
\newcommand{\numberofmodestriangle}{4}
\newcommand{\sfpcasamplesize}{7,840}
\newcommand{\testbedmediansplinenummodes}{three}
\newcommand{\testbedminfinalprob}{0.92}
\newcommand{\testbedmaxfinalprob}{0.94}
\newcommand{\probleftmode}{0.69}
\newcommand{\probcentremode}{0.74}
\newcommand{\probrightmode}{0.99}

\newcommand{\ohtaniexplorationprobthreemodes}{77\%}
\newcommand{\ohtaniexplorationprobfourmodes}{23\%}
\newcommand{\ohtaniminfinalprob}{0.93}
\newcommand{\ohtanimaxfinalprob}{0.98}
\newcommand{\numohtanipitches}{2,626}
\newcommand{\ohtanimcmcsamplesize}{120}
\newcommand{\ohtanisfpcasamplesize}{7,437}
\newcommand{\ohtanimediansplinenummodes}{four}
\newcommand{\ohtanisplinedimension}{32}

\newcommand{\rlanguage}{\textsf{R}}
\newcommand{\baseballsavant}{\textit{BaseballSavant}}
\newcommand{\bigoh}[1]{\mathcal{O}(#1)}

\newcommand{\genindexi}{i}
\newcommand{\genindexj}{j}
\newcommand{\genindexmu}{\mu}
\newcommand{\genindexnu}{\nu}

\newcommand{\kronecker}{\delta}
\newcommand{\kroneckerdd}[2]{\kronecker_{#1#2}}

\newcommand{\numberofmodes}{k}
\newcommand{\samplesize}{n}
\newcommand{\numobservations}{N}
\newcommand{\numbootstrapsamples}{B}

\newcommand{\function}{f}
\newcommand{\funcarg}{\cdot}
\newcommand{\x}{x}
\newcommand{\y}{y}

\newcommand{\mappingelements}[2]{#1 \mapsto #2}
\newcommand{\mappingsets}[2]{#1 \rightarrow #2}
\newcommand{\functiondef}[3]{#1 : \mappingsets{#2}{#3}}

\newcommand{\indicator}[2]{\mathbbm{1}_{#1}(#2)}

\newcommand{\mixtureweight}{w}

\newcommand{\estimateof}[1]{\hat{#1}}
\newcommand{\extendedparam}[1]{\bar{#1}}

\DeclareMathOperator*{\argmin}{arg\,min}
\DeclareMathOperator*{\argmax}{arg\,max}

\newcommand{\reals}{\mathbb{R}}
\newcommand{\naturals}{\mathbb{N}}

\newcommand{\lpspace}[2]{L^{#1}(#2)}
\newcommand{\hilbertltwospace}{\lpspace{2}{\intervalab}}

\newcommand{\setcardinality}[1]{|#1|}

\newcommand{\normalpdf}{\phi}
\newcommand{\normalcdf}{\Phi}

\newcommand{\kernel}{\normalpdf}

\newcommand{\bandwidth}{h}
\newcommand{\criticalbandwidth}{\bandwidth_{\mathrm{crit}}}

\newcommand{\kdeof}[1]{\estimateof{\pdf}_{#1}}
\newcommand{\kde}{\kdeof{\bandwidth}}

\newcommand{\kdederivativeorder}{r}
\newcommand{\piselector}[1]{\mathrm{\acrshort*{pi}}_{#1}}
\newcommand{\lscvselector}[1]{\mathrm{\acrshort*{lscv}}_{#1}}

\newcommand{\excessmasslevelmuller}{\lambda}
\newcommand{\empiricalexcessmass}[2]{E_{#1}(#2)}

\newcommand{\tautstringradius}{\varepsilon}

\newcommand{\dataset}{\mathcal{D}}
\newcommand{\sampleobs}{x}

\newcommand{\pdf}{f}
\newcommand{\pdfsecondary}{g}

\newcommand{\logpdf}{\log\pdf}

\newcommand{\intervala}{a}
\newcommand{\intervalb}{b}
\newcommand{\intervalab}{[\intervala, \intervalb]}

\newcommand{\intbetweenab}{\int_{\intervala}^{\intervalb}}
\newcommand{\differential}[1]{d#1}

\newcommand{\transpose}[1]{#1^{\top}}

\newcommand{\mode}{\hat{\mathsf{m}}}
\newcommand{\antimode}{\check{\mathsf{m}}}
\newcommand{\modeneighbourepsilon}{\epsilon}

\newcommand{\prob}{\mathrm{Pr}}
\newcommand{\priorprob}[1]{\prob(#1)}
\newcommand{\condrv}[2]{#1 | #2}
\newcommand{\condprob}[2]{\prob(\condrv{#1}{#2})}
\newcommand{\parameter}{\theta}
\newcommand{\parameters}{\parameter_1, \dots, \parameter_{\numparameters}}
\newcommand{\parametervalue}{\vartheta}
\newcommand{\parametervec}{\boldsymbol{\parameter}}
\newcommand{\numparameters}{m}
\newcommand{\priorprobparameters}{\priorprob{\parameters}}
\newcommand{\likelihood}{\condprob{\dataset}{\parameters}}
\newcommand{\nummcmcsteps}{s}

\newcommand{\bayes}{\mathcal{B}}
\newcommand{\bayesspace}{\bayes\intervalab}

\newcommand{\bayesperturbation}{\oplus}
\newcommand{\bigbayesperturbation}{\bigoplus}
\newcommand{\bayesminus}{\ominus}
\newcommand{\bayespowering}{\odot}
\newcommand{\bayesscalar}{\gamma}

\newcommand{\clroperator}{\mathrm{clr}}
\newcommand{\clr}[1]{\clroperator[#1]}
\newcommand{\clrinv}[1]{\clroperator^{-1}[#1]}

\newcommand{\clrpdf}{\clr{\pdf}}

\newcommand{\logpdfvar}{p}

\newcommand{\hilbertltwospacezerointegral}{L_0^2(\intervalab)}

\newcommand{\spline}{s}
\newcommand{\splinedegree}{r}
\newcommand{\zbsplinegeneric}{z}
\newcommand{\zbspline}{\zbsplinegeneric_{\zbsplineparams}}
\newcommand{\zbsplinepdf}{\zeta}
\newcommand{\zbsplineparam}{\theta}
\newcommand{\zbsplineparams}{\boldsymbol{\zbsplineparam}}
\newcommand{\approxzbsplineparams}{\estimateof{\zbsplineparams}}
\newcommand{\approxzbsplineparamsof}[2]{\approxzbsplineparams(#1, #2)}
\newcommand{\zbsplineknot}{\kappa}
\newcommand{\zbsplinedim}{d}
\newcommand{\realszbsplinedim}{\reals^{\zbsplinedim}}
\newcommand{\zbsplinepdfspace}{\mathcal{Z}_{\zbsplinedim}\intervalab}
\newcommand{\zbasisspline}{Z}
\newcommand{\numsplinepoints}{m}
\newcommand{\splinepointx}{t}
\newcommand{\oneminuscurvaturepenalty}{\alpha}
\newcommand{\zbsplinelossof}[3]{\mathcal{L}(#1; #2, #3)}
\newcommand{\zbsplineloss}{\zbsplinelossof{\zbsplineparams}{\pdf}{\oneminuscurvaturepenalty}}
\newcommand{\zerointegralconstraint}[1]{\intbetweenab #1(\x) \ \differential{\x} = 0}
\newcommand{\compositionalsplineapprox}[2]{[#2]_{#1}}
\newcommand{\truncatedpdf}{\pdf^{\star}}
\newcommand{\rminusonecontinuousclass}{\mathcal{C}^{\splinedegree - 1}(\intervalab)}
\newcommand{\bayesinnerproductmatrix}{\mathbf{M}}

\newcommand{\distance}[2]{d(#1, #2)}
\newcommand{\innerproduct}[2]{\langle #1, #2 \rangle}
\newcommand{\bayesinnerproduct}[2]{\innerproduct{#1}{#2}_{\bayes}}
\newcommand{\ltwoinnerproduct}[2]{\innerproduct{#1}{#2}_{2}}

\newcommand{\expectedvalue}[1]{\mathbb{E}[#1]}
\newcommand{\variance}[1]{\mathrm{Var}(#1)}
\newcommand{\followsdistr}{\sim}

\newcommand{\lognormaldistribution}[2]{\mathrm{LogNormal}(\upmu = #1, \upsigma = #2)}
\newcommand{\betaparamalpha}{\upalpha}
\newcommand{\betadistribution}[2]{\mathrm{Beta}(\betaparamalpha = #1, \upbeta = #2)}
\newcommand{\poissondistribution}[1]{\mathrm{Poisson}(\uplambda = #1)}
\newcommand{\exponentialparam}{\uplambda}
\newcommand{\exponentialdistribution}[1]{\mathrm{Exponential}(\exponentialparam = #1)}

\newcommand{\btsexplorationmodel}{\condprob{\funcarg}{\bandwidth, \oneminuscurvaturepenalty}}
\newcommand{\btsexplorationprior}{\priorprob{\bandwidth, \oneminuscurvaturepenalty}}

\newcommand{\approxzbsplineparamshalpha}{\approxzbsplineparamsof{\bandwidth}{\oneminuscurvaturepenalty}}
\newcommand{\halphapair}{(\bandwidth, \oneminuscurvaturepenalty)}

\newcommand{\smoothingparamssample}{\mathcal{S}}
\newcommand{\smoothingsamplesize}{\nu}
\newcommand{\compositionalsplinesample}{%
\{\zbsplinepdf_{\zbsplineparams_{\genindexi}}\}_{\genindexi = 1}^{\smoothingsamplesize}
}

\newcommand{\nummodesparam}{\numberofmodes}
\newcommand{\curvatureparam}{\xi}

\newcommand{\bandwidthlocationhyperparam}{\mu_{\bandwidth}}
\newcommand{\bandwidthdispersionhyperparam}{\sigma_{\bandwidth}}
\newcommand{\penaltyhyperparam}{\beta_{1 - \oneminuscurvaturepenalty}}
\newcommand{\curvaturehyperparam}{\lambda_{\curvatureparam}}
\newcommand{\bandwidthsigma}{\sigma}

\newcommand{\haskmodes}{\mathrm{has} \ \nummodesparam \ \mathrm{modes}}

\newcommand{\posteriorhalpha}{\condprob{\bandwidth, \oneminuscurvaturepenalty}{\dataset}}

\newcommand{\genrvx}{X}

\newcommand{\sfpcaparam}{\delta}
\newcommand{\sfpcaparamnum}[1]{\sfpcaparam_{#1}}
\newcommand{\btssfpcamodel}{\condprob{\funcarg}{\sfpcaparam}}
\newcommand{\priorsfpcaparam}{\priorprob{\sfpcaparam}}

\newcommand{\sfpcaparammin}{\sfpcaparam_{\min}}
\newcommand{\sfpcaparammax}{\sfpcaparam_{\max}}
\newcommand{\sfpcamean}{\upmu}
\newcommand{\sfpcastdev}{\upsigma}
\newcommand{\sfpcastdevnum}[1]{\sfpcastdev_{#1}}
\newcommand{\sfpcamodel}[1]{%
    \sfpcamean
    \bayesperturbation
    #1
    \bayespowering
    \sfpcastdev
}
\newcommand{\sfpcapceigenfun}[1]{\mathrm{\acrshort*{pc}}_{#1}}
\newcommand{\sfpcaeigenvalue}[1]{\lambda_{#1}}
\newcommand{\sfpcascore}{s}
\newcommand{\firststandarddeviation}{\sqrt{\sfpcaeigenvalue{1}}}

\newcommand{\priorprobnummodesparam}{\priorprob{\nummodesparam}}
\newcommand{\sfpcaparamset}{\Delta}
\newcommand{\sfpcaparamsubset}[1]{\sfpcaparamset_{#1}}

\newcommand{\sfpcacentreddatum}{\zbsplinepdf_{\zbsplineparams_{\genindexi}} \bayesminus \sfpcamean}

\newcommand{\sfpcanumpcs}{\ell}

\newcommand{\sfpcanummodessubset}{\sfpcaparamsubset{\nummodesparam}}
\newcommand{\sfpcanummodesset}{\mathcal{K}}

\newcommand{\priorsfpcaparamrestricted}{\condprob{\sfpcaparam}{\nummodesparam}}
\newcommand{\likelihoodsfpcaparam}{\condprob{\dataset}{\sfpcaparam}}
\newcommand{\marginallikelihoodrestricted}{\condprob{\dataset}{\nummodesparam}}
\newcommand{\marginallikelihood}{\priorprob{\dataset}}

\newcommand{\priorjeffreysmassk}{%
    \int_{\sfpcanummodessubset}
    \priorsfpcaparam
    \ \differential{\sfpcaparam}
}

\newcommand{\posteriorjeffreysmassk}{%
    \int_{\sfpcanummodessubset}
    \posteriorsfpcaparam
    \ \differential{\sfpcaparam}
}

\newcommand{\sfpcabayesfactor}{%
    \frac
    {\posteriorjeffreysmassk}
    {\priorjeffreysmassk}
}

\newcommand{\posteriornummodesparam}{\condprob{\nummodesparam}{\dataset}}
\newcommand{\posterioralternativehypothesis}{\condprob{\alternativehypothesis}{\dataset, \nummodesparam, \genindexi}}

\newcommand{\posteriorsfpcaparam}{\condprob{\sfpcaparam}{\dataset}}
\newcommand{\posteriorsfpcaparamk}{\condprob{\sfpcaparam}{\nummodesparam, \dataset}}

\newcommand{\sfpcacoordinatesvec}[1]{\mathbf{b}_{#1}}
\newcommand{\sfpcacoordinate}[2]{\sfpcacoordinatesvec{#1}^{#2}}
\newcommand{\sfpcasamplemat}{\mathbf{C}}
\newcommand{\sfpcasamplecoordinates}[2]{\sfpcasamplemat_{#1#2}}
\newcommand{\sfpcauniteigenvector}[1]{\mathbf{u}_{#1}}
\newcommand{\sqrtbayesinnerproductmatrix}{\bayesinnerproductmatrix^{1/2}}

\newcommand{\excessmassregion}{\mathcal{M}}
\newcommand{\excessmassregionof}[2]{\excessmassregion_{#1, #2}}
\newcommand{\excessmassregionofki}{\excessmassregionof{\nummodesparam}{\genindexi}}
\newcommand{\excessmasslevel}{\eta}

\newcommand{\selectedmodel}[1]{\tilde{\pdf_{#1}}}
\newcommand{\selectedmodelnummodesparam}{\selectedmodel{\nummodesparam}}

\newcommand{\cutpdf}{g}
\newcommand{\cutpdfki}{\cutpdf_{\nummodesparam, \genindexi}}

\newcommand{\savagedickeyparam}{\tau}
\newcommand{\savagedickeymodel}[4]{%
    (#2
    \bayespowering
    \selectedmodel{#3})(#1) \cdot
    \indicator{\excessmassregionof{#3}{#4}}{#1}
}

\newcommand{\hypothesis}{\mathcal{H}}
\newcommand{\nullhypothesis}{\hypothesis_0}
\newcommand{\alternativehypothesis}{\hypothesis_1}

\newcommand{\odds}{\mathrm{Odds}}
\newcommand{\condodds}[2]{\odds(\condrv{#1}{#2})}

\newcommand{\btstestingmodel}{\condprob{\x}{\savagedickeyparam, \nummodesparam, \genindexi}}
\newcommand{\priornullhypothesis}{\condprob{\nullhypothesis}{\nummodesparam, \genindexi}}
\newcommand{\prioralternativehypothesis}{\condprob{\alternativehypothesis}{\nummodesparam, \genindexi}}
\newcommand{\marginallikelihoodalternative}{\condprob{\dataset}{\alternativehypothesis, \nummodesparam, \genindexi}}
\newcommand{\marginallikelihoodnull}{\condprob{\dataset}{\nullhypothesis, \nummodesparam, \genindexi}}
\newcommand{\prioroddsalternative}{\condodds{\alternativehypothesis}{\nummodesparam, \genindexi}}
\newcommand{\prioroddsnull}{\condodds{\nullhypothesis}{\nummodesparam, \genindexi}}
\newcommand{\posteriorsavagedickeyzero}{\condprob{\savagedickeyparam = 0}{\dataset, \nummodesparam, \genindexi}}
\newcommand{\likelihoodsavagedickey}{\condprob{\dataset}{\savagedickeyparam, \nummodesparam, \genindexi}}
\newcommand{\priorsavagedickey}{\condprob{\savagedickeyparam}{\nummodesparam, \genindexi}}

\newcommand{\significancescore}{\condprob{\alternativehypothesis}{\nummodesparam, \dataset}}

\newcommand{\posteriorsavagedickey}{\condprob{\savagedickeyparam}{\dataset, \nummodesparam, \genindexi}}
\newcommand{\posteriorbarsavagedickey}{\condprob{\extendedparam{\savagedickeyparam}}{\dataset, \nummodesparam, \genindexi}}

\newcommand{\harmonicmeanodds}{\condodds{\alternativehypothesis}{\nummodesparam}}

\newcommand{\maxnummodes}{K}

\newcommand{\btszero}{\texttt{BTS0}}
\newcommand{\btsonesample}{\texttt{BTS1S}}
\newcommand{\btsonejeffreys}{\texttt{BTS1J}}
\newcommand{\btsoneuniform}{\texttt{BTS1U}}
\newcommand{\btstwosample}{\texttt{BTS2S}}
\newcommand{\btstwojeffreys}{\texttt{BTS2J}}
\newcommand{\btstwouniform}{\texttt{BTS2U}}
\newcommand{\btsselected}{\btstwouniform}
\newcommand{\kdepizero}{\texttt{PI0}}
\newcommand{\kdepione}{\texttt{PI1}}
\newcommand{\kdepitwo}{\texttt{PI2}}
\newcommand{\kdelscvzero}{\texttt{LSCV0}}
\newcommand{\kdelscvone}{\texttt{LSCV1}}
\newcommand{\kdelscvtwo}{\texttt{LSCV2}}
\newcommand{\kdelscv}{\texttt{SCV}}
\newcommand{\kdeste}{\texttt{STE}}
\newcommand{\gaussianmixture}{\texttt{GM}}
\newcommand{\tautstring}{\texttt{TS}}
\newcommand{\silverman}{\texttt{SI}}
\newcommand{\fishermarron}{\texttt{FM}}
\newcommand{\telepathicbootstrap}{\texttt{EIG}}

\newcommand{\nummodesestimator}{\estimateof{\nummodesparam}}
\newcommand{\btsnummodesestimator}[1]{\nummodesestimator_{\mathrm{\acrshort*{bts}},#1}}
\newcommand{\btsnummodesestimatorzero}{\btsnummodesestimator{0}}
\newcommand{\btsnummodesestimatorone}{\btsnummodesestimator{1}}
\newcommand{\btsnummodesestimatortwo}{\btsnummodesestimator{2}}

\newcommand{\nummixturecomponents}{K}
\newcommand{\mixturecomponentmean}{\mu}
\newcommand{\mixturecomponentmeans}{\boldsymbol{\mixturecomponentmean}}
\newcommand{\mixturecomponentstd}{\sigma}
\newcommand{\mixturecomponentvariances}{\boldsymbol{\mixturecomponentstd}^2}
\newcommand{\mixtureweights}{\boldsymbol{\mixtureweight}}

\newcommand{\firstmodel}{\texttt{M21}}
\newcommand{\secondmodel}{\texttt{M22}}
\newcommand{\thirdmodel}{\texttt{M23}}
\newcommand{\fourthmodel}{\texttt{M24}}
\newcommand{\fifthmodel}{\texttt{M25}}

\newcommand{\numexperimentreplicationsparam}{m}
\newcommand{\numexperimentreplications}{200}
\newcommand{\method}{\mathtt{M}}
\newcommand{\numberofmethods}{M}
\newcommand{\nummodesestimatorof}[2]{\nummodesestimator_{#1,#2}}
\newcommand{\outcome}{\omega}
\newcommand{\outcomeof}[2]{\outcome_{#1,#2}}
\newcommand{\accuracy}{\varpi}
\newcommand{\significancelevel}{\alpha}
\newcommand{\mcnemarsignificancelevel}{0.01}

\newcommand{\numtestbeds}{T}
\newcommand{\numsamplesizes}{S}

\newcommand{\numtestbedstimessamplesizes}{\numtestbeds \times \numsamplesizes}
\newcommand{\ranking}{\mathcal{R}}
\newcommand{\finalranking}{\ranking}
\newcommand{\intermediaterankings}{%
    \ranking_1, \dots, \ranking_{\numtestbedstimessamplesizes}
}

\input{01-3-configuration.tex}

\newacronym{nom}{NoM}{number of modes}

\newacronym[
	plural=pdfs,
	firstplural=probability density functions]{pdf}{pdf}
{probability density function}

\newacronym[
	plural=cdfs,
	firstplural=cumulative distribution function]{cdf}{cdf}
{cumulative distribution function}

\newacronym{iid}{i.i.d.}{independent and identically distributed}

\newacronym[
	plural=KDEs,
	firstplural=kernel density estimators]{kde}{KDE}
{\textit{kernel density estimator}}

\newacronym{bts}{BTS}{\textit{Bayesian taut spline}}

\newacronym{clr}{CLR}{\textit{centred log-ratio}}

\newacronym{mcmc}{MCMC}{Markov chain Monte Carlo}

\newacronym[
	plural=PCs,
	firstplural=principal components]{pc}{PC}
{principal component}

\newacronym{sfpca}{SFPCA}
{\textit{simplicial functional principal component analysis}}

\newacronym{dic}{DIC}{\textit{deviance information criterion}}

\newacronym{pi}{PI}{\textit{plug-in}}

\newacronym{lscv}{LSCV}{\textit{least-squares cross-validation}}

\newacronym{ste}{STE}{\textit{solve-the-equation}}

\newacronym{scv}{SCV}{\textit{smoothed cross-validation}}

\newacronym{mlb}{MLB}{\textit{Major League Baseball}}

\newacronym{mph}{mph}{miles per hour}

\newacronym{sm}{SM}{supplementary material}

\begin{document}

\ifnum\journalflag=0
    \ifnum\journalflag=0
    \title{\papertitle}
\fi

\date{}

\hypersetup{
    pdftitle={\papertitle},
}

    \ifnum\journalflag=0
    \author{\joseechacon\footjoseechacon}
    \address{\footaddressjoseechaon}
    \email{\footjoseechacon\emailjoseechacon \ \Letter}
    \thanks{\footorcidjoseechacon}

    \author{\javierfdezserrano\footjavierfdezserrano}
    \address{\footaddressjavierfdezserrano}
    \email{\footjavierfdezserrano\emailjavierfdezserrano}
    \thanks{\footorcidjavierfdezserrano}

    \ifnum\materialflag=1
        \makeatletter
        \def\blfootnote{\xdef\@thefnmark{}\@footnotetext}
        \makeatother

        \AtEndDocument{
            \blfootnote{
                \begin{flushleft}
                    \hspace{1em}\mbox{\textit{Authors}:\space} \textsc{\authors}.
                    \\
                    \bigskip
                    \hspace{1em}\textsc{\footaddressjoseechaon}
                    \\
                    \hspace{1em}\textsc{\footaddressjavierfdezserrano}
                    \\
                    \hspace{1em}\mbox{\textit{E-mail addresses}:\space} \texttt{\emails}.
                    \\
                    \hspace{1em}\footorcidjoseechacon.
                    \\
                    \hspace{1em}\footorcidjavierfdezserrano.
                \end{flushleft}
            }
        }
    \fi
\fi

\hypersetup{
    pdfauthor={\joseechacon, \javierfdezserrano}
}

    \newcommand{\abstractmeta}{%
    The number of modes in a probability density function is representative of the complexity of a model and can also be viewed as the number of subpopulations.
    Despite its relevance, there has been limited research in this area.
    A novel approach to estimating the number of modes in the univariate setting is presented, focusing on prediction accuracy and inspired by some overlooked aspects of the problem: the need for structure in the solutions, the subjective and uncertain nature of modes, and the convenience of a holistic view that blends local and global density properties.
    The technique combines flexible kernel estimators and parsimonious compositional splines in the Bayesian inference paradigm, providing soft solutions and incorporating expert judgment.
    The procedure includes feature exploration, model selection, and mode testing, illustrated in a sports analytics case study showcasing multiple companion visualisation tools.
    A thorough simulation study also demonstrates that traditional modality-driven approaches paradoxically struggle to provide accurate results.
    In this context, the new method emerges as a top-tier alternative, offering innovative solutions for analysts.
}

\begin{abstract}
    \abstractmeta
\end{abstract}

\hypersetup {
    pdfsubject={\abstractmeta},
}

    \ifnum\journalflag=0

    \subjclass[2020]{
        \mscnonparamestimation \ (Primary),
        \mscdensityestimation,
        \mscbayesianinference,
        \mscbayesianproblems,
        \mscfuzziness
    }
\fi

\hypersetup{
    pdfkeywords={%
            \keywordnumberofmodes,
            \keywordbayesianinference,
            \keywordcompositionalspline,
            \keywordkde,
            \keywordmodelselection,
            \keywordmodetesting
        },
}

    \maketitle
\fi

\begin{refsection}
    \section{Introduction}

The concept of \textit{mode} has regained the attention of the research community in recent years~\citep{Chacon2020}.
Density modes are defined as local maxima of a \gls*{pdf}.
As such, they mark regions of relatively high concentration of probability mass.
Consequently, they are essential features in \textit{exploratory data analysis}, pointing out new phenomena~\citep{AriasCastro2022, AmeijeirasAlonso2018}.
Notable application examples include the silica composition of meteors in geology~\citep{Good1980}, the distribution of net income in econometrics~\citep{Marron1992}, and the thickness of stamps in philately~\citep{Izenman1988}.
Even so, since the true \gls*{pdf} is unknown, estimating it from possibly scarce data is prone to generating \textit{spurious} modes that can be confused with actual discoveries~\citep{Good1980, Minnotte1997}.

Concerning modes, two primary considerations emerge: the determination of their quantity and their spatial locations.
The latter subsumes the former, but the techniques and hypotheses vary~\citep{Chacon2020}.
The \gls*{nom} is an integer-valued statistical functional, deceptively leading one to believe that \textit{counting} them is more straightforward and less compelling than \textit{locating} them.
However, a result by~\citet{Donoho1988} states the impossibility of bounding from above this quantity with certain confidence if the underlying finite sample comes from a \textit{genuinely} nonparametric distribution.
Moreover, from a practical standpoint, the \gls*{nom} corresponds to the number of groups in many clustering methodologies~\citep{Chacon2018, Cuevas2000}.

\begin{figure}
    \centering
    \begin{subfigure}[b]{\figurewidth}
        \centering
        \includegraphics[width=\subfigurewidth]{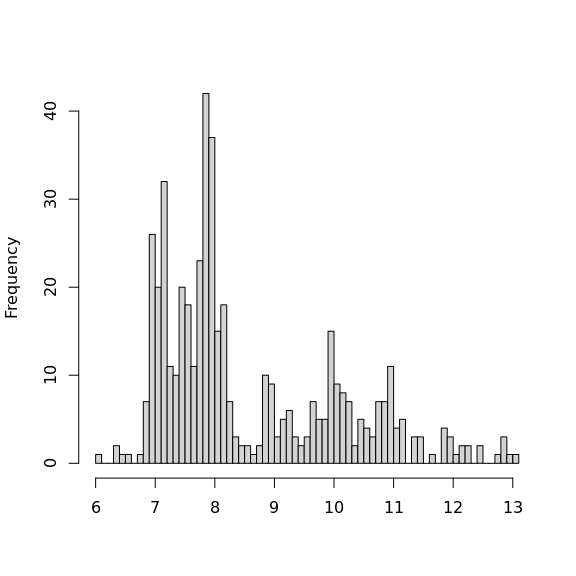}
        \caption{Hidalgo sample}
        \label{fig:hidalgo-example-sample}
    \end{subfigure}
    \begin{subfigure}[b]{\figurewidth}
        \centering
        \includegraphics[width=\subfigurewidth]{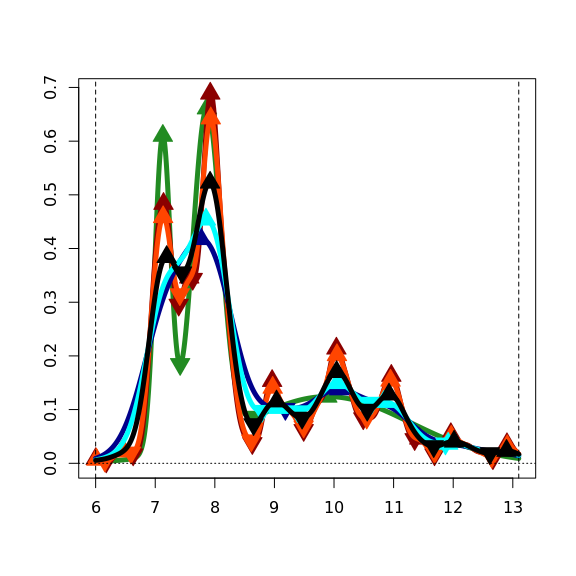}
        \caption{Several \glspl*{pdf}}
        \label{fig:hidalgo-example-pdfs}
    \end{subfigure}
    \begin{subfigure}[b]{\figurewidth}
        \centering
        \includegraphics[width=\subfigurewidth]{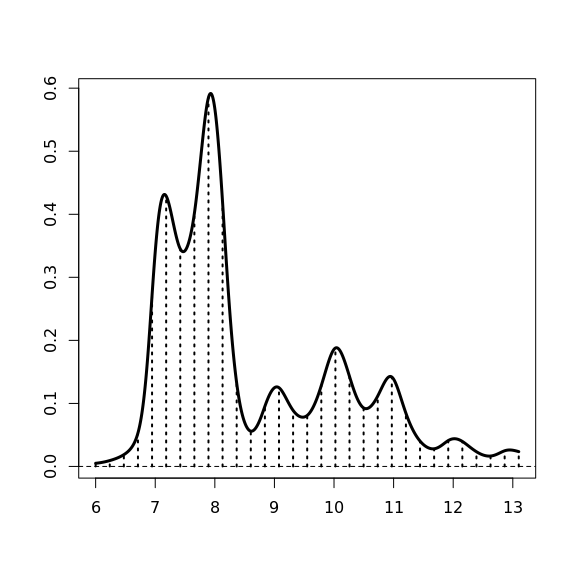}
        \caption{\gls*{bts} solution}
        \label{fig:hidalgo-example-bts}
    \end{subfigure}
    \caption{%
        The Hidalgo stamps data~\citep{Izenman1988}.
        The bar chart on the left displays the sample, which comprises \numhidalgostamps{} measurements of stamp thickness in hundredths of a millimetre.
        Several \glspl*{pdf} for that sample are shown on the right.
        There are some \glsfirstplural*{kde} with different bandwidth selectors, such as \glsfirst*{pi}, \glsfirst*{lscv} and \glsfirst*{ste}, the first two of which count on variations targeting the $\kdederivativeorder$-th \gls*{pdf} derivative~\citep{Chacon2013}: $\piselector{\kdederivativeorder}$ and $\lscvselector{\kdederivativeorder}$.
        Namely, the \glspl*{pdf} are $\piselector{0}$ (black, 7 modes), $\piselector{1}$ (cyan, 5 modes), $\piselector{2}$ (dark blue, 2 modes), \acrshort*{ste} (orange, 9 modes), $\lscvselector{0}$ (red, 11 modes), and a Gaussian mixture (green, 3 modes).
        The bottom picture shows our \gls*{bts} solution with seven modes based on 32 spline basis functions.
    }
    \label{fig:hidalgo-example}
\end{figure}

Consider the Hidalgo stamps data, consisting of \numhidalgostamps{} measurements of stamp thicknesses~\citep{Izenman1988}.
The raw data histogram is depicted in \figurename~\ref{fig:hidalgo-example-sample}.
The issue from 1872 comprises a mix of thicknesses deriving from several extinct paper types and manufacturing processes in various factories~\citep{Fisher2001}.
Philatelists are willing to trace the number of sources of that collection to measure the value of the stamp, which calls for estimating the \gls*{nom}.
Interestingly, the problem is open to this day.
As \figurename~\ref{fig:hidalgo-example-pdfs} shows, many sensible \gls*{pdf} estimators, such as the \gls*{kde}, provide different solutions.
Although each new modality approach is traditionally tested on this dataset, no consensus answer exists~\citep{AmeijeirasAlonso2018}.
The \gls*{bts} method introduced in this paper outputs the \gls*{pdf} in \figurename~\ref{fig:hidalgo-example-bts} with seven modes, quantifying the uncertainty of several alternatives.

\ifnum\journalflag=0
    \paragraph{Goals}
\fi

This paper addresses the classical problem of estimating the \gls*{nom} in the univariate setting.
We advance the development of new ideas targeting some overlooked themes in modality.
Namely, we stress the convenience of structure in the solutions, the uncertain and subjective nature of modes, and the desirability of a holistic view combining global and local properties.

The resulting novel \gls*{bts} proposal allows incorporating prior knowledge regarding the number and significance of modes while balancing data fitting and model complexity in \textit{soft} solutions, gaining valuable insights along the way.
The method will be tested and compared with existing alternatives from the literature in a thorough simulation study based on accuracy.

\ifnum\journalflag=0
    \paragraph{Highlights}

    \begin{itemize}[itemsep=0pt, leftmargin=2em]
        \item Combining kernel estimators and compositional splines benefits mode exploration
\item Dimensionality reduction with one principal component captures the essential modes
\item Bayesian inference admits expert knowledge regarding modality in soft solutions
\item The new proposal and other generic methods outperform classic modality approaches

    \end{itemize}
\fi

\ifnum\journalflag=0
    \paragraph{Related work}
\fi

Modality research has evolved along several branches.
Although \gls*{bts} is influenced by many of the approaches in this section, its philosophy is reminiscent of the \textit{penalised likelihood} method by~\citet{Good1980}, which arguably had no straight continuation.

One of the all-time classics is the \textit{critical bandwidth} test by~\citet{Silverman1981}.
The critical bandwidth $\criticalbandwidth$ is the smallest $\bandwidth$ at which a \gls*{kde} with a Gaussian kernel has at most $\numberofmodes$ modes.
If $\criticalbandwidth$ turns out large after \textit{bootstrapping}, the null hypothesis of $\numberofmodes$ modes is rejected in favour of more than $\numberofmodes$.
\citet{Mammen1992} refined some asymptotic results by~\citeauthor{Silverman1981}, adjusted the scaling of the bootstrap to increase the power of the test~\citep[see also][]{Fisher1994}, and later provided an extension~\citep{Fisher2001}.
Lastly,~\citet{Hall2001} also suggested a recalibration of the p-values.

\begin{sloppypar*}
    An alternative perspective based on the \textit{mass} of the mode was presented by~\citet{Muller1991}.
    Assuming $\numberofmodes$ modes, the \textit{excess mass} $\empiricalexcessmass{\numberofmodes}{\excessmasslevelmuller}$ is the maximal sum of deviations between the empirical distribution and $\excessmasslevelmuller$ times the Lebesgue measure, where the maximum is taken over all sequences of $\numberofmodes$ disjoint sets~\citep{Cheng1998}.
    A large value of $\sup_{\excessmasslevelmuller > 0} \empiricalexcessmass{\numberofmodes}{\excessmasslevelmuller} - \empiricalexcessmass{\numberofmodes - 1}{\excessmasslevelmuller}$ rejects $\numberofmodes - 1$ modes in favour of $\numberofmodes$.
    This test statistic for $\numberofmodes = 2$ is equivalent in the univariate case to the \textit{dip} test by~\citet{Hartigan1985}~\citep[see also][]{Cheng1998, Cheng1999, Hall2001}.
    \citet{Polonik1995, Polonik1995a} further developed the idea of excess mass.
    Since then, several calibration methods have been proposed~\citep{Cheng1998,Cheng1999, AmeijeirasAlonso2018, AmeijeirasAlonso2021}.
\end{sloppypar*}

\citet{Minnotte1993} introduced \textit{mode trees}.
Using a Gaussian kernel, the sequence of \glspl*{kde} as $\bandwidth$ decreases depicts the \textit{branching} of minor modes from larger ones, forming a tree-like diagram.
The original version can be extended with visual hints of the location of antimodes and \textit{bumps}, and the mass and significance of each mode~\citep{Minnotte1997}.
\citet{Minnotte1998} investigated the use of jittering and bootstrapping techniques to build a \textit{mode forest}, which helps to overcome sampling variability.

\citet{Chaudhuri1999} developed SiZer, a graphical tool to examine \glspl*{kde} at different scales.
SiZer explores modes attending to statistically significant sign changes in the first derivative of the \gls*{kde} at each $\bandwidth$.
\citet{Chaudhuri2002} later promoted evaluating second derivatives to complement vanilla SiZer.
The multi-scale view was also explored by~\citet{Duembgen2008}.
\citet{Genovese2016} introduced a mode testing technique, which estimates the \gls*{nom} as the maximum number of significant ones at any scale $\bandwidth$.
Finally,~\citet{Sommerfeld2017} used \textit{topological data analysis} to assess modes from a twofold scale-\textit{lifetime} perspective.

Following~\citet[p. 79]{Hartigan1985},~\citet{Davies2004} proposed the \textit{taut string}, a piecewise-linear \gls*{cdf} confined on an $\tautstringradius$-radius tube surrounding the empirical \gls*{cdf}~\citep[see also][]{Davies2009}.
The hyperparameter $\tautstringradius > 0$ controls the amount of smoothing, with the \gls*{nom} monotonically increasing as $\tautstringradius$ decreases.
Interestingly, the taut string minimises the \gls*{nom} under the $\tautstringradius$-radius constraint.
A more recent proposal fitting \glspl*{pdf} with a fixed \gls*{nom} in a dynamical programming scheme is developed in~\citet{AriasCastro2022}.

\ifnum\journalflag=0
    \paragraph{Outline}
\fi

Section~\ref{sec:preliminaries} provides some technical preliminaries.
Then, our novel proposal is presented in Section~\ref{sec:method}.
A real-world sports analytics application is explored in Section~\ref{sec:case-study} for illustrative purposes.
The simulation study testing and comparing the performance of the new method to existing alternatives comes in Section~\ref{sec:simulation-study}.
Finally, Section~\ref{sec:discussion} discusses the achievements, points of improvement and possibilities of the new method.

    \section{Preliminaries}

\label{sec:preliminaries}

The following lines define concepts such as modes, antimodes and modal regions, making our assumptions explicit.
We will also introduce the Bayesian inference notation used across all the stages of \gls*{bts}.
Finally, we will recall some elements of \glspl*{kde}, Bayes spaces and compositional splines necessary for our construction.

In what follows, assume a non-empty and finite dataset $\dataset = \{\sampleobs_1, \dots, \sampleobs_{\samplesize}\} \subset \reals$ consisting of $\samplesize$ \gls*{iid} random variable realisations.

\ifnum\journalflag=0
    \paragraph{Modes}
\fi

All intermediate and final univariate \glspl*{pdf} of \gls*{bts} are assumed to be \textit{Morse functions}, i.e., functions whose critical points are nondegenerate~\citep{Chacon2015}, avoiding \glspl*{pdf} with flat parts where defining modes is more involved~\citep[Section 4.1]{Donoho1988}.
We will also suppose that all the \glspl*{pdf} $\pdf$ are as smooth as required and have compact support $\intervalab$, i.e., $\pdf(\x) > 0$ if $\x \in \intervalab$, and $\pdf(\x) = 0$ for all $\x \notin \intervalab$.
Consequently, $\pdf$ will have finitely many critical points~\citep[p. 530]{Chacon2015}.

We define \textit{modes} as \textit{local maxima} of the \gls*{pdf}, i.e., points $\mode \in \intervalab$ such that $\pdf(\mode) > \pdf(\x)$ for all $\x \in [\mode - \modeneighbourepsilon, \mode + \modeneighbourepsilon] \setminus \{\mode\}$, for some $\modeneighbourepsilon > 0$.
Such definition includes, but is not limited to, critical points $\x$ where $\pdf'(\x) = 0$ and $\pdf''(\x) < 0$.
In particular, the boundaries $\intervala$ and $\intervalb$ could be modes despite not obeying the latter derivative constraints (remember that $\pdf$ is defined over $\reals$).
Similarly, we define \textit{antimodes} as \textit{local minima} of the \gls*{pdf}, i.e., points $\antimode \in \intervalab$ such that $\pdf(\antimode) < \pdf(\x)$ for all $\x \in [\antimode - \modeneighbourepsilon, \antimode + \modeneighbourepsilon] \setminus \{\antimode\}$, for some $\modeneighbourepsilon > 0$.

Under the previous assumptions, the minimum \gls*{nom} is one.
In general, if there are $\numberofmodes \geq 1$ modes, the number of antimodes will be $\numberofmodes - 1$.
For $\numberofmodes > 1$, the modes and antimodes alternate as
$
    \intervala
    \leq
    \mode_1
    <
    \antimode_1
    <
    \mode_2
    <
    \dots
    <
    \mode_{\numberofmodes - 1}
    <
    \antimode_{\numberofmodes - 1}
    <
    \mode_{\numberofmodes}
    \leq
    \intervalb
$.
Hence, we see that $\intervalab$ can be expressed as a union of intervals containing exactly one mode: the \textit{modal regions}.
More formally, if we write $\antimode_0 = \intervala$ and $\antimode_{\numberofmodes} = \intervalb$, the $\genindexi$-th modal region containing $\mode_{\genindexi}$ is $[\antimode_{\genindexi - 1}, \antimode_{\genindexi}]$, where $\mode_{\genindexi}$ is allowed to coincide with one of the boundaries only if $\genindexi = 1$ or $\genindexi = \numberofmodes$.
With that notation, when $\numberofmodes = 1$, the unique modal region is $[\antimode_0, \antimode_1] = [\intervala, \intervalb]$.

\ifnum\journalflag=0
    \paragraph{Bayesian inference}
\fi

Continuous and discrete parameters and hypotheses are treated as random variables in Bayesian inference.
We will use the same notation $\prob(\funcarg)$ for \glspl*{pdf} and probability mass functions, inferring the continuous/discrete nature from the context.

Consider a parameter vector $\parametervec = (\parameters) \in \reals^{\numparameters}$.
We will refer to its Bayesian prior as $\priorprobparameters$.
The model \gls*{pdf} conditioning on $\parametervec$ will be denoted as $\condprob{\funcarg}{\parameters}$.
Then,
$
    \likelihood
    =
    \prod_{\sampleobs \in \dataset}
    \condprob{\sampleobs}{\parameters}
$
is the likelihood of $\parametervec$ given data $\dataset$, and
$
    \condprob{\parameters}{\dataset}
    \propto
    \likelihood
    \times
    \priorprobparameters
$
is the posterior of $\parametervec$ given $\dataset$, where ``$\propto$'' indicates proportionality, i.e., equality except for a normalising constant making the left-hand side a \gls*{pdf} or a probability mass function.
To specify a value $\parametervalue \in \reals$ for some parameter $\parameter_{\genindexi}$, we will replace the parameter ``$\parameter_{\genindexi}$'' with ``$\parameter_{\genindexi} = \parametervalue$'' in the notations above.

\ifnum\journalflag=0
    \paragraph{Kernel density estimation}
\fi

The \gls*{kde} for the dataset $\dataset$ based on a Gaussian kernel $\kernel$ and bandwidth $\bandwidth > 0$ is the function $\mappingsets{\reals}{(0, \infty)}$ defined by
\begin{equation}
    \label{eq:kernel-estimator}
    \kde(\x)
    =
    \frac{1}{\samplesize\bandwidth}
    \sum_{\genindexi = 1}^{\samplesize}
    \kernel \left(
    \frac{\x - \sampleobs_{\genindexi}}{\bandwidth}
    \right)
    \,.
\end{equation}
The parameter $\bandwidth$ controls the amount of smoothing.
Fixing $\dataset$, a large value of $\bandwidth$ hides prominent features, while a small one produces spurious \textit{wiggles}.
A Gaussian kernel is needed to ensure continuous tracking of modes~\citep{Silverman1981,Minnotte1993}.

\ifnum\journalflag=0
    \paragraph{Bayes spaces}
\fi

The \textit{Bayes space} $\bayesspace$ of positive \glspl*{pdf} with bounded support $\intervalab$ and square-integrable logarithm has a Hilbert space structure~\citep{Machalova2020} originating from mapping $\pdf \in \bayesspace$ to the household $\hilbertltwospace$ via the \gls*{clr} transformation
\begin{equation}
    \label{eq:clr-transformation}
    \clrpdf(\x)
    =
    \logpdf(\x)
    -
    \frac{1}{\intervalb - \intervala}
    \intbetweenab
    \logpdf(\y) \ \differential{\y}
    \,.
\end{equation}
Since $\zerointegralconstraint{\clrpdf}$, if we consider the subset complying with that zero-integral constraint, $\hilbertltwospacezerointegral = \{\logpdfvar \in \hilbertltwospace : \zerointegralconstraint{\logpdfvar}\}$, then $\bayesspace$ is isometric to $\hilbertltwospacezerointegral$ with inverse
$
    \clrinv{\logpdfvar}(\x)
    =
    \exp \logpdfvar(\x) /
    \intbetweenab \exp \logpdfvar(\y) \ \differential{\y}
$.
The vector space operations of sum (perturbation) and scalar multiplication (powering) are, respectively,
$
    (\pdf \bayesperturbation \pdfsecondary) (\x)
    =
    \pdf(\x) \pdfsecondary(\x) /
    \intbetweenab \pdf(\y) \pdfsecondary(\y) \differential{\y}
$,
and
$
    (\bayesscalar \bayespowering \pdf)(\x)
    =
    \pdf(\x)^{\bayesscalar} /
    \intbetweenab \pdf(\y)^{\bayesscalar} \differential{\y}
$,
where $\pdf, \pdfsecondary \in \bayesspace$ and $\bayesscalar \in \reals$.
Let us denote $\pdf \bayesminus \pdfsecondary = \pdf \bayesperturbation (-1) \bayespowering \pdfsecondary$.
Similarly, the inner product between $\pdf, \pdfsecondary \in \bayesspace$ is given by
\begin{equation}
    \label{eq:inner-product}
    \bayesinnerproduct{\pdf}{\pdfsecondary}
    =
    \frac{1}{2(\intervalb - \intervala)}
    \intbetweenab
    \intbetweenab
    \log \frac{\pdf(\x)}{\pdf(\y)}
    \log \frac{\pdfsecondary(\x)}{\pdfsecondary(\y)}
    \ \differential{\x}
    \hspace*{0.3ex} \differential{\y}
    \,,
\end{equation}
that is,
$
    \bayesinnerproduct{\pdf}{\pdfsecondary}
    =
    \intbetweenab
    \clr{\pdf}(\x) \clr{\pdfsecondary}(\x)
    \ \differential{\x}
    =
    \ltwoinnerproduct{\clr{\pdf}}{\clr{\pdfsecondary}}
$.

\ifnum\journalflag=0
    \paragraph{Compositional splines}
\fi

Let us fix a polynomial degree $\splinedegree \geq 3$, enabling a non-flat second derivative.
Given $\zbsplinedim \geq \splinedegree$, we can construct a $\zbsplinedim$-dimensional vector subspace $\zbsplinepdfspace \subset \bayesspace$ consisting of \glspl*{pdf} whose \gls*{clr} transformations~\eqref{eq:clr-transformation} are $\splinedegree$-th degree spline functions $\functiondef{\spline}{\intervalab}{\reals}$ such that $\zerointegralconstraint{\spline}$.
The elements in $\zbsplinepdfspace$ are known as \textit{compositional splines}.

Explicitly constructing compositional splines relies on a variant of the usual B-splines, known as ZB-splines, obeying the zero-integral constraint~\citep{Machalova2020}.
Let us split $\intervalab$ through arbitrary knots $\intervala = \zbsplineknot_0 < \dots < \zbsplineknot_{\zbsplinedim - \splinedegree + 1} = \intervalb$.
Then, we can define $\zbsplinedim$ ZB-spline basis functions $\functiondef{\zbasisspline_1, \dots, \zbasisspline_{\zbsplinedim}}{\intervalab}{\reals}$ built up from $\splinedegree$-th degree polynomials joining at the knots with maximal $\rminusonecontinuousclass$ smoothness.

Let us define the ZB-spline $\zbspline = \sum_{\genindexi = 1}^{\zbsplinedim} \zbsplineparam_{\genindexi} \zbasisspline_{\genindexi}$, for $\zbsplineparams = (\zbsplineparam_1, \dots, \zbsplineparam_{\zbsplinedim}) \in \realszbsplinedim$.
The \gls*{clr} back-transformed \gls*{pdf} $\zbsplinepdf_{\zbsplineparams} = \clrinv{\zbspline} = \bigbayesperturbation_{\genindexi = 1}^{\zbsplinedim} (\zbsplineparam_{\genindexi} \bayespowering \clrinv{\zbasisspline_{\genindexi}})$ is a compositional spline.
Therefore, we can define $\zbsplinepdfspace = \{\zbsplinepdf_{\zbsplineparams} : \zbsplineparams \in \realszbsplinedim\}$.
Note that $\zbsplinepdfspace$ is isometric to $\realszbsplinedim$ via $\mappingelements{\zbsplinepdf_{\zbsplineparams}}{\zbsplineparams}$ with inner product $\bayesinnerproduct{\zbsplinepdf_{\zbsplineparams_1}}{\zbsplinepdf_{\zbsplineparams_2}} = \transpose{\zbsplineparams_1} \bayesinnerproductmatrix \zbsplineparams_2$, for $\zbsplineparams_1, \zbsplineparams_2 \in \realszbsplinedim$, where the symmetric matrix $\bayesinnerproductmatrix$ has entries
$
    \bayesinnerproductmatrix_{\genindexi\genindexj}
    =
    \ltwoinnerproduct{\zbasisspline_{\genindexi}}{\zbasisspline_{\genindexj}}
$.

Every $\pdf \in \bayesspace$ can be approximated in the least squares sense within $\zbsplinepdfspace$ provided equidistant knots and a sufficiently large $\zbsplinedim$ are used.
The value $\splinedegree$ is less critical, making $\splinedegree = 3$ a widespread choice.
Let us fix a fine grid with evenly spaced points $\intervala = \splinepointx_1 < \dots < \splinepointx_{\numsplinepoints} = \intervalb$.
For all $\pdf \in \bayesspace$ and every smoothing penalty factor $\oneminuscurvaturepenalty \in (0, 1)$, we define the curvature-penalised quadratic loss of a parameter vector $\zbsplineparams \in \realszbsplinedim$ as
\begin{equation}
    \label{eq:zbspline-loss}
    \zbsplineloss
    =
    \oneminuscurvaturepenalty
    \sum_{\genindexi = 1}^{\numsplinepoints}
    \{\clrpdf(\splinepointx_{\genindexi}) - \zbspline(\splinepointx_{\genindexi})\}^2
    +
    (1 - \oneminuscurvaturepenalty)
    \intbetweenab \zbspline''(\x)^2 \ \differential{\x}
    \,.
\end{equation}
The sum term relates to data fitting, whereas the integral value amounts to the \textit{total curvature} of the spline, representing its complexity.
Then, the compositional spline $\compositionalsplineapprox{\oneminuscurvaturepenalty}{\pdf}$ approximating $\pdf \in \bayesspace$ with smoothing penalty factor $\oneminuscurvaturepenalty \in (0, 1)$ is
\begin{equation}
    \label{eq:compositional-spline-approx}
    \compositionalsplineapprox{\oneminuscurvaturepenalty}{\pdf}
    =
    \zbsplinepdf_{\approxzbsplineparamsof{\pdf}{\oneminuscurvaturepenalty}}
    \,, \ \text{where} \ \
    \approxzbsplineparamsof{\pdf}{\oneminuscurvaturepenalty}
    =
    \argmin_{\zbsplineparams \in \realszbsplinedim}
    \zbsplineloss
    \,.
\end{equation}

Problem~\eqref{eq:compositional-spline-approx} has a straightforward linear algebra solution~\citep{Machalova2020}.
The larger $\oneminuscurvaturepenalty$, the closer~\eqref{eq:compositional-spline-approx} is to $\pdf$.
In contrast, smaller values of $\oneminuscurvaturepenalty$ produce smoother solutions.
If $\pdf$ is positive but does not have bounded support $\intervalab$, we will assume that $\compositionalsplineapprox{\oneminuscurvaturepenalty}{\pdf} \equiv \compositionalsplineapprox{\oneminuscurvaturepenalty}{\truncatedpdf}$, where the \gls*{pdf} $\truncatedpdf$ satisfies $\truncatedpdf(\x) \propto \pdf(\x) \cdot \indicator{\intervalab}{\x}$, since $\clrpdf$ and $\clr{\truncatedpdf}$ coincide over $\intervalab$.

    \section{The Bayesian taut spline (BTS) method}

\label{sec:method}

This section introduces the new \gls*{bts} method, encompassing several steps.
First, an \textit{exploration} phase scans the \gls*{kde} mode tree while building spline models.
Next, an \textit{analysis} phase summarises the splines into a linear one-parameter model.
Then, a \textit{selection} phase obtains probabilities and \glspl*{pdf} for each $\numberofmodes$-mode hypothesis.
Finally, a \textit{testing} phase evaluates the modes of each representative $\numberofmodes$-modal \gls*{pdf}.

We refer the reader to the \mysupplement{} for implementation details and discussing the pillars of the proposal: model structure, Bayesian inference and a holistic view joining global and local density properties.

\begin{figure}
    \centering
    \begin{subfigure}[b]{\figurewidth}
        \centering
        \includegraphics[width=\subfigurewidth]{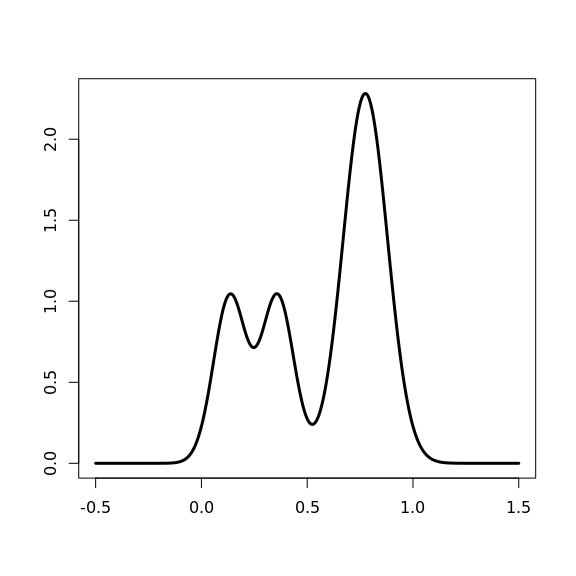}
        \caption{Theoretical \gls*{pdf}}
        \label{fig:theoretical-pdf}
    \end{subfigure}
    \begin{subfigure}[b]{\figurewidth}
        \centering
        \includegraphics[width=\subfigurewidth]{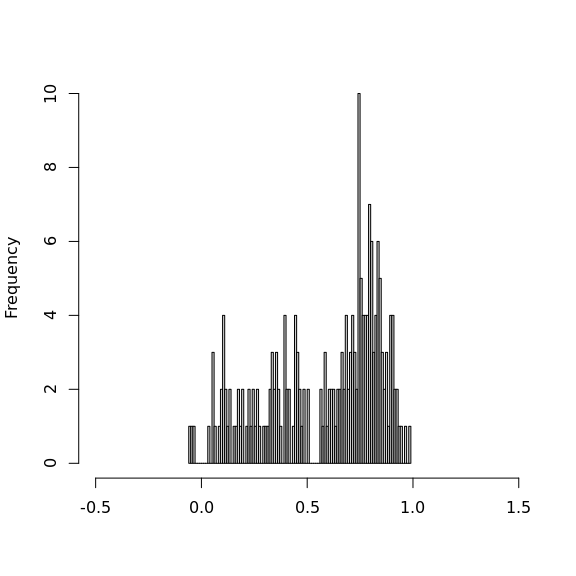}
        \caption{Random sample}
        \label{fig:underlying-sample}
    \end{subfigure}
    \caption{
        An illustrative test-bed for estimating the \gls*{nom}.
        The \testbedid{} mixture model \gls*{pdf} from~\citet{AmeijeirasAlonso2018} is shown on the left.
        A histogram of a random sample of size \testbedsamplesize{} from that model is displayed on the right.
    }
    \label{fig:test-bed-example}
\end{figure}

We will illustrate the different stages through the example in \figurename~\ref{fig:test-bed-example}.
\figurename~\ref{fig:theoretical-pdf} depicts a complicated case with three modes, where two are little pronounced.
The \gls*{nom} in the resulting random sample in \figurename~\ref{fig:underlying-sample} is debatable, for one of the two apparent modes on the left block seems isolated and weak.
We shall see how \gls*{bts} handles the situation.

\subsection{Exploration}

\label{sec:exploration}

\Gls*{bts} starts off exploring modes through a combination of \glspl*{kde} and compositional splines.
Informally speaking, for every \gls*{kde} bandwidth $\bandwidth > 0$ and smoothing penalty factor $\oneminuscurvaturepenalty \in (0, 1)$, we first build the \gls*{kde} $\kde$ from $\dataset$ and then fit a compositional spline to $\kde$ using least squares with penalty $\oneminuscurvaturepenalty$.
More formally, for every pair of smoothing parameters $\halphapair \in (0, \infty) \times (0, 1)$, we define a \gls*{pdf} model for $\dataset$ conditioning on an uncertain pair $\halphapair$ as
\begin{equation}
    \label{eq:kde-spline-model}
    \btsexplorationmodel
    =
    \compositionalsplineapprox{\oneminuscurvaturepenalty}{\kde}
    \,.
\end{equation}
Since $\kde$ in~\eqref{eq:kde-spline-model} has unbounded support, we will set $\intervala = \min(\dataset)$ and $\intervalb = \max(\dataset)$ in practice.
In this context, let us abbreviate for later use $\approxzbsplineparamshalpha \equiv \approxzbsplineparamsof{\kde}{\oneminuscurvaturepenalty}$, where the right-hand side is as in~\eqref{eq:compositional-spline-approx}.

The \gls*{kde} is known to be sensitive to outliers in the tails~\citep{Chacon2013}.
Spurious modes, almost invisible to the human eye, might appear despite the curvature penalisation of \gls*{bts} in cases with severely isolated points.
We propose removing the outliers before building the \gls*{kde} in~\eqref{eq:kde-spline-model}, similarly to~\citet[Section 5.2]{Chacon2013}.
Using~\eqref{eq:kernel-estimator} as a pilot, one can calculate the probability masses of all the modal regions and remove those $\sampleobs \in \dataset$ linked to underrepresented modes.
After removal, the resulting subsample is again fed to~\eqref{eq:kernel-estimator}.
This preprocessing step shall be implicitly assumed in~\eqref{eq:kde-spline-model} in the upcoming sections.

Model~\eqref{eq:kde-spline-model} is not a typical parametric \gls*{pdf}, for $\halphapair$ are a \textit{proxy} of the underlying spline coordinates $\zbsplineparams$.
Since direct likelihood-based estimation tends to \textit{overfit} data, especially with discretised data, we suggest imposing strong regularity conditions through a prior $\btsexplorationprior$.
Then, exploring the spline space and the subsequent modes comes down to evaluating the posterior $\posteriorhalpha$.
This can be achieved via \gls*{mcmc} simulation~\citep{Kass1995}, securing a posterior sample $\smoothingparamssample = \{(\bandwidth_{\genindexi}, \oneminuscurvaturepenalty_{\genindexi})\}_{\genindexi = 1}^{\smoothingsamplesize}$.

We propose an \textit{improper} prior
$
    \btsexplorationprior
    \equiv
    \priorprob{\bandwidth}
    \times
    \priorprob{\oneminuscurvaturepenalty}
    \times
    \priorprobnummodesparam
    \times
    \priorprob{\curvatureparam}
$
factoring the shape parameters alongside quantities representing the complexity of the true \gls*{pdf}, such as $\nummodesparam \equiv \nummodesparam(\bandwidth, \oneminuscurvaturepenalty)$, the \gls*{nom} of $\btsexplorationmodel$, and $\curvatureparam \equiv \curvatureparam(\bandwidth, \oneminuscurvaturepenalty)$, the total curvature term in~\eqref{eq:zbspline-loss} with $\zbsplineparams = \approxzbsplineparamshalpha$.
See~\citet[p. 45]{Good1980} for a similar use of an ``improper'' prior in regularisation.
A design with well-known parametric distributions is
\begin{equation}
    \label{eq:exploration-prior-distributions}
    \begin{aligned}
        \bandwidth
         & \followsdistr
        \lognormaldistribution
        {\bandwidthlocationhyperparam}
        {\bandwidthdispersionhyperparam}
        \,,
        \\
        1 - \oneminuscurvaturepenalty
         & \followsdistr
        \betadistribution{1}{\penaltyhyperparam}
        \,,
        \\
        \nummodesparam
         & \followsdistr
        \poissondistribution{1}
        \,,
        \\
        \curvatureparam
         & \followsdistr
        \exponentialdistribution{\curvaturehyperparam}
        \,.
    \end{aligned}
\end{equation}
See the \mysupplement{} for justification of the previous scheme and the selection of $\bandwidthlocationhyperparam$, $\bandwidthdispersionhyperparam$, $\penaltyhyperparam$ and $\curvaturehyperparam$, depending on the case.

The first \gls*{bts} estimator for the \gls*{nom} shall be called the \textit{raw} variant of \gls*{bts}.
It is the most frequent modality in $\smoothingparamssample$, i.e.,
\begin{equation}
    \label{eq:bts0-estimator}
    \btsnummodesestimatorzero
    =
    \argmax_{\nummodesparam \in \naturals}
    \setcardinality{%
        \{\genindexi \in \{1, \dots, \smoothingsamplesize\} : \condprob{\funcarg}{\bandwidth = \bandwidth_{\genindexi}, \oneminuscurvaturepenalty = \oneminuscurvaturepenalty_{\genindexi}} \ \haskmodes\}
    }
    \,,
\end{equation}
where the lower $\nummodesparam$ should be taken for parsimony in case of a draw.
The same criterion shall also apply to the upcoming estimators.
The counts in~\eqref{eq:bts0-estimator} can also be normalised to obtain probabilities for each $\nummodesparam$ hypothesis, gaining more insight.
We will further process $\smoothingparamssample$ in the analysis phase.

\begin{figure}
    \centering
    \includegraphics[width=\figurewidth]{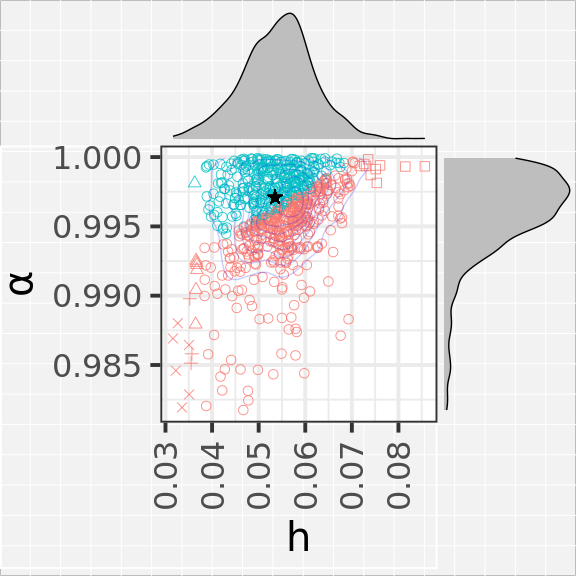}
    \caption{
        \gls*{mcmc} sample from $\posteriorhalpha$ consisting of \testbedmcmcsamplesize{} observations.
        The horizontal and vertical axes represent the $\bandwidth$ and $\oneminuscurvaturepenalty$ components, respectively.
        The points are coloured according to the \gls*{nom} of model~\eqref{eq:kde-spline-model}: red (\percentagered{} of the total, \numberofmodesred{} modes) and blue (\percentageblue{} of the total, \numberofmodesblue{} modes).
        The shape of each point represents the \gls*{nom} of the underlying \gls*{kde}, thus only depending on $\bandwidth$: square (\numberofmodessquare{} modes), circle (\numberofmodescircle{} modes) and triangle (\numberofmodestriangle{} modes), among others.
        The average point is the black star in the middle of the point cloud.
        \Glspl*{kde} for the margins are also provided.
    }
    \label{fig:posterior-h-alpha}
\end{figure}

\figurename~\ref{fig:posterior-h-alpha} shows the results from the exploration phase of \gls*{bts} on the random sample in \figurename~\ref{fig:underlying-sample}.
The two main scenarios, of two and three modes, are considered in the \gls*{mcmc} sample.
The former hypothesis has an inconclusive advantage with \percentagered{} of the \textit{ballots}.
Because of the intended conservatism of this stage, \gls*{bts} sticks to two modes without overlooking three.
\figurename~\ref{fig:posterior-h-alpha} displays some compelling features of \gls*{bts}.
The spline structure and regularisation leave a fourth and a fifth mode in the underlying \gls*{kde} with no effect, restraining excessive complexity.
Similarly, many points correspond to partially \textit{ironed} three-mode \glspl*{kde}.
We can graphically see a neat oblique line separating the two modality hypotheses, meaning \glspl*{kde} and smoothing splines benefit from each other to test modes.

Revisiting the classic Hidalgo example in \figurename~\ref{fig:hidalgo-example}, the above prior design oriented to discrete data leads in~\eqref{eq:bts0-estimator} to $\btsnummodesestimatorzero = 7$ with 100\% probability.
This unanimity ensures that this seven-mode result stands through the upcoming stages.

\subsection{Analysis}

\label{sec:analysis-stage}

The exploration phase of \gls*{bts} could benefit from some improvement.
Since $\posteriorhalpha$ is not genuinely parametric, we depend on finite bivariate \gls*{mcmc} samples.
Robust approximations require long \gls*{mcmc} chains, and solving~\eqref{eq:kde-spline-model} at each iteration is time-consuming.
On the other hand, although $\posteriorhalpha$ captures modes thoroughly, mode trees count on just one parameter, $\bandwidth$, making them easier to interpret~\citep{Minnotte1993}.
The analysis stage of \gls*{bts} will address both issues.

Analysing $\smoothingparamssample$ is problematic, for $\halphapair$ does not reflect the spline structure.
Instead, let us define $\zbsplineparams_{\genindexi} = \approxzbsplineparamsof{\bandwidth_{\genindexi}}{\oneminuscurvaturepenalty_{\genindexi}}$ for every $(\bandwidth_{\genindexi},\oneminuscurvaturepenalty_{\genindexi}) \in \smoothingparamssample$.
We propose a dimensionality reduction on the \gls*{pdf} sample $\compositionalsplinesample \subset \zbsplinepdfspace$ using \gls*{sfpca}~\citep{Hron2016}.
\gls*{sfpca} works similarly to its non-functional counterpart, expressing the compositional spline space in terms of orthonormal axes $\sfpcapceigenfun{1}, \dots, \sfpcapceigenfun{\zbsplinedim} \in \zbsplinepdfspace$ (each one, a \gls*{pc}) with respective variances $\sfpcaeigenvalue{1} \geq \dots \geq \sfpcaeigenvalue{\zbsplinedim}$.
Then, the usual next step is to simplify a sample by projecting each centred functional datum over a few \glspl*{pc} that retain a considerable proportion of the original variability.
See the \mysupplement{} for a brief note on \gls*{sfpca} in our context.

Beyond convenience and simplicity, a single \gls*{pc} provides enough power in the \gls*{bts} context.
See the \mysupplement{} for justification.
We propose a refined \gls*{pdf} model for $\dataset$ depending on a single parameter $\sfpcaparam$ through
\begin{equation}
    \label{eq:sfpca-model}
    \btssfpcamodel
    =
    \sfpcamodel{\sfpcaparam}
    \,,
\end{equation}
where $\sfpcamean = (1 / \smoothingsamplesize) \bayespowering \bigbayesperturbation_{\genindexi = 1}^{\smoothingsamplesize} \zbsplinepdf_{\zbsplineparams_{\genindexi}}$ is a sample average \gls*{pdf}, and $\sfpcastdev = \firststandarddeviation \bayespowering \sfpcapceigenfun{1}$ is a standard deviation \gls*{pdf} along the first \gls*{pc}.
Even though $\sfpcaparam$ could range over the whole $\reals$, extrapolation beyond certain limits leads to nonsensical solutions.
We can assess sensible values for $\sfpcaparam$ by inspecting the centred projections along $\sfpcapceigenfun{1}$, i.e., the scores $\sfpcascore_{\genindexi} = \bayesinnerproduct{\sfpcacentreddatum}{\sfpcapceigenfun{1}}$.
If we write $\sfpcaparam_{\genindexi} = \sfpcascore_{\genindexi} / \firststandarddeviation$, we can build a \textit{support} for $\sfpcaparam$ as $\sfpcaparamset = [\sfpcaparammin, \sfpcaparammax]$, where $\sfpcaparammin = \min_{1 \leq \genindexi \leq \smoothingsamplesize} \sfpcaparam_{\genindexi}$ and $\sfpcaparammax = \max_{1 \leq \genindexi \leq \smoothingsamplesize} \sfpcaparam_{\genindexi}$.
Provided $\sfpcaparam \in \sfpcaparamset$,~\eqref{eq:sfpca-model} represents the modal features in $\posteriorhalpha$.

Since~\eqref{eq:sfpca-model} captures the essential parts of~\eqref{eq:kde-spline-model} by construction, one can adopt a more \textit{objective} approach for the prior distribution.
The Jeffreys prior is a standard alternative in a univariate setting~\citep{Bernardo1994}.
It assigns equal probabilities to regions of the statistical manifold~\eqref{eq:sfpca-model} with the same \textit{volume}, conveying the \textit{principle of indifference}.
Additionally, the Jeffreys prior is invariant to reparametrisation.
Also, note that since $\sfpcaparamset$ is bounded, the Jeffreys prior is \textit{proper}, integrating up to one.
Finally, straightforward calculations for this type of prior yield, for $\sfpcaparam \in \sfpcaparamset$,
\begin{equation}
    \label{eq:jeffreys-prior}
    \priorsfpcaparam
    \propto
    \sqrt{\variance{\log \sfpcastdev(\genrvx_{\sfpcaparam})}}
    \,,
    \ \mathrm{where} \
    \genrvx_{\sfpcaparam} \followsdistr \btssfpcamodel
    \,,
\end{equation}
and $\variance{\funcarg}$ denotes the variance of a random variable.
We can see that $\priorsfpcaparam > 0$ for all $\sfpcaparam \in \sfpcaparamset$, since $\sfpcastdev$ can never be constant and $\condprob{\x}{\sfpcaparam} > 0$ for all $\x \in \intervalab$.

We will refer to the combination of ~\eqref{eq:sfpca-model} and~\eqref{eq:jeffreys-prior} as the \gls*{sfpca} model of \gls*{bts}.
\Gls*{bts} considers a second Bayesian inference on the \gls*{sfpca} model to estimate the \gls*{nom}.

\begin{figure}
    \centering
    \begin{subfigure}[b]{\figurewidth}
        \centering
        \includegraphics[width=\subfigurewidth]{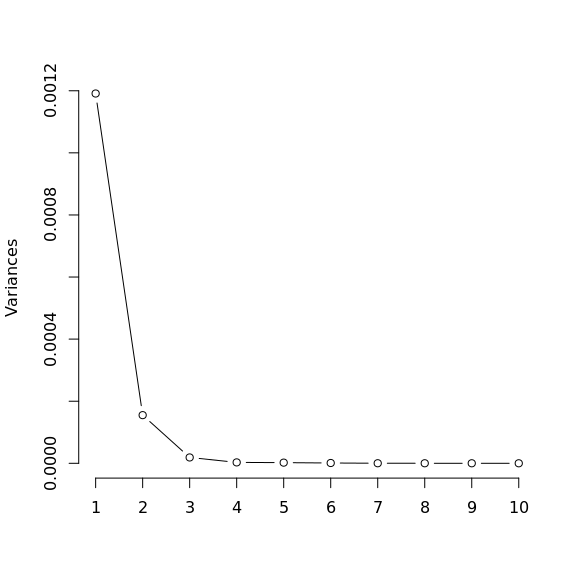}
        \caption{\gls*{sfpca} variances scree plot}
        \label{fig:sfpca-scree-plot}
    \end{subfigure}
    \begin{subfigure}[b]{\figurewidth}
        \centering
        \includegraphics[width=\subfigurewidth]{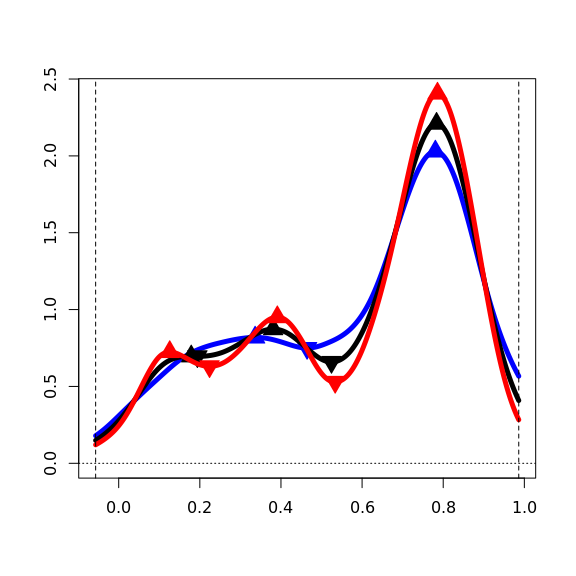}
        \caption{Mode of variation of $\sfpcapceigenfun{1}$}
        \label{fig:sfpca-splines}
    \end{subfigure}
    \caption{
        \gls*{sfpca} analysis phase results.
        The scree plot of the ordered \glspl*{pc} against their variances is presented on the left.
        Some representative \glspl*{pdf} in the \gls*{sfpca} model~\eqref{eq:sfpca-model} are displayed on the right: the mean $\sfpcamean$ (black, 3 modes), the lower bound $\sfpcamodel{\sfpcaparammin}$ (blue, 2 modes) and the upper bound $\sfpcamodel{\sfpcaparammax}$ (red, 3 modes).
    }
    \label{fig:sfpca-results}
\end{figure}

\figurename~\ref{fig:sfpca-results} shows the results of the analysis stage after the exploration in \figurename~\ref{fig:posterior-h-alpha}.
First, \figurename~\ref{fig:sfpca-scree-plot} plots the \glspl*{pc} against their variances.
The sudden drop is characteristic, justifying keeping one dimension.
Then, \figurename~\ref{fig:sfpca-splines} sheds some light on the shape of the splines in the sample in \figurename~\ref{fig:posterior-h-alpha}.
All the uncertainty gravitates around whether the left block splits into two modes.
The mean $\sfpcamean$ has three modes, one of which is barely noticeable as an incipient \textit{shoulder} that finally emerges as the left-most mode of $\sfpcamodel{\sfpcaparammax}$.

\subsection{Selection}

\label{sec:model-selection}

\begin{sloppypar*}
    The \gls*{sfpca} model encompasses several hypotheses about the \gls*{nom} of the \gls*{pdf}.
    Indeed, the prior~\eqref{eq:jeffreys-prior} is an \textit{encompassing prior}~\citep{Klugkist2005}, expressing the relative uncertainty of $\sfpcaparam$ under different domain restrictions.
    More precisely, let us define, for $\nummodesparam \in \naturals$,
    $
        \sfpcanummodessubset
        =
        \{
        \sfpcaparam \in \sfpcaparamset
        :
        \btssfpcamodel
        \ \haskmodes
        \}
    $,
    i.e., the set of parameters $\sfpcaparam$ producing $\nummodesparam$ modes,
    and let $\sfpcanummodesset = \{\nummodesparam \in \naturals : \sfpcanummodessubset \neq \emptyset\}$ be the set of all the reachable modality numbers.
    For every $\nummodesparam \in \sfpcanummodesset$, we can define a prior ensuring $\nummodesparam$ modes as
    \begin{equation}
        \label{eq:jeffreys-prior-restricted}
        \priorsfpcaparamrestricted
        \propto
        \priorsfpcaparam
        \cdot
        \indicator{\sfpcanummodessubset}{\sfpcaparam}
        \,.
    \end{equation}
    Combining the new priors~\eqref{eq:jeffreys-prior-restricted} with the original model~\eqref{eq:sfpca-model}, we obtain a class of \gls*{sfpca} $\nummodesparam$-modal \glspl*{pdf}.
    This way, \gls*{bts} translates estimating the \gls*{nom} into a Bayesian model selection problem.
\end{sloppypar*}

Note that $\sfpcanummodesset$ is finite since the \glspl*{pdf}~\eqref{eq:sfpca-model} are derived from splines.
From our experience, we can virtually assure that every $\nummodesparam \in \sfpcanummodesset$ shows up as the modality of some \gls*{pdf} in the exploratory sample $\smoothingparamssample$.
However, it is not uncommon for some modalities observed in the exploration phase to be absent in $\sfpcanummodesset$.
This loss of information, which is welcomed for parsimony, roots in dimensionality reduction.

The \textit{marginal likelihood} of the $\nummodesparam$-modal prior, $\nummodesparam \in \sfpcanummodesset$, and the total prior, respectively, are
$
    \marginallikelihoodrestricted
    =
    \int_{\sfpcaparamset}
    \likelihoodsfpcaparam
    \times
    \priorsfpcaparamrestricted
    \ \differential{\sfpcaparam}
$
and
$
    \marginallikelihood
    =
    \int_{\sfpcaparamset}
    \likelihoodsfpcaparam
    \times
    \priorsfpcaparam
    \ \differential{\sfpcaparam}
$.
After assigning prior probabilities $\priorprobnummodesparam$ to each hypothesis $\nummodesparam \in \naturals$, we can calculate the posterior probabilities~\citep{Wasserman2000} as
$
    \posteriornummodesparam
    \propto
    \marginallikelihoodrestricted
    \times
    \priorprobnummodesparam
$,
assuming that $\posteriornummodesparam = 0$ whenever $\nummodesparam \notin \sfpcanummodesset$.

Calculating $\marginallikelihoodrestricted$ is generally intractable~\citep{Kass1995}.
Instead, we can use the fact that, in the case of an encompassing prior~\citep[p. 60]{Klugkist2005}, for every $\nummodesparam \in \sfpcanummodesset$, we have
$
    \marginallikelihoodrestricted / \marginallikelihood
    =
    \posteriorjeffreysmassk / \priorjeffreysmassk
$,
where the denominator on the right-hand side is positive.
This ratio of marginal likelihoods is the \textit{Bayes factor}.
Plugging it into the equation for $\posteriornummodesparam$ above, we get
\begin{equation}
    \label{eq:posterior-prob-k-simplified}
    \posteriornummodesparam
    \propto
    \sfpcabayesfactor
    \times
    \priorprobnummodesparam
    \,.
\end{equation}

Based on~\eqref{eq:posterior-prob-k-simplified}, a sensible choice for the prior is $\priorprobnummodesparam = \priorjeffreysmassk$.
However, $\priorprobnummodesparam \propto 1$ is usually preferred~\citep{Kass1995,Klugkist2005}.
Furthermore, we can consider the prior probabilities from the posterior sample $\smoothingparamssample$ in the exploration phase.

Maximising the posterior probability~\eqref{eq:posterior-prob-k-simplified}, we finally get the \textit{processed} estimator variant of \gls*{bts}:
\begin{equation}
    \label{eq:bts1-estimator}
    \btsnummodesestimatorone
    =
    \argmax_{\nummodesparam \in \naturals}
    \posteriornummodesparam
    \,.
\end{equation}
In addition to the new insights gained through the process, the latter estimator will generally produce better results than~\eqref{eq:bts0-estimator}, as demonstrated later in Section~\ref{sec:simulation-study}.
The testing phase of \gls*{bts}, introduced in Section~\ref{sec:testing} below, will further refine~\eqref{eq:bts1-estimator}, providing yet more information.

One can easily verify that, for all $\nummodesparam \in \sfpcanummodesset$, the Bayesian update to the $\nummodesparam$-modality prior~\eqref{eq:jeffreys-prior-restricted} is
$
    \posteriorsfpcaparamk
    \propto
    \posteriorsfpcaparam
    \cdot
    \indicator{\sfpcanummodessubset}{\sfpcaparam}
$.
Hence, the prior $\posteriorsfpcaparam$ in~\eqref{eq:posterior-prob-k-simplified} can be used to obtain a representative \gls*{pdf} of the $\nummodesparam$-modality hypothesis, typically summarising $\posteriorsfpcaparamk$~\citep{Bernardo1994}.
However, neither the \textit{posterior predictive distribution} nor the \textit{posterior mean} preserves the \gls*{nom}.
A common choice in such circumstances is the \textit{posterior median}, which belongs to $\sfpcanummodessubset$.
The median Bayes estimator can then be interpreted as the minimiser of the average distance induced by~\eqref{eq:inner-product}.

\begin{figure}
    \centering
    \begin{subfigure}[b]{\figurewidth}
        \centering
        \includegraphics[width=\subfigurewidth]{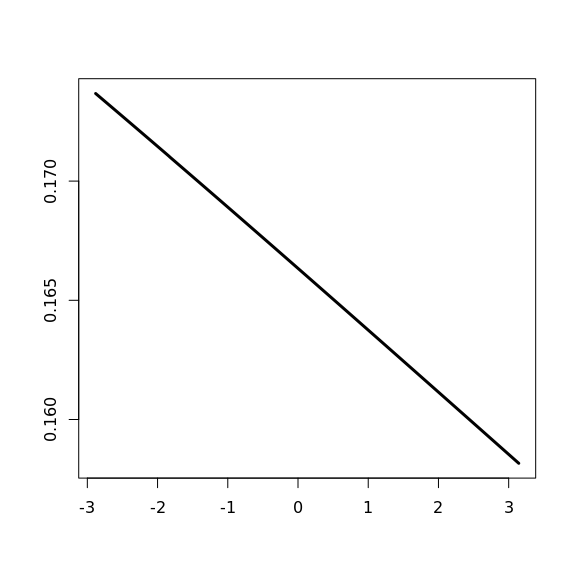}
        \caption{Jeffreys prior}
        \label{fig:sfpca-jeffreys-prior}
    \end{subfigure}
    \begin{subfigure}[b]{\figurewidth}
        \centering
        \includegraphics[width=\subfigurewidth]{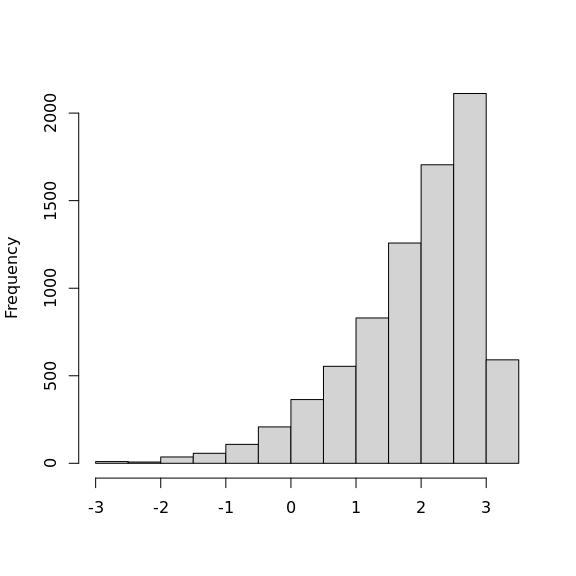}
        \caption{Posterior sample}
        \label{fig:sfpca-posterior}
    \end{subfigure}
    \caption{
        Second Bayesian inference on the \gls*{sfpca} model.
        The Jeffreys prior \gls*{pdf}~\eqref{eq:jeffreys-prior} is shown on the left.
        A histogram of a sample from $\posteriorsfpcaparam$ consisting of \sfpcasamplesize{} observations is shown on the right.
    }
    \label{fig:sfpca-prior-posterior}
\end{figure}

\begin{figure}
    \centering
    \begin{subfigure}[b]{\figurewidth}
        \centering
        \includegraphics[width=\subfigurewidth]{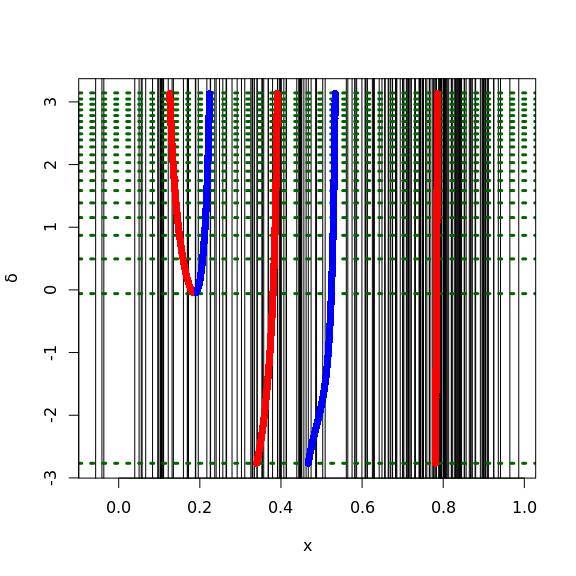}
        \caption{Mode tree}
        \label{fig:sfpca-mode-tree}
    \end{subfigure}
    \begin{subfigure}[b]{\figurewidth}
        \centering
        \includegraphics[width=\subfigurewidth]{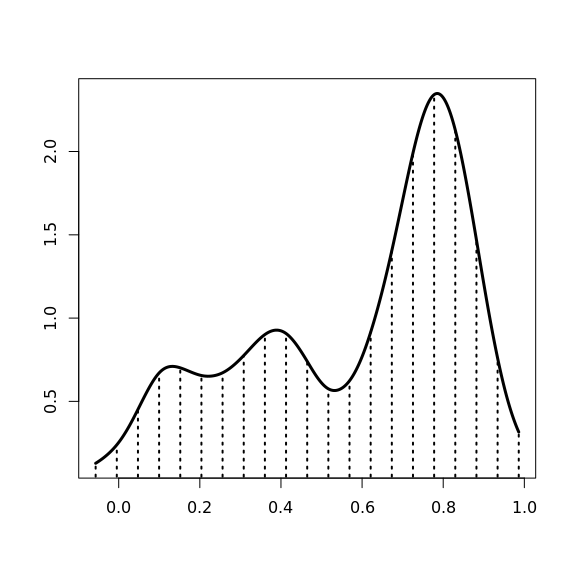}
        \caption{Posterior median spline}
        \label{fig:sfpca-final-spline}
    \end{subfigure}
    \caption{
        Results of the \gls*{bts} method.
        The left-hand side picture shows a mode tree with $\sfpcaparam$ values on the vertical axis.
        Red points correspond to modes, whereas the blue ones are antimodes.
        The black vertical lines highlight the original sample observations in \figurename~\ref{fig:underlying-sample}.
        The dashed green horizontal lines represent uniform quantiles of the posterior sample in \figurename~\ref{fig:sfpca-posterior}.
        The figure on the right displays the posterior median model, which has \testbedmediansplinenummodes{} modes and a \gls*{clr} dimension of \testbedsplinedimension{}.
    }
    \label{fig:model-selection}
\end{figure}

\figurename~\ref{fig:sfpca-prior-posterior} shows the before and after of the second inference.
\figurename~\ref{fig:sfpca-jeffreys-prior} depicts the Jeffreys prior~\eqref{eq:jeffreys-prior}, which exhibits a slight slope, mildly penalising the more complex \glspl*{pdf}.
Then, \figurename~\ref{fig:sfpca-posterior} shows a histogram of the posterior sample displaying a varying slope in the opposite direction.
The Bayesian update for the \gls*{sfpca} model favours the three-mode hypothesis more than the exploratory inference, as confirmed by \figurename~\ref{fig:model-selection}.
Regardless of the choice of $\priorprobnummodesparam$ (among those proposed here), \gls*{bts} correctly selects three modes based on posterior probabilities ranging between \testbedminfinalprob{} and \testbedmaxfinalprob{}.

\subsection{Testing}

\label{sec:testing}

The \gls*{nom} in \gls*{bts} measured \textit{model complexity} in Section~\ref{sec:exploration} and was a \textit{feature} encompassing a range of parameter values in Section~\ref{sec:model-selection}.
Both correspond to \textit{global} views.
However, modes are defined locally, which calls for testing their significance with nearby data.
We thus invoke the concept of \textit{excess mass region}.

For every $\nummodesparam \in \sfpcanummodesset$ modality hypothesis, let us call $\selectedmodelnummodesparam$ the median $\nummodesparam$-modal \gls*{pdf} selected in Section~\ref{sec:model-selection} with probability $\posteriornummodesparam$.
Then, for the $\genindexi$-th mode $\mode_{\genindexi}$ of the model $\selectedmodelnummodesparam$, we define the $\genindexi$-th excess mass region as
$
    \excessmassregionofki
    =
    \{\x \in [\antimode_{\genindexi - 1}, \antimode_{\genindexi}] :
    \selectedmodelnummodesparam(\x)
    \geq
    \excessmasslevel_{\genindexi}
    \}
$,
where $\antimode_{\genindexi - 1}$ and $\antimode_{\genindexi}$ are the minimum and the maximum, respectively, of the modal region corresponding to $\mode_{\genindexi}$, and
$
    \excessmasslevel_{\genindexi}
    =
    \max\{
    \selectedmodelnummodesparam(\x)
    :
    \x \in \{\antimode_{\genindexi - 1},
    \antimode_{\genindexi}\}
    \setminus \{\mode_{\genindexi}\}
    \}
$.
Naturally, $\mode_{\genindexi} \in \excessmassregionofki$.
These excess mass regions are a slight variation of those considered in~\citet{Minnotte1993} to allow modes to appear at modal region boundaries.

The rationale behind excess mass regions is as follows.
Let $\distance{\pdf_1}{\pdf_2}$ denote the distance between two functions $\pdf_1, \pdf_2 \in \lpspace{1}{\intervalab}$.
Then, define, for $\x \in \intervalab$,
\begin{equation*}
    \cutpdfki(\x)
    =
    \begin{cases}
        \excessmasslevel_{\genindexi},   & \text{if} \ \x \in \excessmassregionofki                       \\
        \selectedmodelnummodesparam(\x), & \text{if} \ \x \in \intervalab \setminus \excessmassregionofki
    \end{cases}
    \,.
\end{equation*}
As mentioned in~\citet{Minnotte1993}, for any continuous $\function$ without local maxima in $[\antimode_{\genindexi - 1}, \antimode_{\genindexi}]$, we have $\distance{\selectedmodelnummodesparam}{\cutpdfki} \leq \distance{\selectedmodelnummodesparam}{\function}$.
In other words, $\cutpdfki$ is the least distinguishable function from $\selectedmodelnummodesparam$ with such constraints.
This suggests that, to test the existence of $\mode_{\genindexi}$, assuming $\selectedmodelnummodesparam$ is true, the most conservative null hypothesis of non-existence should be based on $\cutpdfki$.
Since $\cutpdfki$ only differs from $\selectedmodelnummodesparam$ in being constant over $\excessmassregionofki$, we see an identification between removing $\mode_{\genindexi}$ and being \textit{uniform} over $\excessmassregionofki$.

Given a $\nummodesparam$-modality hypothesis, we can \textit{zoom in} on the $\genindexi$-th excess mass region to test the strength of evidence in favour of $\mode_{\genindexi}$ locally.
Let us define a one-parameter power variation of $\selectedmodelnummodesparam$ around $\excessmassregionofki$ through
\begin{equation}
    \label{eq:savage-dickey-model}
    \btstestingmodel
    \propto
    \savagedickeymodel{\x}{\savagedickeyparam}{\nummodesparam}{\genindexi}
    \,,
\end{equation}
where $\savagedickeyparam \in [0, \infty)$.
We note that~\eqref{eq:savage-dickey-model} coincides with the $\savagedickeyparam$-powering of the restriction of $\selectedmodelnummodesparam$ to $\excessmassregionofki$, which is retrieved for $\savagedickeyparam = 1$.
Powering has the effect of intensifying or weakening the mode at $\mode_{\genindexi}$.
For a large $\savagedickeyparam$, the probability mass concentrates around $\mode_{\genindexi}$, while for a small $\savagedickeyparam$, the mass is spread over $\excessmassregionofki$, reaching uniformity for $\savagedickeyparam = 0$.
This observation motivates a mode significance test in which the null hypothesis $\nullhypothesis$ of the non-existence of the mode is paired with $\savagedickeyparam = 0$, while $\savagedickeyparam \neq 0$ represents the alternative hypothesis $\alternativehypothesis$ of existence.
This way, testing the mode becomes a Bayesian single-parameter value testing problem.

The Savage-Dickey method~\citep{Wagenmakers2010} assigns probabilities to $\nullhypothesis$ and $\alternativehypothesis$, weighing the evidence in data $\dataset$ with prior knowledge.
Namely, we can transform prior odds into posterior odds via
\begin{equation}
    \label{eq:prior-to-posterior-odds}
    \frac
    {\posterioralternativehypothesis}
    {\condprob{\nullhypothesis}{\dataset, \nummodesparam, \genindexi}}
    =
    \frac
    {\marginallikelihoodalternative}
    {\marginallikelihoodnull}
    \times
    \frac
    {\prioralternativehypothesis}
    {\priornullhypothesis}
    \,.
\end{equation}
We will denote the prior odds for $\alternativehypothesis$ as $\prioroddsalternative = \prioralternativehypothesis / \priornullhypothesis$, and those in favour of $\nullhypothesis$ as $\prioroddsnull = \prioroddsalternative^{-1}$.
If $\prioroddsalternative = 0$, we shall assume $\posterioralternativehypothesis = 0$.
At this point, one usually takes
$
    \priornullhypothesis
    =
    \prioralternativehypothesis
$,
making the posterior odds equal a ratio of marginal likelihoods, the \textit{Bayes factor}, where
\begin{equation}
    \label{eq:savage-dickey-marginal-likelihood}
    \marginallikelihoodalternative
    =
    \int_0^{\infty}
    \likelihoodsavagedickey
    \times
    \priorsavagedickey
    \ \differential{\savagedickeyparam}
    \,,
\end{equation}
and
$
    \marginallikelihoodnull
    =
    \condprob{\dataset}{\savagedickeyparam = 0, \nummodesparam, \genindexi}
$.
The likelihood $\likelihoodsavagedickey$ is evaluated only with data in the excess mass region, i.e.,
$
    \likelihoodsavagedickey
    =
    \prod\nolimits_{\x \in \dataset \cap \excessmassregionofki}
    \btstestingmodel
$.
Whenever $\dataset \cap \excessmassregionofki = \emptyset$, we can assume $\posterioralternativehypothesis = 0$.
A natural prior density choice is
$
    \condrv{\savagedickeyparam}{\nummodesparam, \genindexi}
    \followsdistr
    \exponentialdistribution{1}
$,
regardless of $\nummodesparam$ and $\genindexi$.
Hence, $\savagedickeyparam$ has mean one and a global mode at zero, favouring the null hypothesis and leaning the average scenario towards $\selectedmodelnummodesparam$.

Computing~\eqref{eq:savage-dickey-marginal-likelihood} is generally complicated~\citep{Kass1995}.
However, defining the posterior
$
    \posteriorsavagedickey
    \propto
    \likelihoodsavagedickey
    \times
    \priorsavagedickey
$,
the ratio of marginal likelihoods can be expressed as the ratio between the prior and the posterior at $\savagedickeyparam = 0$, i.e.,
\begin{equation}
    \label{eq:savage-dickey-bayes-factor}
    \frac
    {\marginallikelihoodalternative}
    {\marginallikelihoodnull}
    =
    \frac
    {\condprob{\savagedickeyparam = 0}{\nummodesparam, \genindexi}}
    {\posteriorsavagedickeyzero}
    =
    \posteriorsavagedickeyzero^{-1}
    \,,
\end{equation}
assuming an exponential prior in the numerator.
Plugging~\eqref{eq:savage-dickey-bayes-factor} into~\eqref{eq:prior-to-posterior-odds} and solving, we finally get
\begin{equation}
    \label{eq:savage-dickey-alternative-probability}
    \posterioralternativehypothesis
    =
    \left[
        1 + \frac{\posteriorsavagedickeyzero}{\prioroddsalternative}
        \right]^{-1}
    \,.
\end{equation}

The probability~\eqref{eq:savage-dickey-alternative-probability} expresses the significance of the mode $\mode_{\genindexi}$ in the $\nummodesparam$-modal \gls*{pdf} $\selectedmodelnummodesparam$.
The classical theory of Bayes factors for hypothesis testing establishes some reference values to interpret~\eqref{eq:savage-dickey-alternative-probability}.
See~\citet[Section 3.2]{Kass1995} for a scale in terms of~\eqref{eq:savage-dickey-bayes-factor}.
This methodology gives the benefit of the doubt to the null hypothesis $\nullhypothesis$, requiring a value of~\eqref{eq:savage-dickey-alternative-probability} well above $0.50$ to reject it.
Such conservatism would also be justified in our case for parsimony.

The probabilities~\eqref{eq:savage-dickey-alternative-probability} for the left, centre and right modes in \figurename~\ref{fig:sfpca-final-spline} are \probleftmode{}, \probcentremode{} and \probrightmode{}, respectively, assuming all the prior odds equal one.
These results agree with our intuition that the left mode is relatively weak, and only that to the right is beyond doubt.

Applying the same procedure on the \gls*{pdf} in \figurename~\ref{fig:hidalgo-example-bts} of the Hidalgo problem, we obtain probabilities from left to right: $0.63$, $0.99$, $0.83$, $0.96$, $0.91$, $0.51$ and $0.75$.
According to the scale in~\citet{Kass1995}, only the second to fifth modes would be significant, yielding four modes instead of the original seven.
It is easy to see why by looking at \figurename~\ref{fig:hidalgo-example-sample}.
The modes at 7 and 12 are \textit{fractured}, while few data points support the one at 13.
From this one-dimensional mode testing perspective, \gls*{bts} would agree with the four-mode solution by~\citet{AmeijeirasAlonso2018}.

Notwithstanding, limiting our analysis to the probabilities~\eqref{eq:savage-dickey-alternative-probability} has several drawbacks.
Setting a decision threshold might be too rigid and arbitrary, and, more importantly, discarding modes on the grounds of significance equates to forgetting the global results making the local analysis~\eqref{eq:savage-dickey-alternative-probability} possible.
In fact, for the Hidalgo problem, the previous stages of \gls*{bts} led unequivocally to seven modes.

We propose to round \gls*{bts} off by combining the \textit{global} probabilities obtained in Section~\ref{sec:model-selection} with new \textit{scores} for each $\nummodesparam$ based on~\eqref{eq:savage-dickey-alternative-probability}.
Let $\significancescore$ be the new score for the $\nummodesparam$ hypothesis, representing its \textit{overall} significance.
If the latter plays the part of the likelihood and we take $\posteriornummodesparam$ as prior probabilities, the ``Bayesian update'' yields the posterior
$
    \condprob{\nummodesparam}{\alternativehypothesis, \dataset}
    \propto
    \significancescore
    \times
    \posteriornummodesparam
$,
assuming that $\condprob{\nummodesparam}{\alternativehypothesis, \dataset} = 0$ whenever $\posteriornummodesparam = 0$.
The latter posterior can be seen as the sum of two compositional vectors.
If $\significancescore = 0$ for all $\nummodesparam$, we should take $\condprob{\nummodesparam}{\alternativehypothesis, \dataset} = \posteriornummodesparam$.
See~\citet{Egozcue2018} for a similar treatment of likelihoods as general \textit{evidence functions}.

Finally, the \textit{refined} estimator variant of \gls*{bts} is defined analogously to~\eqref{eq:bts1-estimator} as
\begin{equation}
    \label{eq:bts2-estimator}
    \btsnummodesestimatortwo
    =
    \argmax_{\nummodesparam \in \naturals}
    \condprob{\nummodesparam}{\alternativehypothesis, \dataset}
    \,.
\end{equation}
If there is only one $\nummodesparam \in \naturals$ such that $\posteriornummodesparam > 0$, the estimators~\eqref{eq:bts1-estimator} and~\eqref{eq:bts2-estimator} coincide.
Therefore, in the case of the Hidalgo problem, despite some modes being non-significant, we have $\btsnummodesestimatortwo = \btsnummodesestimatorone = 7$.

There are many ways to aggregate all the probabilities~\eqref{eq:savage-dickey-alternative-probability} into a single score.
We propose taking the \textit{harmonic mean}.
Amongst the Pythagorean means, it is the most sensitive to the lower outliers while stable against the higher ones.
Let us define $\harmonicmeanodds$ as the harmonic mean of the prior odds for each of the $\nummodesparam$ modes in $\selectedmodelnummodesparam$, i.e., $\harmonicmeanodds = \nummodesparam / \sum_{\genindexi = 1}^{\nummodesparam} \prioroddsalternative^{-1}$.
Also, define the mixture
\begin{equation*}
    \condprob{\savagedickeyparam}{\nummodesparam, \dataset}
    =
    \sum_{\genindexi = 1}^{\nummodesparam}
    \mixtureweight_{\genindexi} \
    \posteriorsavagedickey
    \,,
    \ \mathrm{where} \
    \mixtureweight_{\genindexi}
    =
    \frac
    {\prioroddsnull}
    {\sum_{\genindexj = 1}^{\nummodesparam} \condodds{\nullhypothesis}{\nummodesparam, \genindexj}}
    \,.
\end{equation*}
Then, the harmonic mean $\significancescore = \nummodesparam / \sum_{\genindexi = 1}^{\nummodesparam} \posterioralternativehypothesis^{-1}$ satisfies
\begin{equation*}
    \significancescore
    =
    \left[
        1 +
        \frac
        {\condprob{\savagedickeyparam = 0}{\nummodesparam, \dataset}}
        {\harmonicmeanodds}
        \right]^{-1}
    \,,
\end{equation*}
making it a generalisation of~\eqref{eq:savage-dickey-alternative-probability} for $\nummodesparam$ higher than one.

    \section{Case study}

\label{sec:case-study}

Pitchers are a central part of baseball.
They develop a comprehensive throw repertoire that varies in speed, spin and target.
Thus, classifying a pitch is challenging, even for a well-trained eye.
Nonetheless, to a first approximation, speed takes considerable variability while widely recognised as the main asset of a pitcher.
Therefore, let us study pitching from the univariate perspective of speed.

Knowing the \textit{arsenal} of the opposing pitcher increases the chances of a batter hitting the ball.
In particular, we can gain valuable knowledge from the speed modes.
The larger the \gls*{nom}, the greater the unpredictability of the pitcher.
A scouting report could advise that batters focus their pre-game training and in-game strategy on specific ball speeds corresponding to the modes.

\begin{figure}
    \centering
    \begin{subfigure}[b]{\figurewidth}
        \centering
        \includegraphics[width=\subfigurewidth]{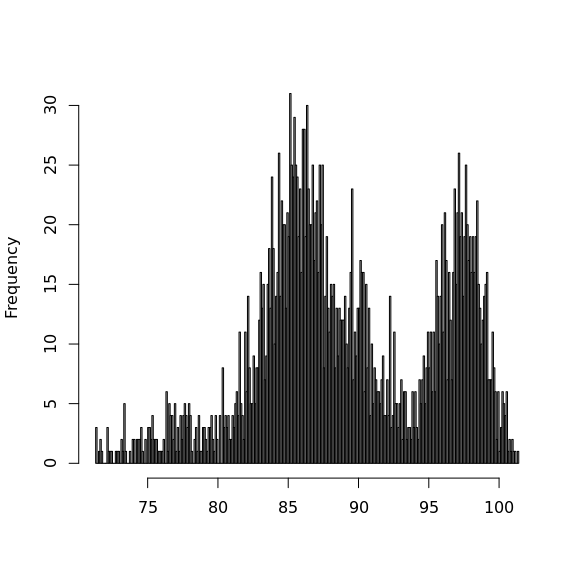}
        \caption{Ohtani's 2022 pitches}
        \label{fig:ohtani-sample}
    \end{subfigure}
    \begin{subfigure}[b]{\figurewidth}
        \centering
        \includegraphics[width=\subfigurewidth]{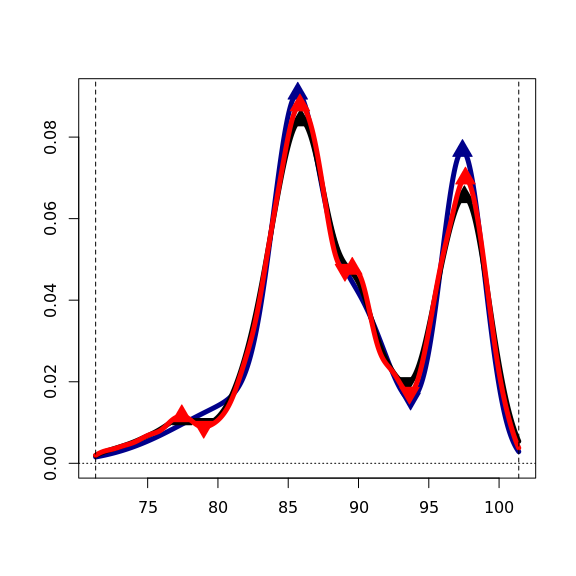}
        \caption{Several \glspl*{pdf}}
        \label{fig:ohtani-pdfs}
    \end{subfigure}
    \caption{%
        Shohei Ohtani's 2022 pitching season data.
        The left-hand side picture shows a bar chart of the sample, consisting of \numohtanipitches{} pitch speeds greater than or equal to 70 \gls*{mph}.
        Several \glspl*{pdf} for that sample are shown on the right.
        Namely, the \glspl*{pdf} are $\piselector{0}$ (red, 4 modes), $\piselector{1}$ (black, 3 modes), and a Gaussian mixture (blue, 2 modes).
    }
\end{figure}

We will look at the pitching speeds of \gls*{mlb} top player Shohei Ohtani in the 2022 season.
See the \mysupplement{} for further details on the data.
\figurename~\ref{fig:ohtani-sample} is a bar chart of the underlying pitches, omitting some outliers below 70 \gls*{mph}.
The \numohtanipitches{}-point sample consists of speed values reported up to 0.1 \gls*{mph}, making up a discretised dataset.
Hence, extra smoothing will be induced in \gls*{bts}, as mentioned in the \mysupplement{}.
A large compositional spline space dimension $\zbsplinedim = 32$ will be employed, as well.

The official interpretation of the pitches by the \gls*{mlb} comprises four modes.
Three can be identified with specific pitch types: the \textit{curveball} (with an average of 78 \gls*{mph}), the \textit{slider} (85 \gls*{mph}) and the \textit{fastball} (97 \gls*{mph}).
The last mode, located at approximately 90 \gls*{mph}, emerges from a mix of \textit{changeups} and \textit{cutters}.
Several preliminary \gls*{pdf} models are depicted in \figurename~\ref{fig:ohtani-pdfs}.
The $\piselector{0}$-variant of the \gls*{kde} points out four modes, whereas $\piselector{1}$ eludes the 90-\gls*{mph} mode.
The parametric Gaussian mixture model is the least expressive with two modes, missing the latter plus the curveball.

\begin{figure}
    \centering
    \begin{subfigure}[b]{\figurewidth}
        \centering
        \includegraphics[width=\subfigurewidth]{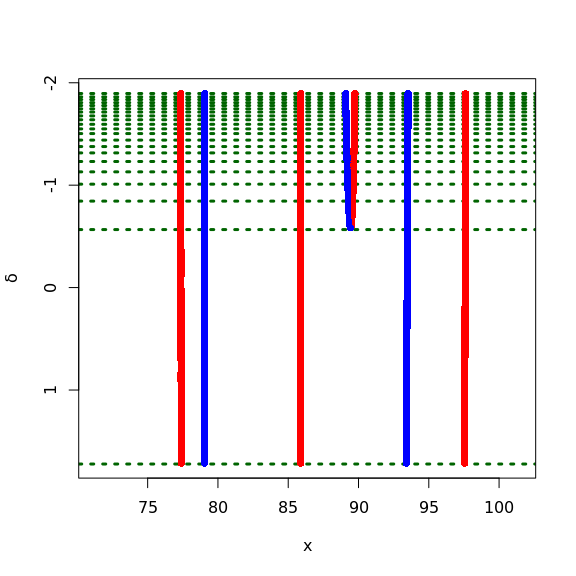}
        \caption{Mode tree}
        \label{fig:ohtani-sfpca-mode-tree}
    \end{subfigure}
    \begin{subfigure}[b]{\figurewidth}
        \centering
        \includegraphics[width=\subfigurewidth]{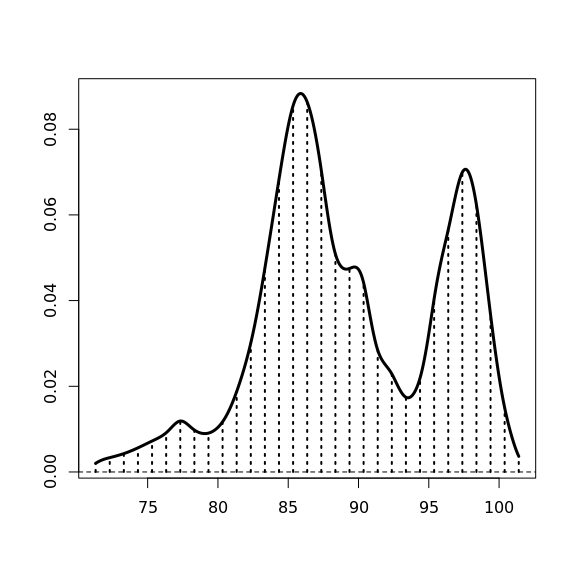}
        \caption{Posterior median spline}
        \label{fig:ohtani-sfpca-final-spline}
    \end{subfigure}
    \caption{%
        Results of the \gls*{bts} method for the \gls*{mlb} case study with the same structure as \figurename~\ref{fig:model-selection}.
        The black vertical lines are omitted from the mode tree on the left to enhance readability under a larger sample size.
        The dimension of the posterior median model with \ohtanimediansplinenummodes{} modes is \ohtanisplinedimension{}.
    }
    \label{fig:ohtani-model-selection}
\end{figure}

The intermediate results of \gls*{bts} are qualitatively very similar to the examples in Section~\ref{sec:method}.
We will go straight to the final results from the selection and testing stages and refer the reader to the \mysupplement{} for the rest.
Depending on the prior probabilities $\priorprobnummodesparam$, we have both $\condprob{\nummodesparam = 4}{\dataset}$ and $\condprob{\nummodesparam = 4}{\alternativehypothesis, \dataset}$ ranging between \ohtaniminfinalprob{} and \ohtanimaxfinalprob{}.
The mode tree weighted with the posterior sample is in \figurename~\ref{fig:ohtani-sfpca-mode-tree}, while the posterior median spline for $\nummodesparam = 4$ is in \figurename~\ref{fig:ohtani-sfpca-final-spline}.
The modes of the latter \gls*{pdf} are located at 77.4, 85.9, 89.7 and 97.6 \gls*{mph}, having significance probabilities of $0.85$, $0.99$, $0.62$ and $0.99$, respectively.

    \section{Simulation study}

\label{sec:simulation-study}

This section demonstrates the effectiveness of our proposal in a thorough comparison with other well-established techniques.
As we will see, \gls*{bts} is a top-tier method according to an overall ranking aggregating results from a broad array of test-beds.
We will also examine under what circumstances each procedure performs best and worst and analyse the distribution of the predictions.
The reader is referred to the \mysupplement{} for extra auxiliary results and setup details.

\subsection{Setup}

\label{sec:setup}

The experimental design is fully described in the following lines.

\ifnum\journalflag=0
    \paragraph{Methods}
\fi

Seven variants of \gls*{bts} are tested in this simulation study:

\begin{itemize}[itemsep=0pt, leftmargin=2em]
    \item \btszero{}: The \textit{raw} \gls*{bts} estimator~\eqref{eq:bts0-estimator}.
    \item \btsonesample{}: The \textit{processed} \gls*{bts} estimator~\eqref{eq:bts1-estimator} with $\priorprobnummodesparam$ estimated from $\smoothingparamssample$.
    \item \btsonejeffreys{}: The \textit{processed} \gls*{bts} estimator~\eqref{eq:bts1-estimator} with $\priorprobnummodesparam = \priorjeffreysmassk$.
    \item \btsoneuniform{}: The \textit{processed} \gls*{bts} estimator~\eqref{eq:bts1-estimator} with $\priorprobnummodesparam \propto 1$.
    \item \btstwosample{}: The \textit{refined} \gls*{bts} estimator~\eqref{eq:bts2-estimator} with $\priorprobnummodesparam$ estimated from $\smoothingparamssample$.
    \item \btstwojeffreys{}: The \textit{refined} \gls*{bts} estimator~\eqref{eq:bts2-estimator} with $\priorprobnummodesparam = \priorjeffreysmassk$.
    \item \btstwouniform{}: The \textit{refined} \gls*{bts} estimator~\eqref{eq:bts2-estimator} with $\priorprobnummodesparam \propto 1$.
\end{itemize}

When comparing alternative methods, we will focus on \btsselected{}, which encompasses all four \gls*{bts} stages and provides competitive results.
We refer the reader to the \mysupplement{} to compare all the different \gls*{bts} variants.
Standard \gls*{bts} configurations with $\zbsplinedim = 22$ basis functions were used in all cases.

The alternative methods considered in this simulation study, which are described and justified in the \mysupplement{}, are the following:

\begin{itemize}[itemsep=0pt, leftmargin=2em]
    \item \kdepizero{}: The \gls*{nom} of the \gls*{kde} with \gls*{pi} bandwidth for $\kdederivativeorder = 0$.
    \item \kdepione{}: The \gls*{nom} of the \gls*{kde} with \gls*{pi} bandwidth for $\kdederivativeorder = 1$.
    \item \kdepitwo{}: The \gls*{nom} of the \gls*{kde} with \gls*{pi} bandwidth for $\kdederivativeorder = 2$.
    \item \kdelscv{}: The \gls*{nom} of the \gls*{kde} with \gls*{scv} bandwidth.
    \item \kdeste{}: The \gls*{nom} of the \gls*{kde} with \gls*{ste} bandwidth.
    \item \kdelscvzero{}: The \gls*{nom} of the \gls*{kde} with \gls*{lscv} bandwidth for $\kdederivativeorder = 0$.
    \item \kdelscvone{}: The \gls*{nom} of the \gls*{kde} with \gls*{lscv} bandwidth for $\kdederivativeorder = 1$.
    \item \kdelscvtwo{}: The \gls*{nom} of the \gls*{kde} with \gls*{lscv} bandwidth for $\kdederivativeorder = 2$.
    \item \gaussianmixture{}: The \gls*{nom} of a Gaussian mixture model selected via \textit{Bayesian information criterion}.
    \item \tautstring{}: The number of \textit{peaks} of a \textit{taut string} fitted based on Kuiper metrics~\citep{Davies2004}.
    \item \silverman{}: The highest \gls*{nom} that the critical bandwidth test by~\citet{Silverman1981} cannot reject at a 0.05 significance level.
    \item \fishermarron{}: The highest \gls*{nom} that the critical bandwidth test by~\citet{Fisher2001} cannot reject at a 0.05 significance level.
    \item \telepathicbootstrap{}: The maximum number of significant \gls*{kde} modes at a 0.10 significance level, according to~\citet{Genovese2016}.
\end{itemize}

\ifnum\journalflag=0
    \paragraph{Test-beds}
\fi

We propose as test-beds the $\numtestbeds = 5$ three-modal Gaussian mixtures considered in~\citet[pp. 124-125]{AmeijeirasAlonso2017}: \firstmodel{}, \secondmodel{}, \thirdmodel{}, \fourthmodel{} and \fifthmodel{}.
They are the most complex models in~\citet{AmeijeirasAlonso2017,AmeijeirasAlonso2018}.
We refer the reader to the \mysupplement{} for the definition and plotting of these \glspl*{pdf}.

\ifnum\journalflag=0
    \paragraph{Experimental design}
\fi

Each test-bed will be paired with $\numsamplesizes = 3$ sample sizes: $\samplesize  = 100$ (small), $\samplesize = 400$ (medium-sized), and $\samplesize = 1600$ (large), adding up to $\numtestbedstimessamplesizes = 15$ sampling configurations.
Then, $\numexperimentreplicationsparam = \numexperimentreplications$ replications of the experiment, drawing \gls*{iid} observations from each combination of test-bed and sample size, will be carried out to obtain significant results, as in~\citet{Lee2003}.

The results from each sampling configuration will be analysed separately first.
For each method $\method_{\genindexj}$, $\genindexj \in \{1, \dots, \numberofmethods\}$, and sample $\dataset_{\genindexi}$, $\genindexi \in \{1, \dots, \numexperimentreplicationsparam\}$, an estimate $\nummodesestimatorof{\genindexj}{\genindexi} \in \naturals$ of the \gls*{nom} will be recorded.
Comparing the \gls*{bts} variants, we have $\numberofmethods = 7$, whereas comparing $\btsselected$ with the alternative methods yields $\numberofmethods = 14$.
Assuming a ground truth target \gls*{nom} $\numberofmodes = 3$ for every test-bed model, an outcome $\outcomeof{\genindexj}{\genindexi} \in \{0, 1\}$ is derived taking $\outcomeof{\genindexj}{\genindexi} = 1$, if $\nummodesestimatorof{\genindexj}{\genindexi} = \numberofmodes$, and $\outcomeof{\genindexj}{\genindexi} = 0$, otherwise.
Then, the \textit{accuracy} of the $\genindexj$-th method is $\accuracy_{\genindexj} = \numexperimentreplicationsparam^{-1} \sum_{\genindexi = 1}^{\numexperimentreplicationsparam} \outcomeof{\genindexj}{\genindexi}$.

Two methods $\method_{\genindexmu}$ and $\method_{\genindexnu}$ are confronted by comparing the outcome sequences $(\outcomeof{\genindexmu}{\genindexi})_{\genindexi = 1}^{\numexperimentreplicationsparam}$ and $(\outcomeof{\genindexnu}{\genindexi})_{\genindexi = 1}^{\numexperimentreplicationsparam}$.
One method is ranked ahead if (i) its accuracy is greater than that of the other \textit{and} (ii) both methods provide significantly different performances according to~\citeauthor{McNemar1947}'s test~\citep{McNemar1947} using a standard significance level $\significancelevel = \mcnemarsignificancelevel$.
Otherwise, both methods shall be ranked the same.
The $\numberofmethods (\numberofmethods - 1) / 2$ pairwise orderings are then aggregated using the Kemeny distance approach in~\citet{Amodio2016}, which computes a \textit{median} ranking, possibly including ties among the individual method ranks.
If there is no unique solution, we propose aggregating all the resulting rankings via simple component-wise rank averaging, a typical default procedure mentioned in~\citet{Amodio2016}.

After computing the intermediate rankings $\intermediaterankings$, a final aggregation using the previous procedure, consisting of a median consensus ranking and averaging in case of multiple solutions, yields a unique global ranking $\finalranking$, possibly with ties.

\subsection{Results}

\label{sec:results}

We will now look at the results following the previous setup and methodology.
By convention, the best position is 1 in all rankings, as in~\citet{Amodio2016}.

Before diving into the results, we urge the reader to take them cautiously.
As warned by~\citet{Lee2003}, these are only simulations, spanning a small and simplified piece of the problem.
Moreover, there is no canonical way of analysing and aggregating the results~\citep[see, for instance,][Section 3.2]{Cao1994}.

\ifnum\journalflag=0
    \paragraph{Global ranking}
\fi

The global ranking $\finalranking$ is presented in \tablename~\ref{tab:final-ranking}.
As we can see, \btsselected{} belongs in the same top-ranked group as the parametric \gaussianmixture{} and the \glspl*{kde} \kdepizero{}, \kdelscv{} and \kdeste{}, all of them generic direct approaches.
In the second place, we find two \gls*{lscv} variants, \kdelscvzero{} and \kdelscvone{}, and the first-order \gls*{pi} version, \kdepione{}.
Next, \kdelscvtwo{}, the remaining \gls*{lscv}, holds the third rank.
The bottom three positions gather all the methods tailored explicitly for mode estimation.
Coming fourth and fifth, we have the best performer of the two test-based approaches, \fishermarron{}, and the last member of the \gls*{pi} family, \kdepitwo{}.
Finally, ranking at the bottom appears the other test, \silverman{}, alongside two methods genuinely related to modes, \tautstring{} and \telepathicbootstrap{}.

\begin{table}
    \centering
    \footnotesize
    \begin{tabular}{cccccccccccccc}
    \toprule
    \texttt{PI0} & \texttt{PI1} & \texttt{PI2} & \texttt{SCV} & \texttt{STE} & \texttt{LSCV0} & \texttt{LSCV1} & \texttt{LSCV2} & \texttt{GM} & \texttt{TS} & \texttt{SI} & \texttt{FM} & \texttt{EIG} & \texttt{BTS2U} \\
    \midrule
    \textbf{1}   & 2            & 5            & \textbf{1}   & \textbf{1}   & 2              & 2              & 3              & \textbf{1}  & 6           & 6           & 4           & 6            & \textbf{1}     \\
    \bottomrule
\end{tabular}

    \captionsetup{font=footnotesize}
    \caption{%
        Global ranking.
    }
    \label{tab:final-ranking}
\end{table}

The same exact final ranking is obtained for various significance levels in~\citeauthor{McNemar1947}'s test, aside from our reference value $\significancelevel = \mcnemarsignificancelevel$, evidencing the robustness of our conclusions.

The results in \tablename~\ref{tab:final-ranking} may look counterintuitive.
Direct non-specific methods perform better than those specifically designed for modality assessment.
In that sense, customary \gls*{kde} bandwidth selectors for the \gls*{pdf} usually provide the correct answer, matching the performance of \gaussianmixture{}, which, given the nature of the considered test-beds, can be seen as an upper limit in performance.

The \fishermarron{} method by~\citet{Fisher2001} comes as the top-qualified classic approach, outmatching \silverman{}, the original critical bandwidth proposal by~\citet{Silverman1981}.
In turn, the poor results of~\tautstring{} and~\telepathicbootstrap{} are especially striking, demonstrating that too much parsimony might not work well in practice.

\ifnum\journalflag=0
    \paragraph{Intermediate rankings}
\fi

The rankings $\intermediaterankings$ are reported in \tablename~\ref{tab:intermediate-rankings}.
See the \mysupplement{} for an in-depth commentary with plots.

\begin{table}
    \centering
    \ifnum\journalflag=0
        \tiny
    \fi
        \begin{tabular}{lrcccccccccccccc}
    \toprule
                 &      & \texttt{PI0} & \texttt{PI1} & \texttt{PI2} & \texttt{SCV} & \texttt{STE} & \texttt{LSCV0} & \texttt{LSCV1} & \texttt{LSCV2} & \texttt{GM} & \texttt{TS} & \texttt{SI} & \texttt{FM} & \texttt{EIG} & \texttt{BTS2U} \\
    \midrule
    \texttt{M21} & 100  & 2            & 3            & 3            & 2            & \textbf{1}   & \textbf{1}     & \textbf{1}     & 2              & 2           & 3           & 3           & 3           & 3            & 2              \\
    \texttt{M21} & 400  & 2            & 4            & 6            & 2            & \textbf{1}   & \textbf{1}     & 2              & 3              & 2           & 6           & 6           & 5           & 6            & 2              \\
    \texttt{M21} & 1600 & 2            & 4            & 6            & 2            & 2            & 3              & 4              & 5              & \textbf{1}  & 9           & 7           & 5           & 8            & 2              \\
    \addlinespace
    \texttt{M22} & 100  & 3            & 7            & 8            & 5            & 2            & 4              & 6              & 7              & 6           & 8           & 8           & 7           & 8            & \textbf{1}     \\
    \texttt{M22} & 400  & \textbf{1}   & \textbf{1}   & 6            & \textbf{1}   & 2            & 5              & 3              & 4              & \textbf{1}  & 7           & 2           & \textbf{1}  & 7            & 2              \\
    \texttt{M22} & 1600 & 2            & \textbf{1}   & \textbf{1}   & 2            & 4            & 8              & 6              & 6              & \textbf{1}  & 7           & \textbf{1}  & \textbf{1}  & 5            & 3              \\
    \addlinespace
    \texttt{M23} & 100  & 2            & 4            & 6            & 2            & \textbf{1}   & 2              & 2              & 3              & \textbf{1}  & 7           & 7           & 5           & 7            & 2              \\
    \texttt{M23} & 400  & 2            & \textbf{1}   & 3            & 2            & 5            & 7              & 6              & 6              & \textbf{1}  & 6           & 6           & 4           & 7            & 4              \\
    \texttt{M23} & 1600 & 4            & 2            & \textbf{1}   & 5            & 7            & 8              & 8              & 7              & \textbf{1}  & 3           & \textbf{1}  & 3           & 6            & 6              \\
    \addlinespace
    \texttt{M24} & 100  & 3            & 4            & 5            & 3            & \textbf{1}   & 3              & 3              & 3              & 5           & 5           & 5           & 5           & 3            & 2              \\
    \texttt{M24} & 400  & \textbf{1}   & 2            & 3            & \textbf{1}   & \textbf{1}   & 2              & 2              & 2              & 3           & 3           & 3           & 3           & 2            & \textbf{1}     \\
    \texttt{M24} & 1600 & \textbf{1}   & \textbf{1}   & 4            & \textbf{1}   & 2            & 5              & 2              & 2              & 5           & 5           & 7           & 6           & 3            & \textbf{1}     \\
    \addlinespace
    \texttt{M25} & 100  & 2            & 5            & 6            & 2            & \textbf{1}   & 2              & 3              & 4              & 6           & 8           & 8           & 5           & 7            & 3              \\
    \texttt{M25} & 400  & \textbf{1}   & 3            & 8            & \textbf{1}   & \textbf{1}   & 2              & 4              & 6              & 7           & 10          & 9           & 5           & 10           & \textbf{1}     \\
    \texttt{M25} & 1600 & 2            & \textbf{1}   & 2            & 2            & 2            & 3              & 3              & 4              & \textbf{1}  & 6           & 4           & 2           & 5            & 2              \\
    \bottomrule
\end{tabular}

    \captionsetup{font=footnotesize}
    \caption{%
        Intermediate rankings by test-bed and sample size configuration.
    }
    \label{tab:intermediate-rankings}
\end{table}

As we see, no method performs uniformly better than the rest across all settings.
Nonetheless, the intermediate rankings confirm the superiority of the top-ranked methods in \tablename~\ref{tab:final-ranking}: \gaussianmixture{}, \kdepizero{}, \kdelscv{}, \kdeste{} and our \btsselected{}.
They rank high most of the time and, more importantly, consistently escape from the bottom position.
Indeed, at least one of these methods ranks first in each considered case.
Especially worth mentioning is the \secondmodel{}-100 setting, where \btsselected{} ranks ahead of the rest.

Beyond the top-tier methods, \tablename~\ref{tab:intermediate-rankings} shows that the global ranking is quite unfair with \kdepione{}, as \kdepione{} lands the first position once more than \kdepizero{} and \kdelscv{}.
The reason why \kdepione{} ranks globally worse lies in its uneven performance.
Small datasets particularly harm \kdepione{}.
Also, \kdepione{} struggles with \firstmodel{}, regardless of the sample size.
As we can see in the \mysupplement{}, the slight middle mode of \firstmodel{} poses serious problems for the over-smoothing strategy of \kdepione{}.

    \section{Discussion}

\label{sec:discussion}

\begin{sloppypar*}
    \Gls*{bts} faces estimating the \gls*{nom} inspired by some little-known aspects of the problem.
    The need for structure diagnosed by~\citet{Donoho1988} is implemented through a combination of \glspl*{kde}~\citep{Wand1995} and compositional splines~\citep{Machalova2020}.
    Subsequently, the Bayesian inference machinery~\citep{Bernardo1994} offers standard procedures to balance data fitting and model complexity, incorporate expert knowledge and assess the uncertainty of the results.
    Finally, global model selection~\citep{Wasserman2000} and local testing~\citep{Minnotte1993} allow for a holistic view of modality.
\end{sloppypar*}

We follow a strategic \textit{divide-and-conquer} approach similar to that suggested by~\citet{Good1980}.
An exploration phase via \gls*{mcmc} enforces regularity in the solutions, penalising curvature and multimodality.
A numerous sample of candidate compositional splines is summarised using \gls*{sfpca}~\citep{Hron2016}, obtaining a one-parameter model that retains the essential modal features.
Then, the Bayesian update weighs the probabilities of each $\nummodesparam$-modality hypothesis in an encompassing prior~\citep{Klugkist2005}, producing a characteristic median compositional spline for each $\nummodesparam$.
Next, each representative \gls*{pdf} has its $\nummodesparam$ modes tested using a Bayesian Savage-Dickey scheme~\citep{Wagenmakers2010} with data in the excess mass region of the mode.
Lastly, the global and local scores are \textit{summed} to determine the most likely $\nummodesparam$.

Traditional approaches for modality either offer a lot of sparse information, such as mode trees~\citep{Minnotte1993} or SiZer~\citep{Chaudhuri1999}, or act as a \textit{black box} outputting an actionable but cryptic p-value, such as the critical bandwidth test~\citep{Silverman1981,Fisher2001}.
\Gls*{bts} delivers traceable and interpretable intermediate products, providing a fuzzy but straightforward result.
The $\bandwidth$-$\oneminuscurvaturepenalty$ scatter plot in \figurename~\ref{fig:posterior-h-alpha} studying modality from a twofold perspective represents a legitimate innovation for mode exploration.
Then, \figurename~\ref{fig:sfpca-mode-tree} and \figurename~\ref{fig:ohtani-sfpca-mode-tree} are upgraded mode trees with several advantages: sensible prescribed upper and lower bounds, a well-behaved natural parameter without resorting to the logarithmic scale, and a posterior \gls*{pdf} that allows placing cue quantiles in the parameter axis.
Additionally, the final decision for $\nummodesparam$ is supported on a tangible \gls*{pdf}, as in \figurename~\ref{fig:ohtani-sfpca-final-spline}.

Several practical experiences support \gls*{bts}.
Our seven-mode result for the Hidalgo problem is well-grounded on likelihood.
There are also qualitative reasons, such as the modes approximately appearing at regular integer values, a much human-like trait.
On the other hand, the four-mode solution of~\citet{AmeijeirasAlonso2018}, for instance, demands a similar smoothing to \kdepione{}, which we know does not perform well on medium-sized samples with subtle features, according to Section~\ref{sec:simulation-study}.
Also, such a smoothing level would imply that the stamp manufacturing process had high variability, including a strange \textit{shoulder} in the density of the thickness distribution.
As for the \gls*{mlb} case study, our four-mode solution agrees with the official interpretation of the \gls*{mlb}.
Interestingly, our results grant a small probability of having only three modes, with a weak significance score for the dubious changeup-cutter mode.
Unlike our method, the Gaussian mixture approach failed conservatively in both real settings since, contrary to the simulation study, non-Gaussianity and discretisation were present.

The simulation study unveils highly unexpected findings.
The traditional methods for modality, e.g., \fishermarron{}, \silverman{} or \tautstring{}, fall deeply in the rankings behind generic plug-in \gls*{kde}-based estimators such as \kdepizero{} and \kdeste{}.
Hypothesis testing methods underperform if evaluated far from their restricted theoretical framework, which leads us to question their actual value in practice when they are sequentially employed to estimate the \gls*{nom}.
Moreover, we have not observed the typical asymptotic accuracy boost of nonparametric methods.
In turn, the taut string \tautstring{} is conservatively biased, recalling the approach by~\citet{Donoho1988} for calculating a lower confidence bound for the \gls*{nom}.
As for our proposal, even though the \btsselected{} variant of \gls*{bts} is a top-tier method, there is still room for accuracy improvement, perhaps using auxiliary techniques such as ensembles or bootstrapping.

If the user has to make a \textit{blindfold} decision, our overall recommendation for estimating the \gls*{nom} is \kdepizero{} based on accuracy, robustness, versatility, simplicity and low computational cost.
The empirical evidence generally supports the sound theory behind it.
With large samples, looking at \kdepione{} should pay off, as well.
Nevertheless, analysts usually prefer analysing data rather than simply \textit{hitting a button} and reporting some output.
For them, \gls*{bts} offers valuable resources with similar accuracy to \kdepizero{}.
In this respect, \gls*{bts} would benefit from a graphical interface for the analysts to work interactively and efficiently access all the available information.
Additionally, beyond the \gls*{nom}, \gls*{bts} could serve as a more elaborate and dependable version of \kdepizero{} for \gls*{pdf} estimation in the context of bounded data.

    \newcommand{\acknowledgements}{%
The research of the first author has been supported by the MICINN grants PID2019-109387GB-I00 and PID2021-124051NB-I00.
The second author would like to thank Professor Amparo Ba\'{i}llo Moreno for her advice as a doctoral counsellor at the Autonomous University of Madrid.
Finally, we thank an associate editor and two anonymous reviewers for their helpful comments.
}

\section*{Acknowledgements}
\acknowledgements

\end{refsection}

\renewcommand*{\mkbibcompletename}[1]{\textsc{#1}}
\printbibliography[section=1, filter=references, title={References}, sorting=nyt]
\renewcommand*{\mkbibcompletename}[1]{#1}

\ifnum\materialflag=1
    \appendix

    \begin{refsection}
        \cleardoublepage

\pagenumbering{roman}
\setcounter{page}{1}
\setcounter{equation}{0}
\renewcommand{\theequation}{S\arabic{equation}}

\part*{Supplementary material}

The following \textit{appendices} are provided as \textit{supplementary material} to the manuscript \textit{\papertitle}.
Equation numbers refer to the manuscript.
The numbers for new equations are prefixed.
References are included at the end.
Notation and acronyms are reused from the manuscript.

Appendix~\ref{sec:motivation} and Appendix~\ref{sec:implementation} explain the philosophy and implementation details of \gls*{bts}.
Then, Appendix~\ref{sec:extra-case-study}, Appendix~\ref{sec:extra-setup}, and Appendix~\ref{sec:extra-results} expand on the case and simulation studies.
Finally, Appendix~\ref{sec:dim-reduction} and Appendix~\ref{sec:computing} discuss relevant theoretical and practical issues.

        \section{Motivation}

\label{sec:motivation}

The following lines motivate the design of \gls*{bts} in Section~\ref{sec:method} along three axes.

\paragraph{Structure}

Researchers have mainly tackled modality through nonparametric approaches, for parametric ones are deemed too rigid~\citep{AmeijeirasAlonso2017}.
However, the work by~\citet{Donoho1988} suggests that modality is impossible to \textit{tame} under weak nonparametric regularity assumptions.
For that matter, the quest for modes has the existence of a \gls*{pdf} as a prerequisite, something we cannot test empirically~\citep{Donoho1988}.

Consequently, some structure seems due in the study of modes, balancing data fitting and model complexity.
This policy is exhibited by~\citet{Davies2009} in the context of histograms or~\citet{Good1980} with Fourier series.
In this respect, smoothing splines are the right tool, allowing restrained flexibility~\citep{Eilers1996,Hron2016,Machalova2020}.
\Gls*{bts} thus targets a compromise between both dimensions by blending \glspl*{kde} and compositional splines.

\begin{sloppypar*}
    On the one hand, \glspl*{kde} serve as a \textit{scaffold} for splines, guiding high-dimensional fitting more efficiently than customary histograms~\citep{Eilers1996,Machalova2020}.
    On the other, the comparatively rigid structure of splines and their built-in curvature penalisation~\citep{Machalova2020} represent a \textit{harness} for \glspl*{kde}, preventing spurious modes.
    Additionally, curvature penalisation mitigates the effect of the \gls*{kde} having a single global bandwidth $\bandwidth$~\citep{Minnotte1993}.
    Finally, splines allow for a deeper analysis and simplification of the modal structure via dimensionality reduction~\citep{Hron2016}.
\end{sloppypar*}

\paragraph{Bayesian inference}

\begin{sloppypar*}
    Mode estimation reveals philosophical issues excellently handled by Bayesian inference~\citep{Bernardo1994}.
    Emphasising \textit{frequentist} population-wide properties over actual data is unrealistic in some cases.
    The Hidalgo problem is a paradigmatic example since the stamps are no longer issued.
    In turn, the Bayesian approach is the right choice when data is scarce, and any information, such as philately expert knowledge, could be helpful.
\end{sloppypar*}

\begin{sloppypar*}
    Bayesian inference provides \textit{soft} solutions, quantifying their uncertainty~\citep{Bernardo1994}.
    Current methods cannot assign a probability to each $\numberofmodes$-modality hypothesis.
    For instance, hypothesis testing procedures report p-values limited to the null hypothesis~\citep{Wagenmakers2010}.
    Meanwhile, mode trees~\citep{Minnotte1993,Minnotte1997,Minnotte1998} and SiZer~\citep{Chaudhuri1999}, though incorporating mechanisms to assess the uncertainty, fall short of providing actionable answers.
    By contrast, the Bayesian framework excels at operating with probabilities, offering standard hypothesis selection tools~\citep{Kass1995,Wasserman2000,Klugkist2005,Wagenmakers2010,Spiegelhalter2014}.
\end{sloppypar*}

The subjective nature of modes particularly suits Bayesian methods.
The wide array of graphical methods~\citep{Minnotte1993,Minnotte1997,Minnotte1998,Chaudhuri1999,Chaudhuri2002} evidence that the human eye, aided by a computer, better appreciates such features~\citep{Good1980}.
Consequently, the usual criticism that Bayesian inference is not objective loses force~\citep{Bernardo1994}.
Moreover, Bayesian tools are less prone to overfitting since they examine a range of plausible outcomes rather than isolated optima.
This inherent \textit{parsimony}~\citep{Wagenmakers2010} will be valuable against spurious modes.

\paragraph{Holism}

Modes are challenging for their dual local and global nature.
They are defined via a local property of the \gls*{pdf}, but, at the same time, that \gls*{pdf} is built from disconnected data.
In that sense, modes are \textit{emergent} phenomena.

\Gls*{bts} aims to combine both perspectives.
During the first three stages, the mode concept helps build candidate \glspl*{pdf}.
Then, at the fourth stage, the representative \glspl*{pdf} of each modality hypothesis have their modes tested individually using neighbouring data, yielding \textit{significance} scores.

\begin{sloppypar*}
    The penalised likelihood approach by~\citet{Good1980} and mode trees~\citep{Minnotte1993} also include local testing mechanisms after global fitting.
    Our proposal merges the global and local probabilities into a single result, obtaining a holistic view of modality.
\end{sloppypar*}

        \section{Implementation}

\label{sec:implementation}

The following lines discuss the implementation of the \gls*{bts} method in Section~\ref{sec:method}.

\subsection{Hyperparameter tuning}

\Gls*{bts} requires setting several configurations.
We comment here on how this can be done in practice with attention to the data.

\paragraph{Prior design}

The parametric design \eqref{eq:exploration-prior-distributions} of the prior in the exploration stage of \gls*{bts} was left unexplained.
Let us now go deeper into the underlying principles and experiences.

The choice of beta distribution for $1 - \oneminuscurvaturepenalty$ is conventional when $\oneminuscurvaturepenalty \in (0, 1)$.
Fixing $\betaparamalpha = 1$ makes the \gls*{pdf} diverge at zero, while $\penaltyhyperparam$ controls its expected value through $\expectedvalue{1 - \oneminuscurvaturepenalty} = (1 + \penaltyhyperparam)^{-1}$.
Next, the rationale behind the bandwidth distribution lies in the logarithmic scale, known for improving the appreciation of the \gls*{kde} changes in mode trees~\citep{Minnotte1993}.
Assuming a normal distribution for $\log \bandwidth$ with location $\bandwidthlocationhyperparam$ and scale $\bandwidthdispersionhyperparam$ yields maximum entropy and allows focusing on a suitable region of the mode tree.
Moreover, preliminary simulations studying critical bandwidths and the posterior $\posteriorhalpha$ confirm that the log-normal provides a good approximation.
On the other hand, the distributions of $\nummodesparam$ and $\curvatureparam$ have been selected to penalise complexity.
In the case of $\nummodesparam$, the Poisson distribution with mean one favours unimodality.
For $\curvatureparam$, the exponential pulls the curvature towards zero, leaving control over the mean via $\expectedvalue{\curvatureparam} = \curvaturehyperparam^{-1}$.

Imposing hyperpriors on~\eqref{eq:exploration-prior-distributions} would make the \gls*{mcmc} heavier.
We propose choosing the hyperparameters empirically.
Taking $\penaltyhyperparam = 99$ yields $\expectedvalue{\oneminuscurvaturepenalty} = 0.99$, which works well in practice.
In turn, for $\bandwidthlocationhyperparam$, $\bandwidthdispersionhyperparam$ and $\curvaturehyperparam$, we first recommend estimating two tentative values, say $\bandwidth_1 < \bandwidth_2$, from distinct bandwidth selectors.
Imposing $\log \bandwidth$ to enclose a central probability $\normalcdf(\bandwidthsigma) - \normalcdf(-\bandwidthsigma)$ between $\log \bandwidth_1$ and $\log \bandwidth_2$, where $\normalcdf$ is the standard univariate Gaussian \gls*{cdf} and $\bandwidthsigma > 0$, implies $\bandwidthlocationhyperparam = \log \sqrt{\bandwidth_1 \bandwidth_2}$ and $\bandwidthdispersionhyperparam = \bandwidthsigma^{-1} \log \sqrt{\bandwidth_2 / \bandwidth_1}$.
On the other hand, if $\curvatureparam_1$ and $\curvatureparam_2$ are the curvatures of $\kdeof{\bandwidth_1}$ and $\kdeof{\bandwidth_2}$ in the sense of~\eqref{eq:zbspline-loss}, respectively, taking $\curvaturehyperparam^{-1} = (\curvatureparam_1 + \curvatureparam_2) / 2$ produces a $\curvaturehyperparam$ that is the harmonic mean of the $\exponentialparam$ parameters corresponding to $\curvatureparam_1$ and $\curvatureparam_2$.

The recommended bandwidth selectors for calculating $\bandwidth_1$ and $\bandwidth_2$ belong to the \gls*{pi} family of methods $\piselector{\kdederivativeorder}$, targeting the $\kdederivativeorder$-th derivative for $\kdederivativeorder = 0, 1, 2$~\citep{Chacon2013}.
These are robust in an asymptotic sense, avoiding overfitting to $\dataset$.
Namely, we propose taking $(\bandwidth_1, \bandwidth_2) = (\bandwidth_{\piselector{0}}, \bandwidth_{\piselector{1}})$ in a general setting.
For severely discretised data, $(\bandwidth_1, \bandwidth_2) = (\bandwidth_{\piselector{1}}, \bandwidth_{\piselector{2}})$ offers an extra \textit{shield} against spurious modes.
In both cases, we propose $\bandwidthsigma = 1$ to leave room for exploration beyond $(\bandwidth_1, \bandwidth_2)$.
All in all, the previous configurations are somewhat conservative but conform to the regularising goal of the prior $\btsexplorationprior$.

\paragraph{Other configurations}

\begin{sloppypar*}
    We recommend the \gls*{dic}~\citep{Spiegelhalter2014} to assess the optimal spline dimension $\zbsplinedim$ and knot placement strategy, fixing the spline degree to $\splinedegree = 3$.
    Too small $\zbsplinedim$ will produce too low likelihood values, whereas a too large $\zbsplinedim$ will increase the \textit{effective} number of parameters, i.e., the model complexity.
    The \gls*{dic} will generally advise against both extremes.
    Typical values for $\zbsplinedim$ are 22 or 32 (i.e., 21 or 31 knots), depending on the intricacies of the data, far from the hundreds of Fourier series terms in~\citet{Good1980}.
\end{sloppypar*}

Generally, the grid size $\numsplinepoints$ is far less critical than $\zbsplinedim$ and can be held to a constant value such as $\numsplinepoints = 1001$.
The larger the $\numsplinepoints$, the more accurate the spline approximation but the higher the computational cost.

\subsection{Simulation}

\Gls*{bts} strongly relies on \gls*{mcmc}~\citep{Bernardo1994}.
In all three Bayesian inference steps in Section~\ref{sec:method}, the updated parameters have different support than $\reals$.
Sampling directly from those posteriors would lead to abnormally low acceptance rates in \gls*{mcmc}.

In the case of $\posteriorhalpha$, we recommend applying the change of variables $\extendedparam{\bandwidth} = \log \bandwidth$, $\extendedparam{\oneminuscurvaturepenalty} = \normalcdf^{-1}(\oneminuscurvaturepenalty)$ to obtain the posterior
$
    \condprob{\extendedparam{\bandwidth}, \extendedparam{\oneminuscurvaturepenalty}}{\dataset}
    =
    \condprob{\bandwidth, \oneminuscurvaturepenalty}{\dataset}
    \cdot
    e^{\extendedparam{\bandwidth}}
    \cdot
    \normalpdf(\extendedparam{\oneminuscurvaturepenalty})
$.
Then, we can sample from $(\extendedparam{\bandwidth}, \extendedparam{\oneminuscurvaturepenalty})$ and obtain $\smoothingparamssample$ after undoing the change of variables.
Similarly, for $\posteriorsfpcaparam$, we propose taking $\extendedparam{\sfpcaparam} = \normalcdf^{-1}[(\sfpcaparam - \sfpcaparammin) / (\sfpcaparammax - \sfpcaparammin)]$, which has \gls*{pdf}
$
    \condprob{\extendedparam{\sfpcaparam}}{\dataset}
    \propto
    \posteriorsfpcaparam
    \cdot
    \normalpdf(\extendedparam{\sfpcaparam})
$.
Finally, for $\posteriorsavagedickey$, we suggest $\extendedparam{\savagedickeyparam} = \log \savagedickeyparam$, yielding
$\posteriorbarsavagedickey = \posteriorsavagedickey \cdot e^{\extendedparam{\savagedickeyparam}}$.

Finding a good initial state for \gls*{mcmc} by calculating the \textit{maximum a posteriori} estimator through a small optimisation will ensure the proper behaviour of the posterior.
This will prevent \gls*{mcmc} from including outliers in the posterior sample: the so-called \textit{burn-in} period observations that are usually removed~\citep{Wagenmakers2010}.

\subsection{Time complexity}

\Gls*{bts} is a compound method.
It comprises several stages with algorithms of varied nature, including estimation, optimisation, and simulation.
In addition to the intrinsic complexity and size of the input data, each procedure has its configuration options, affecting both the execution time and the precision of the results.
On the other hand, most of these algorithms are readily implemented as packages, which eases building \gls*{bts} from scratch but may lead to small inefficiencies.
Considering the above, making a comprehensive time complexity analysis of \gls*{bts} that honours reality is not easy.

Regarding \gls*{pdf} evaluations, all the Bayesian inferences have complexity $\bigoh{\numobservations \nummcmcsteps}$, where $\numobservations$ is the number of data points, and $\nummcmcsteps$ is the number of \gls*{mcmc} steps.
Except for the testing stage, where $\numobservations$ is the number of points in the excess mass region, $\numobservations$ coincides with the total sample size $\samplesize$.
Then, all the \glspl*{pdf} are compositional splines, which can be efficiently evaluated at a single point using B-splines with complexity $\bigoh{\zbsplinedim + \splinedegree^2}$~\citep{Boor1972}.
However, building the model sometimes has a different cost.
In the selection and testing stages, the \glspl*{pdf}~\eqref{eq:sfpca-model} and~\eqref{eq:savage-dickey-model} belong to a one-dimensional parametric family, so instantiating the model for each $\sfpcaparam$ and $\savagedickeyparam$ is almost immediate.
By contrast, in the exploration phase, we must build~\eqref{eq:kde-spline-model} for each $\halphapair$, consuming much time.
The latter step involves several hyperparameters, deserving careful examination.

First, building the \gls*{clr} grid of the \gls*{kde} in~\eqref{eq:zbspline-loss} depends on the sample size $\samplesize$.
Naively evaluating~\eqref{eq:kernel-estimator} at $\numsplinepoints$ points has complexity $\bigoh{\samplesize \numsplinepoints}$ regarding Gaussian kernel $\kernel$ evaluations.
Even if $\numsplinepoints$ is held to a moderate constant value, as indicated above, if $\samplesize$ is sufficiently large, the execution time may be unaffordable.
Using the concept of \textit{binning}, implemented in the package~\textit{ks}~\citep{Duong2022}, we can build a discrete approximation to~\eqref{eq:kernel-estimator} in $\bigoh{\samplesize}$ steps that can be evaluated over the grid in $\bigoh{\numsplinepoints \log \numsplinepoints}$ steps via the FFT algorithm~\citep{Gramacki2017}.
Assuming $\numsplinepoints \ll \samplesize$, the dominant term is $\bigoh{\samplesize}$, vastly improving the naive algorithm.

Secondly, solving for the $\zbsplinedim$ compositional spline coordinates in~\eqref{eq:compositional-spline-approx} requires computations that no longer depend on the sample size once the $\numsplinepoints$-size grid is formed.
See~\citet{Machalova2020} for further details.
Constructing the final linear system matrix has complexity $\bigoh{\numsplinepoints \zbsplinedim}$ in terms of B-spline evaluations, plus $\bigoh{\zbsplinedim^2}$ integral calculations for the total curvature penalty component in~\eqref{eq:zbspline-loss}.
Finally, using standard implementations, finding the solution to the $\zbsplinedim \times \zbsplinedim$ linear system has complexity $\bigoh{\zbsplinedim^3}$.

Apart from the Bayesian inferences, there is an analysis phase with \gls*{sfpca} between the exploration and selection stages.
See Section~\ref{sec:sfpca} below for further details on \gls*{sfpca}.
Roughly speaking, \gls*{sfpca} reduces to vanilla PCA and, subsequently, to solving an eigenvalue problem followed by a linear system over $\zbsplinedim \times \zbsplinedim$ matrices, yielding a complexity $\bigoh{\zbsplinedim^3}$ in both cases.

Beyond the sample size and the rest of the hyperparameters, \gls*{bts} hides a strong dependency regarding the intricacies of the data.
The more \textit{ambiguous} the \gls*{nom} in the data and the higher the maximum predicted \gls*{nom}, the more calculations are necessary.
In some cases, the computational cost increase is negligible.
For instance, if many $\nummodesparam$ values have non-null posterior probability~\eqref{eq:posterior-prob-k-simplified}, it is more expensive to compute the maximum~\eqref{eq:bts1-estimator}.
However, letting $\maxnummodes$ be the maximum $\nummodesparam \in \naturals$ with non-zero probability after the selection phase, the testing stage consists of $\maxnummodes(\maxnummodes + 1) / 2$ inferences in the worst-case scenario, a complexity $\bigoh{\maxnummodes^2}$, corresponding to testing the $\nummodesparam$ modes of $\selectedmodelnummodesparam$ for $\nummodesparam \in \{1, \dots, \maxnummodes\}$.

As the last remark reminds us, \gls*{bts} is especially suited for data exploration in an interactive environment, possibly with the assistance of a graphical tool.
Employing \gls*{bts} from start to finish in \textit{batch} mode, as we have done in this paper, evenly demonstrates the value of the proposal compared to other alternatives but misses much of its power.
Many of the time complexity issues above can be solved with an expert \textit{hand} making good decisions behind the algorithms.
For instance, a low-value $\zbsplinedim$ might suffice if spline knots are intelligently placed by hand, dramatically reducing the computational cost.
Also, running \gls*{mcmc} chains with as many steps $\nummcmcsteps$ as we have used here may be unnecessary in practice.
Moreover, of course, if analysts are overwhelmed by the complexity of the data, they may use the graphical tool to omit some stages of \gls*{bts} or discard on the run some modality hypotheses based on expert knowledge.

        \section{Extra case study}

\label{sec:extra-case-study}

This annexe completes the exposition of the case study in Section~\ref{sec:case-study}.

\paragraph{Data}

The pitching data has been retrieved using the \rlanguage{} package~\textit{baseballr}~\citep{Petti2022}.
\figurename~\ref{fig:ohtani2022-baseball-savant} is taken from the \acrshort*{mlb}-supported advanced metrics website~\baseballsavant{}~\citep{Willman2023}.
It shows a mixture \gls*{pdf} model of the pitching speeds by Shohei Ohtani with the four modes advanced in Section~\ref{sec:case-study}.

\begin{figure}
    \centering
    \includegraphics[width=\textwidth]{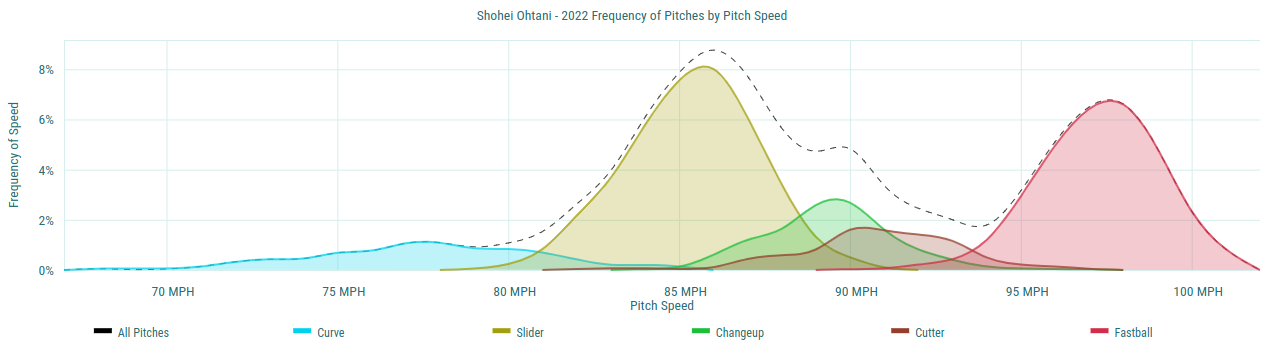}
    \caption{%
        Mixture \gls*{pdf} model of the pitching speeds by Shohei Ohtani in the 2022 season taken from~\baseballsavant{}~\citep{Willman2023}.
        Pitches of different types are modelled in separate mixture components: \textit{curveballs} (blue), \textit{sliders} (yellow), \textit{changeups} (green), \textit{cutters} (brown), and \textit{fastballs} (red).
        The overall \gls*{pdf} shows up as the black dashed line.
    }
    \label{fig:ohtani2022-baseball-savant}
\end{figure}

\paragraph{Intermediate results}

The results of the exploration phase of \gls*{bts} are shown in \figurename~\ref{fig:ohtani-posterior-h-alpha}, which is similar to \figurename~\ref{fig:posterior-h-alpha} in the synthetic mixture example from Section~\ref{sec:method}.
Up to \ohtaniexplorationprobthreemodes{} of the odds favour three modes, while the remaining \ohtaniexplorationprobfourmodes{} correspond to four.
The subsequent analysis phase results are gathered in \figurename~\ref{fig:ohtani-sfpca-results}.
On one side, \figurename~\ref{fig:ohtani-sfpca-scree-plot} shows an even more prominent first \gls*{pc} than \figurename~\ref{fig:sfpca-scree-plot}.
In turn, \figurename~\ref{fig:ohtani-sfpca-splines} is similar to \figurename~\ref{fig:ohtani-pdfs} in that only $\sfpcamodel{\sfpcaparammin}$ captures the elusive 90-\gls*{mph} mode.
In \figurename~\ref{fig:ohtani-sfpca-jeffreys-prior}, the Jeffreys prior for the \gls*{sfpca} model behind \figurename~\ref{fig:ohtani-sfpca-splines} is qualitatively very similar to \figurename~\ref{fig:sfpca-jeffreys-prior}, displaying a mild uniform slope that slightly penalises the fourth mode.
The posterior sample for the selection phase appears in \figurename~\ref{fig:ohtani-sfpca-posterior}, also having a very similar look to \figurename~\ref{fig:sfpca-posterior}, definitely leaning the odds towards four modes.

\begin{figure}
    \centering
    \includegraphics[width=\figurewidth]{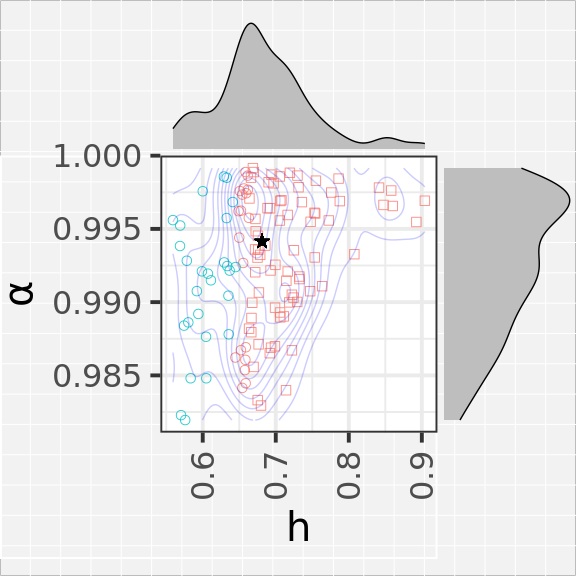}
    \caption{%
        \Gls*{mcmc} sample from the exploration phase of \gls*{bts} consisting of \ohtanimcmcsamplesize{} observations for the \gls*{mlb} case study.
        The figure follows the structure of \figurename~\ref{fig:posterior-h-alpha}.
        Blue corresponds to four modes (\ohtaniexplorationprobfourmodes{} of all points) and red to three (\ohtaniexplorationprobthreemodes{} of the total).
        Squares and circles refer to three and four modes, respectively.
    }
    \label{fig:ohtani-posterior-h-alpha}
\end{figure}

\begin{figure}
    \centering
    \begin{subfigure}[b]{\figurewidth}
        \centering
        \includegraphics[width=\subfigurewidth]{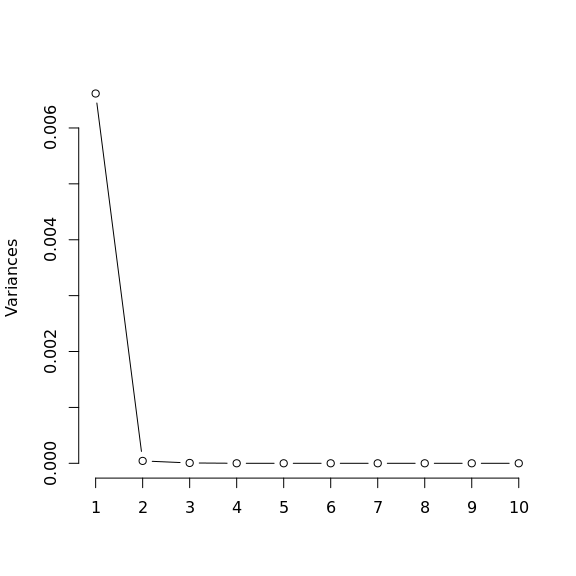}
        \caption{\gls*{sfpca} variances scree plot}
        \label{fig:ohtani-sfpca-scree-plot}
    \end{subfigure}
    \begin{subfigure}[b]{\figurewidth}
        \centering
        \includegraphics[width=\subfigurewidth]{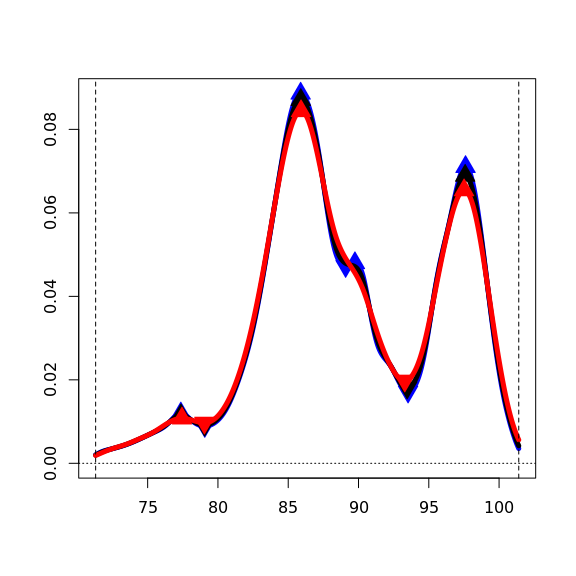}
        \caption{Mode of variation of $\sfpcapceigenfun{1}$}
        \label{fig:ohtani-sfpca-splines}
    \end{subfigure}
    \caption{%
        \gls*{sfpca} analysis phase results for the \gls*{mlb} case study with the same structure as \figurename~\ref{fig:sfpca-results}.
        The upper bound (red) and the mean (black) \glspl*{pdf} have three modes, while the lower bound (blue) has four.
    }
    \label{fig:ohtani-sfpca-results}
\end{figure}

\begin{figure}
    \centering
    \begin{subfigure}[b]{\figurewidth}
        \centering
        \includegraphics[width=\subfigurewidth]{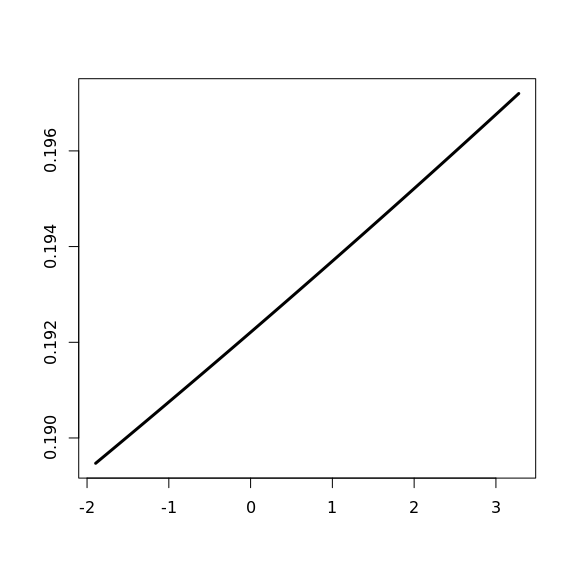}
        \caption{Jeffreys prior}
        \label{fig:ohtani-sfpca-jeffreys-prior}
    \end{subfigure}
    \begin{subfigure}[b]{\figurewidth}
        \centering
        \includegraphics[width=\subfigurewidth]{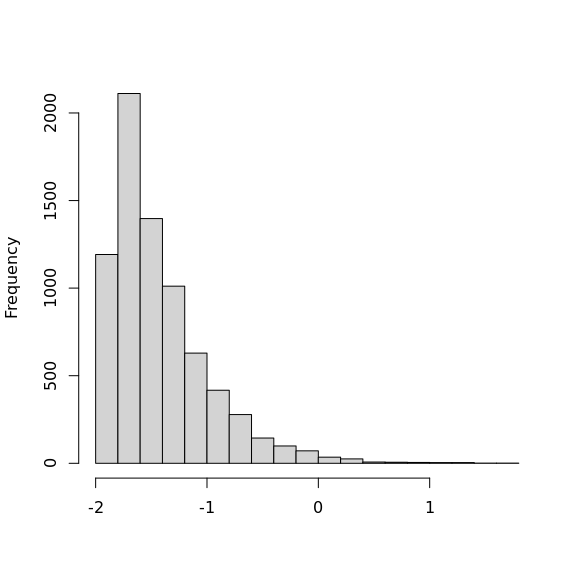}
        \caption{Posterior sample}
        \label{fig:ohtani-sfpca-posterior}
    \end{subfigure}
    \caption{%
        Second Bayesian inference on the \gls*{sfpca} model for the \gls*{mlb} case study with the same structure as \figurename~\ref{fig:sfpca-prior-posterior}.
        The posterior sample to the right comprises \ohtanisfpcasamplesize{} observations.
    }
    \label{fig:ohtani-sfpca-prior-posterior}
\end{figure}

        \section{Extra setup}

\label{sec:extra-setup}

This appendix expands on the simulation study setup in Section~\ref{sec:setup}.

\subsection{Methods}

The following lines justify and explain the methods compared in Section~\ref{sec:setup}.

\paragraph{Theoretical grounds}

The \gls*{kde}-based methods \kdepizero{}, \kdepione{}, \kdepitwo{}, \kdelscv{}, \kdeste{}, \kdelscvzero{}, \kdelscvone{}, \kdelscvtwo{} and the mixture-based \gaussianmixture{} are direct \textit{plug-in} approaches, meaning the \gls*{nom} derives from counting the modes in a fitted \gls*{pdf} model.
To do so, we apply the definition in Section~\ref{sec:preliminaries} of modes as local maxima, restricting the number of \gls*{pdf} evaluations to a sufficiently fine grid over $\intervalab$.
Among these direct approaches, \gaussianmixture{} is parametric, whereas the rest are nonparametric.
None of them is tailored explicitly for mode estimation.

The taut string \tautstring{} method, nonparametric and centred on modality, is also a direct approach but with some additional peculiarities.
The continuous and non-uniform definition of mode does not translate well to histogram-like \glspl*{pdf}.
In this case, modes should be counted as peaks: the midpoints of bins \textit{taller} than their immediate neighbours.
Also, we will interpret a flat taut string, which arguably has no modes, as having modality one, so the minimum \gls*{nom} will be the same across all direct methods.

Some previous \glspl*{kde} have been portrayed in \figurename~\ref{fig:hidalgo-example-pdfs} and \figurename~\ref{fig:ohtani-pdfs}.
The theory behind the different bandwidth selectors targeting the \gls*{pdf} ($\kdederivativeorder = 0$) can be found in~\citet{Wand1995}.
The \gls*{ste} selector is closely related to $\piselector{0}$, minimising the same loss function but with a slightly different solution scheme~\citep[p. 74]{Wand1995}.
The corresponding extensions for density derivative estimation ($\kdederivativeorder > 0$) are described in~\citet{Chacon2018a}.
Those were included because of the connection of the first and second derivatives with local maxima~\citep{Chaudhuri1999,Chaudhuri2002}.

To make a fair comparison with \gls*{bts}, all the \glspl*{kde} are equipped with the same outlier-filtering preprocessing step described for \gls*{bts} in Section~\ref{sec:exploration}, employing an almost negligible mass threshold of 0.001.
Data points belonging to modal regions below that mass are discarded when building the final \gls*{pdf}.
The preliminary findings leading to the design of \gls*{bts} evidenced the extreme sensitivity of \glspl*{kde} to isolated points when estimating the \gls*{nom}, a rare risk but with a high impact on the results.
The vast majority of such spurious modes generated nearly imperceptible modal regions, easily ignored by the human eye.
Therefore, we decided to expand all the outlier-prone methods to match the performance a human would get from them under regular operation.
This extra help is unnecessary for the rest.

A second group comprises the mode hypothesis testing methods \silverman{} and \fishermarron{}.
To obtain the \gls*{nom} estimate, we iteratively test the null hypothesis that the \gls*{nom} is less than or equal to $\numberofmodes$ against the alternative of being greater than $\numberofmodes$, beginning with $\numberofmodes = 1$ and stopping the first time the null hypothesis cannot be rejected.
Such an iterative process is customarily used to convert hypothesis testing procedures into estimation ones~\citep[p. 917]{AmeijeirasAlonso2018}.
The same intermediate significance level $\significancelevel = 0.05$ is used at every iteration, while the number of bootstrap samples is $\numbootstrapsamples = 500$, being both settings as in~\citet{AmeijeirasAlonso2018}.

The excess mass approach is a notable absence among the studied methods because of its high computational cost.
Calculating the excess mass statistic has a high asymptotic complexity as the sample size $\samplesize$ and the tested $\numberofmodes$ grow.
Despite approximations, the execution time under the recommended number of bootstrap samples ($\numbootstrapsamples = 500$) was unworkable for reasonable values $\samplesize \geq 1000$ and $\numberofmodes > 1$.
Moreover, the risks implied are magnified by the iterative use of the test, making it difficult to limit the total time per task.
Therefore, we finally refrained from an exhaustive comparison.
Nonetheless, preliminary experiments suggested a similar performance to \fishermarron{} in terms of accuracy.

Last but not least, the \telepathicbootstrap{} method by~\citet{Genovese2016}, which we named after the role of the eigenvalues in its multivariate version, is based on very different principles, such as testing modes locally and splitting data into train and test sets.
The confidence level for \telepathicbootstrap{} was the same as in~\citet{Genovese2016}.

\paragraph{Software implementation}

All the methods under study are coded in the \rlanguage{} programming language.
The source code for \gls*{bts}, an \rlanguage{} package, shall be distributed under licence and on demand.
Our implementation relies on the package \textit{robCompositions} for compositional data analysis~\citep{Templ2009}.
Our package will also include the source code for \telepathicbootstrap{} by~\citet{Genovese2016} and the data from the case study.

The rest of the methods are publicly available.
\gaussianmixture{} is based on the classical package \textit{mclust}~\citep{Scrucca1999,Scrucca2016}.
\tautstring{} is implemented in~\textit{ftnonpar}~\citep{Davies2012}.
The test-based approaches are provided in the package~\textit{multimode} by~\citet{AmeijeirasAlonso2021a}~\citep[see also][]{AmeijeirasAlonso2021}.
Except for \kdeste{}, which corresponds to the default routine \texttt{bw.SJ} in \rlanguage{} with \texttt{method = "ste"}, all the bandwidth selectors are in the package \textit{ks}~\citep{Duong2022}.

\subsection{Test-beds}

The test-bed \glspl*{pdf} in Section~\ref{sec:setup} condense many different shapes, as seen in \figurename~\ref{fig:mixture-models}.
In particular, \figurename~\ref{fig:fifth-mixture-model} coincides with the example \gls*{pdf} in \figurename~\ref{fig:theoretical-pdf}.
The five \glspl*{pdf} take the form $\x \mapsto \sum_{\genindexi = 1}^{\nummixturecomponents} \mixtureweight_{\genindexi} \ \normalpdf((\x - \mixturecomponentmean_{\genindexi}) / \mixturecomponentstd_{\genindexi})$, where $\normalpdf$ is the standard Gaussian \gls*{pdf}, $\nummixturecomponents$ is the number of mixture components and $\mixturecomponentmean_{\genindexi}$, $\mixturecomponentstd_{\genindexi}$ and $\mixtureweight_{\genindexi}$ are, respectively, the mean, the standard deviation and the weight of the $\genindexi$-th mixture component.
Collecting all the mixture model parameters in vectors $\mixturecomponentmeans = (\mixturecomponentmean_1, \dots, \mixturecomponentmean_{\nummixturecomponents})$, $\mixturecomponentvariances = (\mixturecomponentstd_1^2, \dots, \mixturecomponentstd_{\nummixturecomponents}^2)$ and $\mixtureweights = (\mixtureweight_1, \dots, \mixtureweight_{\nummixturecomponents})$, the five mixtures are defined in \tablename~\ref{tab:mixture-model-parameters}.

\begin{figure}
    \centering
    \begin{subfigure}[b]{\figurewidth}
        \centering
        \includegraphics[width=\subfigurewidth]{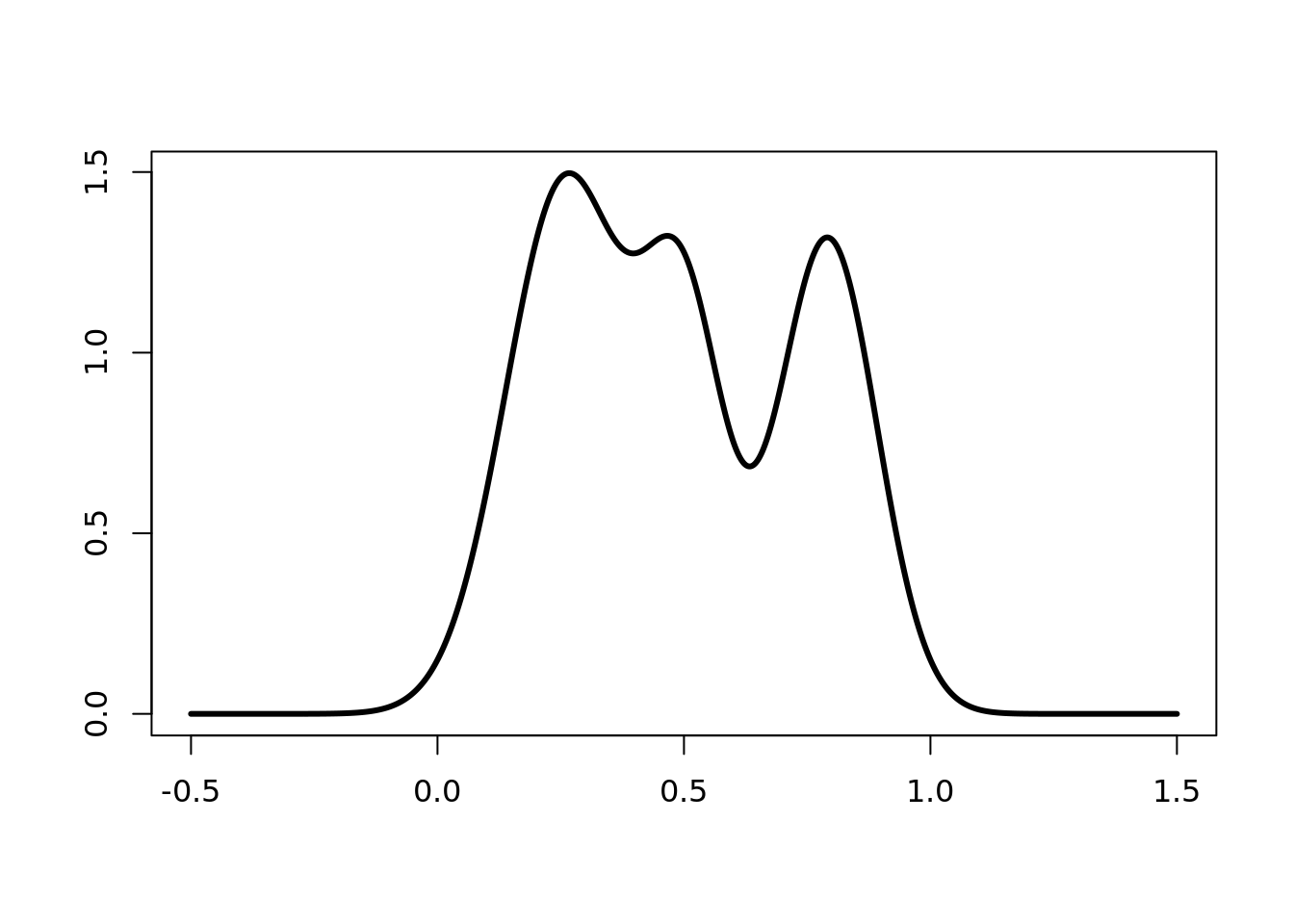}
        \caption{\firstmodel{}}
        \label{fig:first-mixture-model}
    \end{subfigure}
    \begin{subfigure}[b]{\figurewidth}
        \centering
        \includegraphics[width=\subfigurewidth]{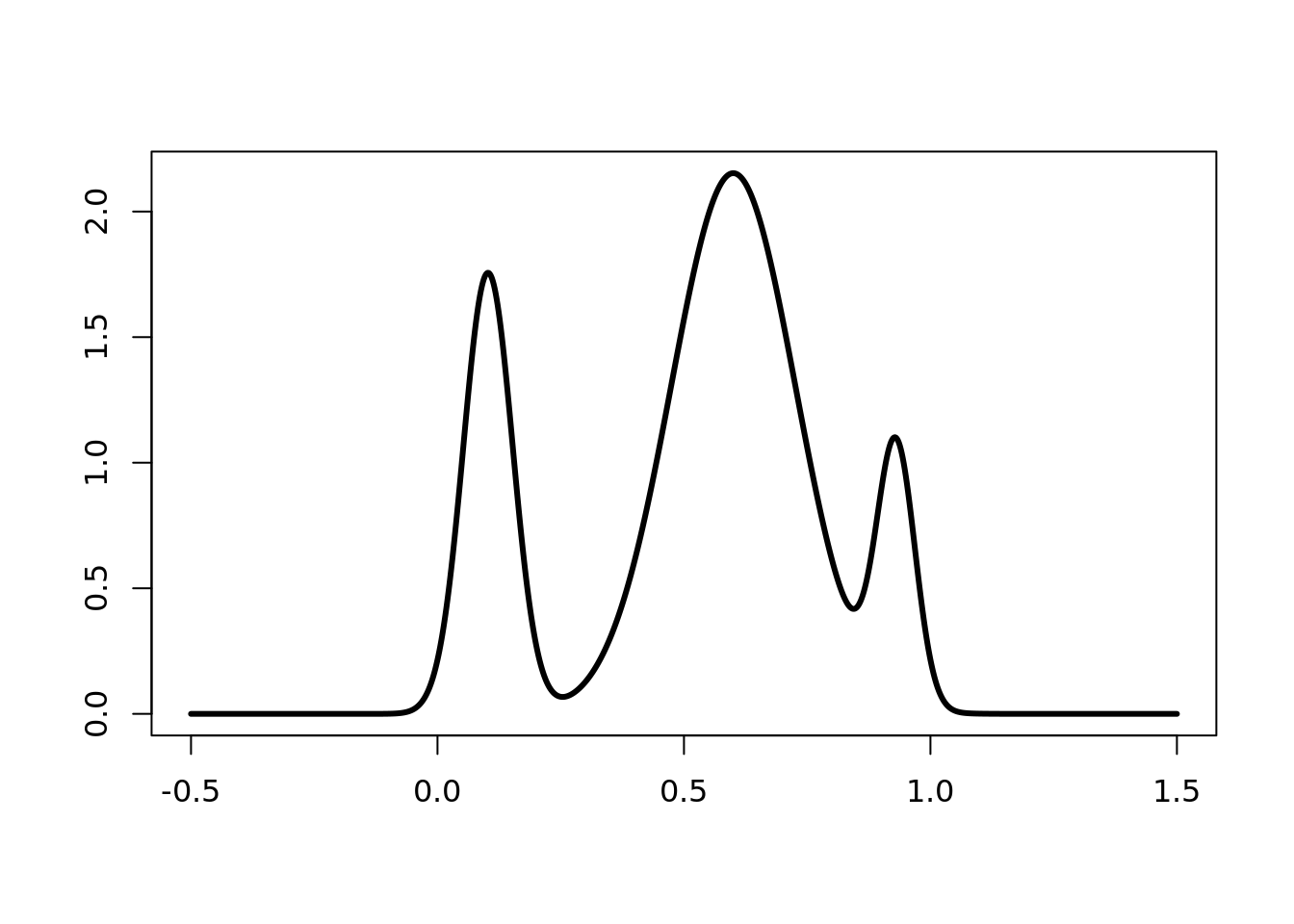}
        \caption{\secondmodel{}}
    \end{subfigure}
    \begin{subfigure}[b]{\figurewidth}
        \centering
        \includegraphics[width=\subfigurewidth]{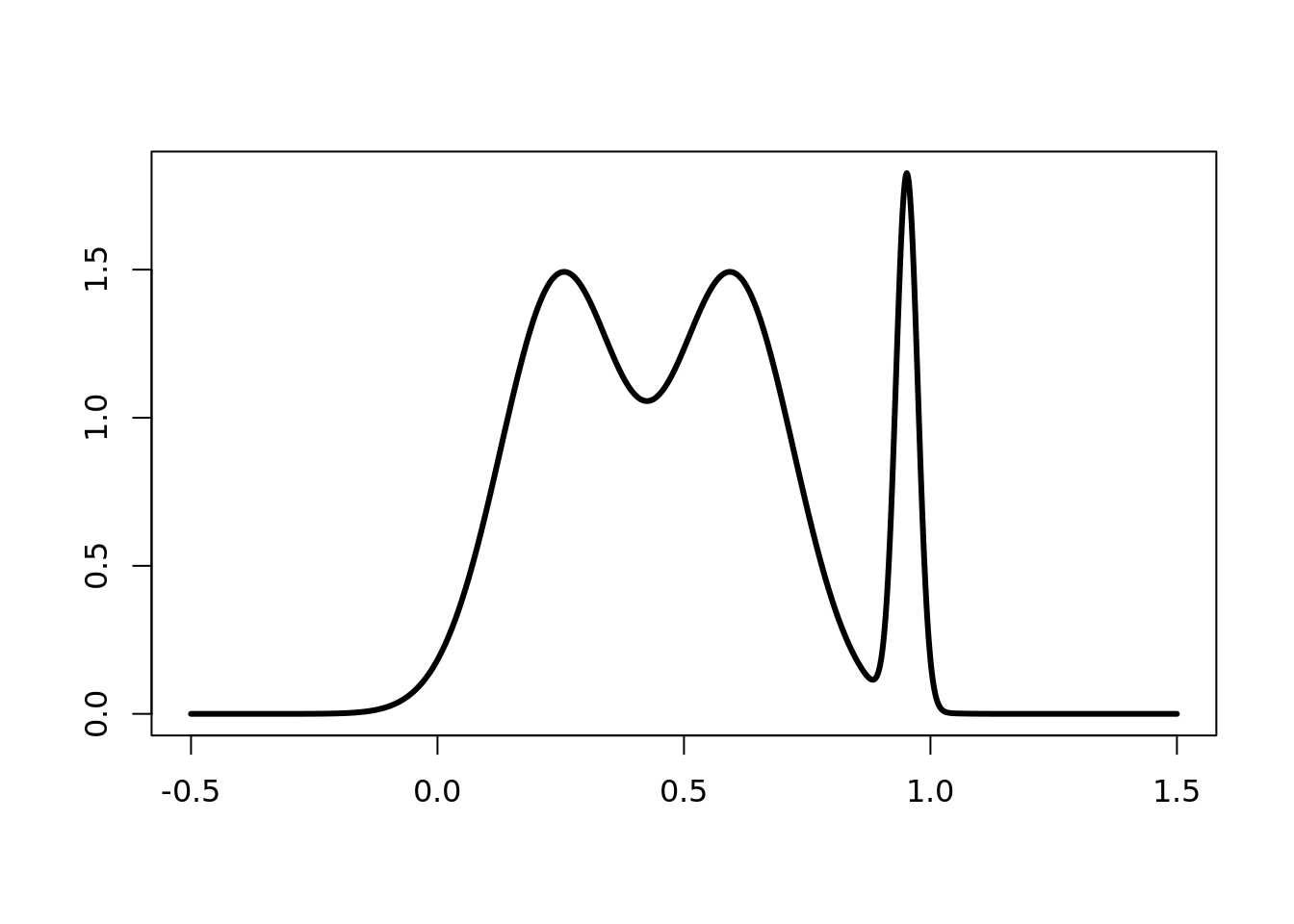}
        \caption{\thirdmodel{}}
    \end{subfigure}
    \begin{subfigure}[b]{\figurewidth}
        \centering
        \includegraphics[width=\subfigurewidth]{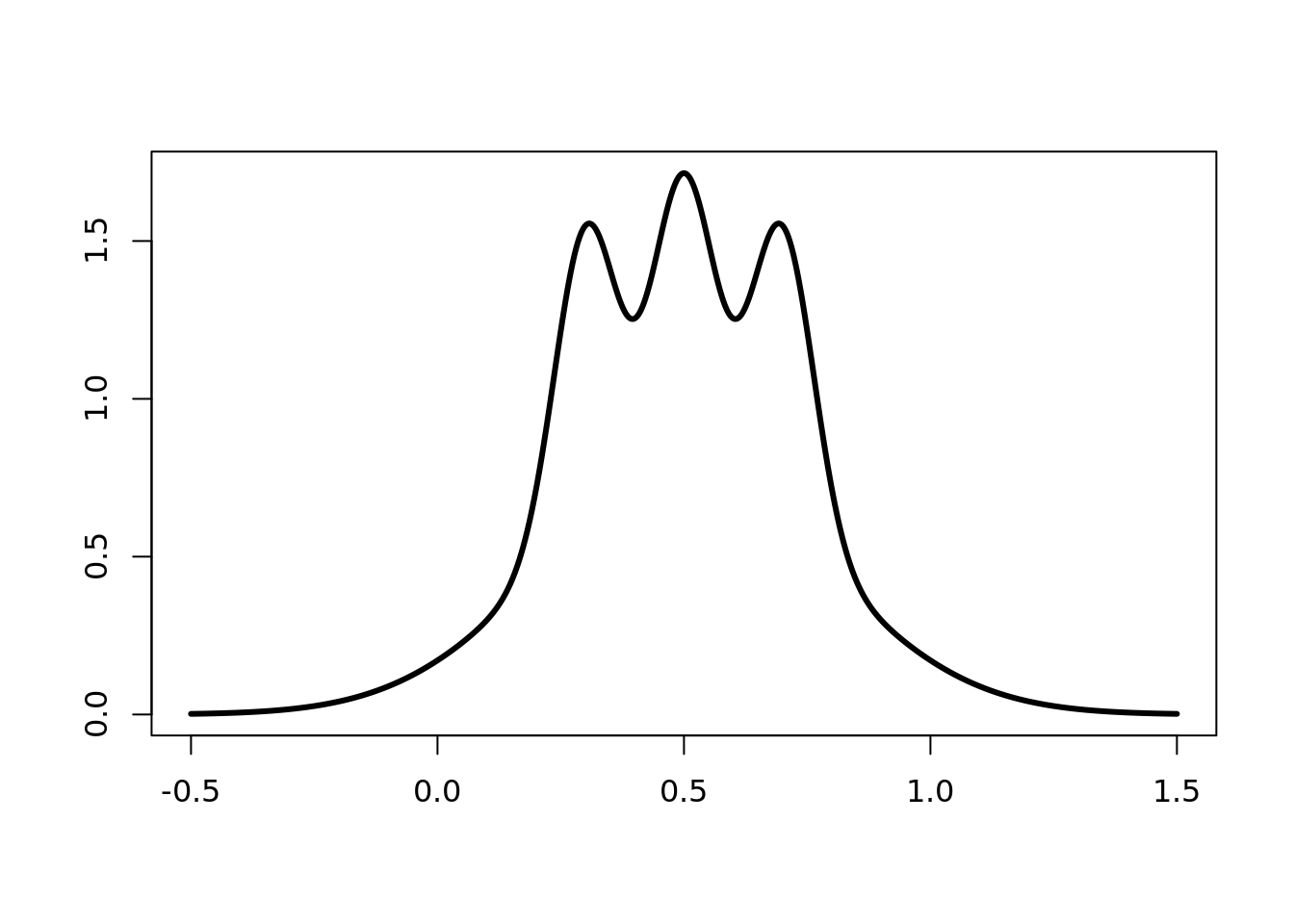}
        \caption{\fourthmodel{}}
    \end{subfigure}
    \begin{subfigure}[b]{\figurewidth}
        \centering
        \includegraphics[width=\subfigurewidth]{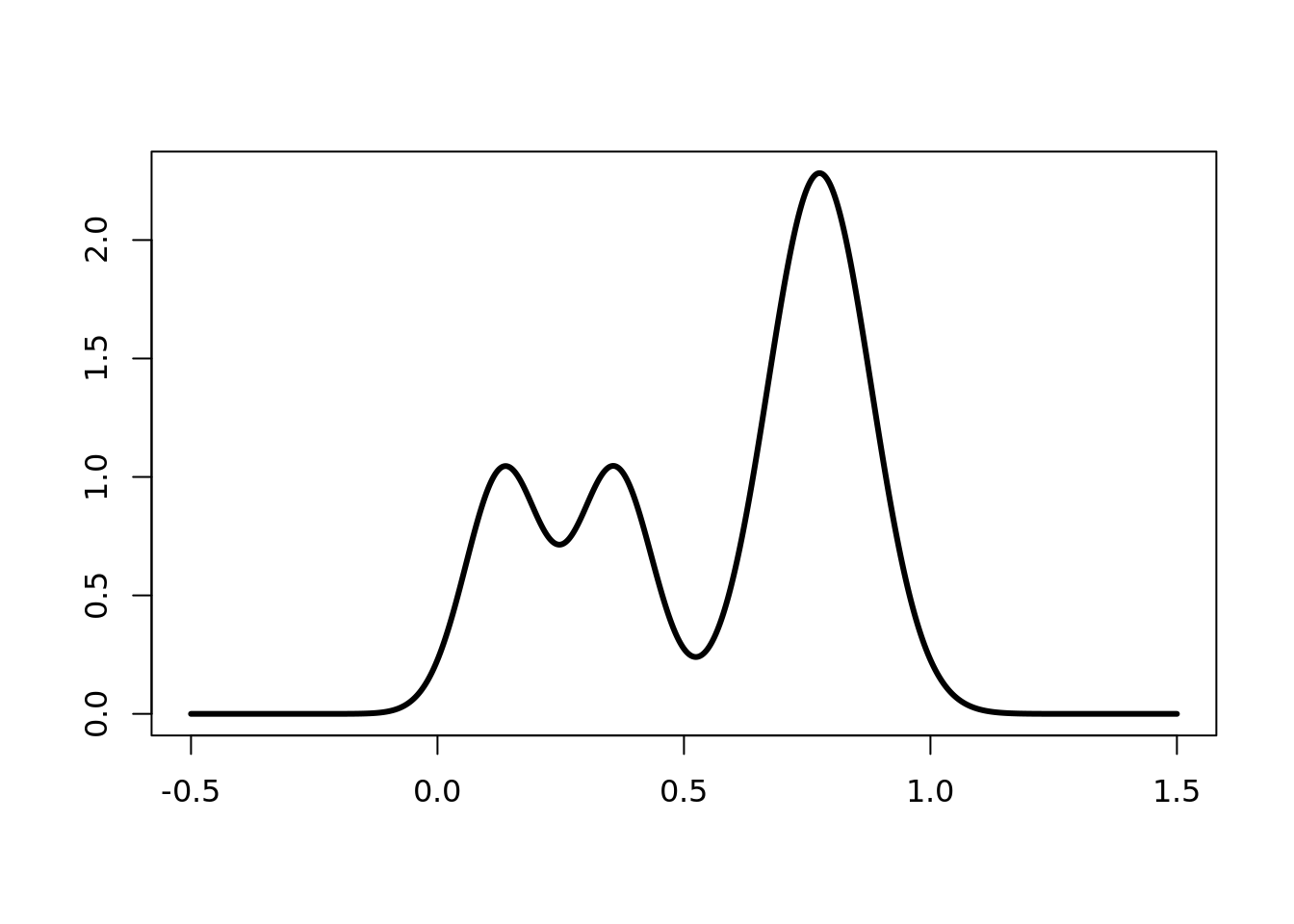}
        \caption{\fifthmodel{}}
        \label{fig:fifth-mixture-model}
    \end{subfigure}
    \caption{%
        Test-bed Gaussian mixture model \glspl*{pdf}.
    }
    \label{fig:mixture-models}
\end{figure}

\begin{table}
    \centering
    \footnotesize
    \begin{tabular}{cccc}
        \toprule
         &
        $\mixturecomponentmeans$
         &
        $\mixturecomponentvariances$
         &
        $\mixtureweights$
        \\
        \midrule
        \firstmodel{}
         &
        $(0.26, 0.79145, 0.5)$
         &
        $(0.01476, 0.01, 0.007)$
         &
        $(0.45, 0.33, 0.22)$
        \\
        \secondmodel{}
         &
        $(0.6, 0.10245, 0.93)$
         &
        $(0.01588, 0.0025, 0.0015)$
         &
        $(0.68, 0.22, 0.1)$
        \\
        \thirdmodel{}
         &
        $(0.25, 0.6, 0.95222)$
         &
        $(0.015, 0.015, 0.00049)$
         &
        $(0.45, 0.45, 0.1)$
        \\
        \fourthmodel{}
         &
        $(0.5, 0.3, 0.5, 0.7)$
         &
        $(0.08425, 0.004, 0.004, 0.004)$
         &
        $(0.55, 0.15, 0.15, 0.15)$
        \\
        \fifthmodel{}
         &
        $(0.7749, 0.1345, 0.36)$
         &
        $(0.011, 0.006, 0.006)$
         &
        $(0.6, 0.2, 0.2)$
        \\
        \bottomrule
    \end{tabular}
    \captionsetup{font=footnotesize}
    \caption{
        Test-bed Gaussian mixture model parameters.
    }
    \label{tab:mixture-model-parameters}
\end{table}

\subsection{Rankings}

The global and intermediate rankings in Table~\ref{tab:final-ranking} and Table~\ref{tab:intermediate-rankings}, respectively, are computed with the assistance of the \rlanguage{} package~\textit{ConsRank}~\citep{DAmbrosio2015}.
Namely, we used the routine \texttt{consrank} with all the default parameters except one for suppressing screen output.
In particular, we selected \texttt{algorithm = "BB"}, corresponding to the \textit{branch-and-bound} algorithm~\citep{Amodio2016}, and \texttt{full = FALSE}, meaning ties were allowed among ranks.

        \section{Extra results}

\label{sec:extra-results}

This section is a follow-up of the simulation study in Section~\ref{sec:results} of the manuscript, including further results and comments.

\subsection{Intermediate rankings}

The rankings $\intermediaterankings$ in \tablename~\ref{tab:intermediate-rankings} are depicted in \figurename~\ref{fig:intermediate-rankings}.
The Kendall correlations between ranks are presented in \figurename~\ref{fig:rank-correlations}.
With these auxiliary representations, let us further analyse the results in Section~\ref{sec:results} focusing on the less successful methods.

\begin{figure}
    \centering
    \begin{subfigure}[b]{\figurewidth}
        \centering
        \includegraphics[width=\subfigurewidth]{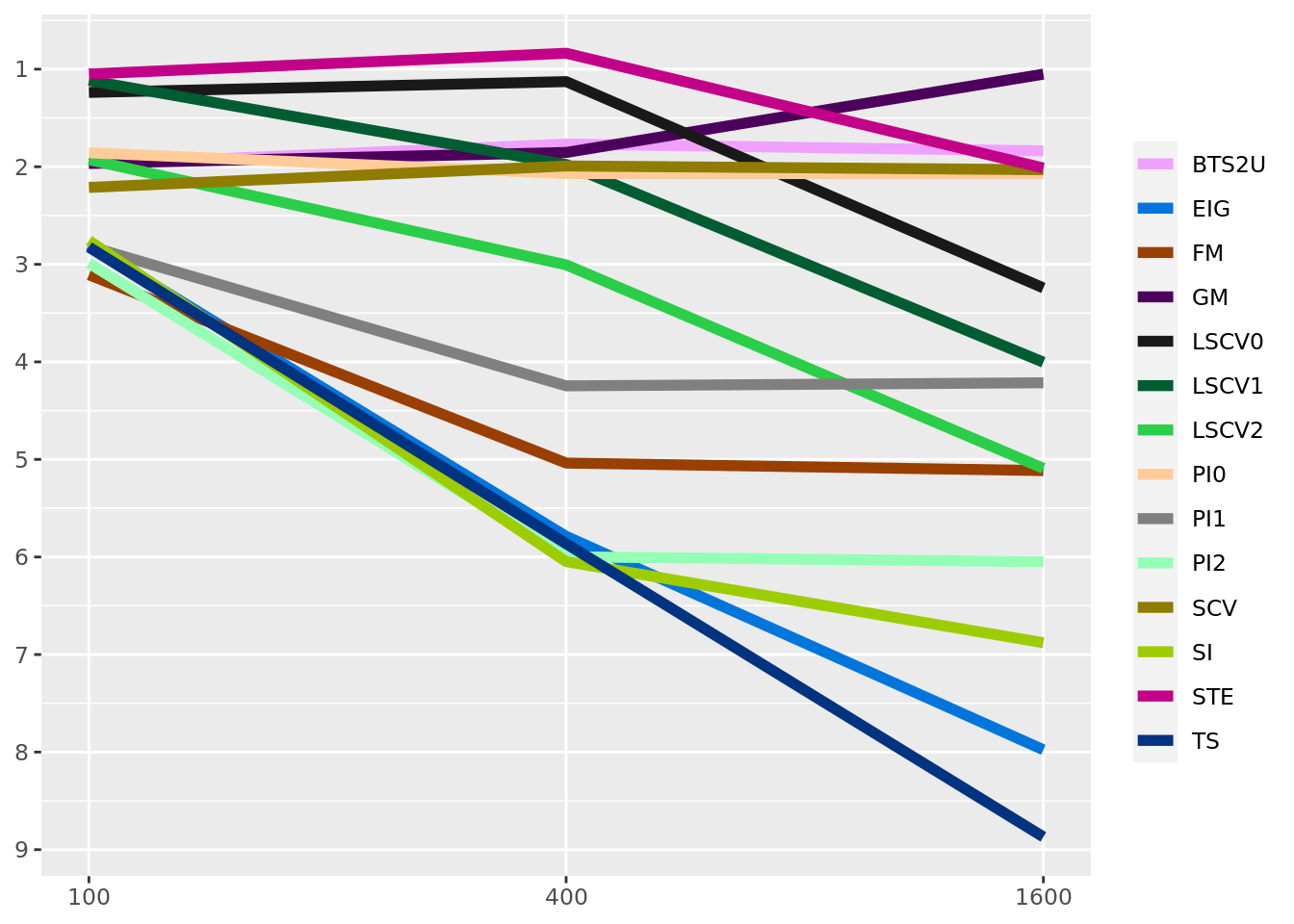}
        \caption{\firstmodel{}}
        \label{fig:rankings-m21}
    \end{subfigure}
    \begin{subfigure}[b]{\figurewidth}
        \centering
        \includegraphics[width=\subfigurewidth]{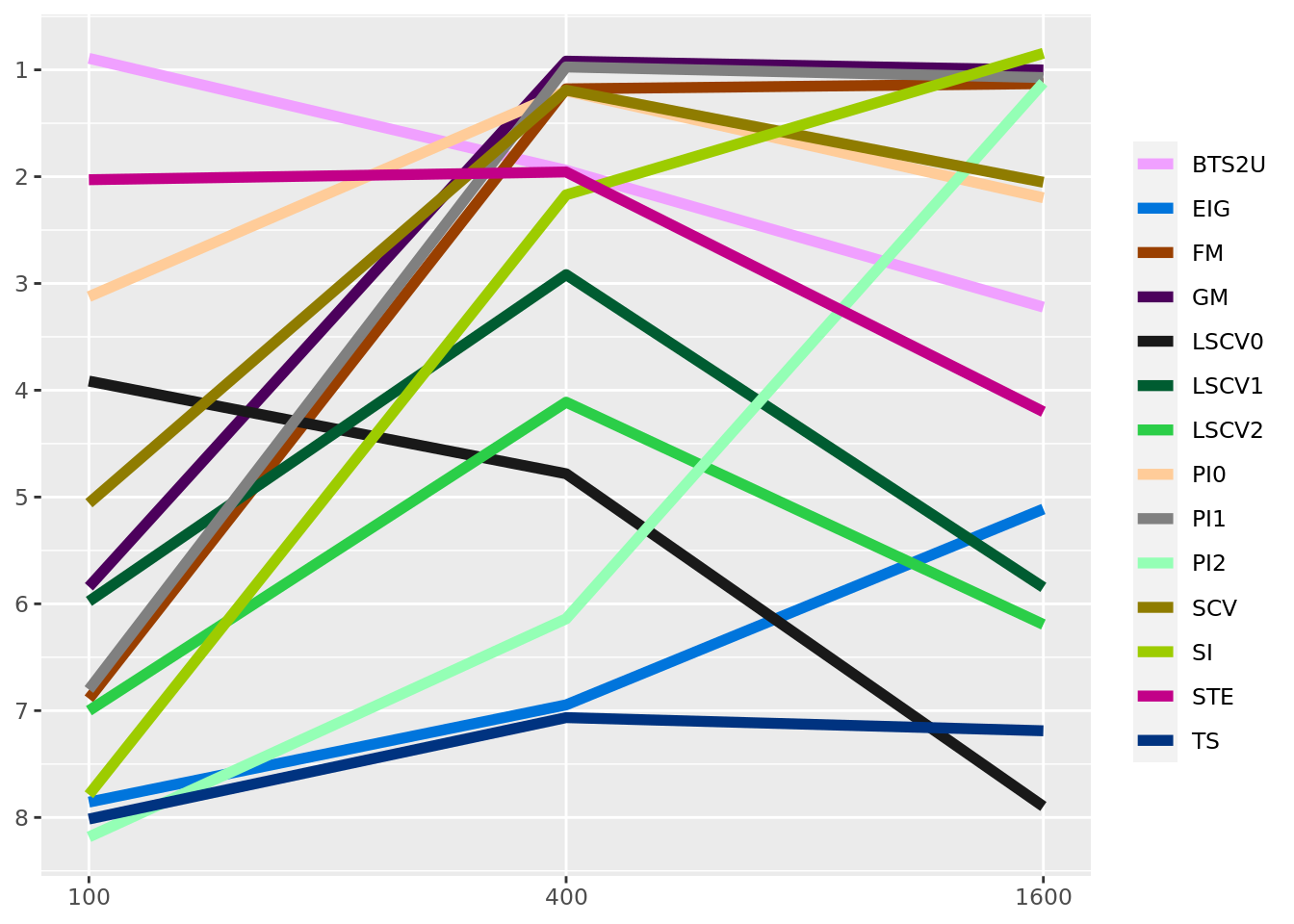}
        \caption{\secondmodel{}}
        \label{fig:rankings-m22}
    \end{subfigure}
    \begin{subfigure}[b]{\figurewidth}
        \centering
        \includegraphics[width=\subfigurewidth]{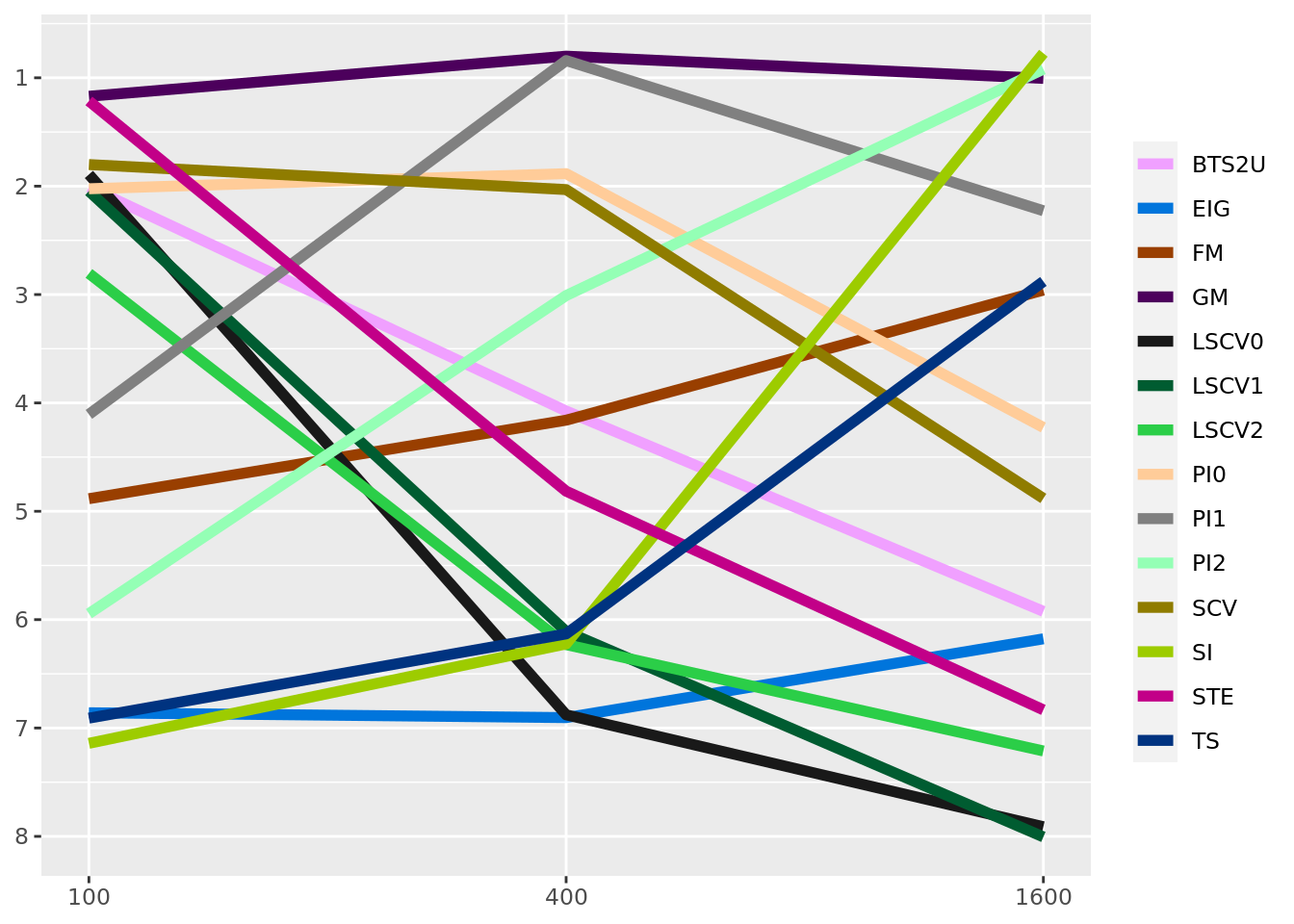}
        \caption{\thirdmodel{}}
        \label{fig:rankings-m23}
    \end{subfigure}
    \begin{subfigure}[b]{\figurewidth}
        \centering
        \includegraphics[width=\subfigurewidth]{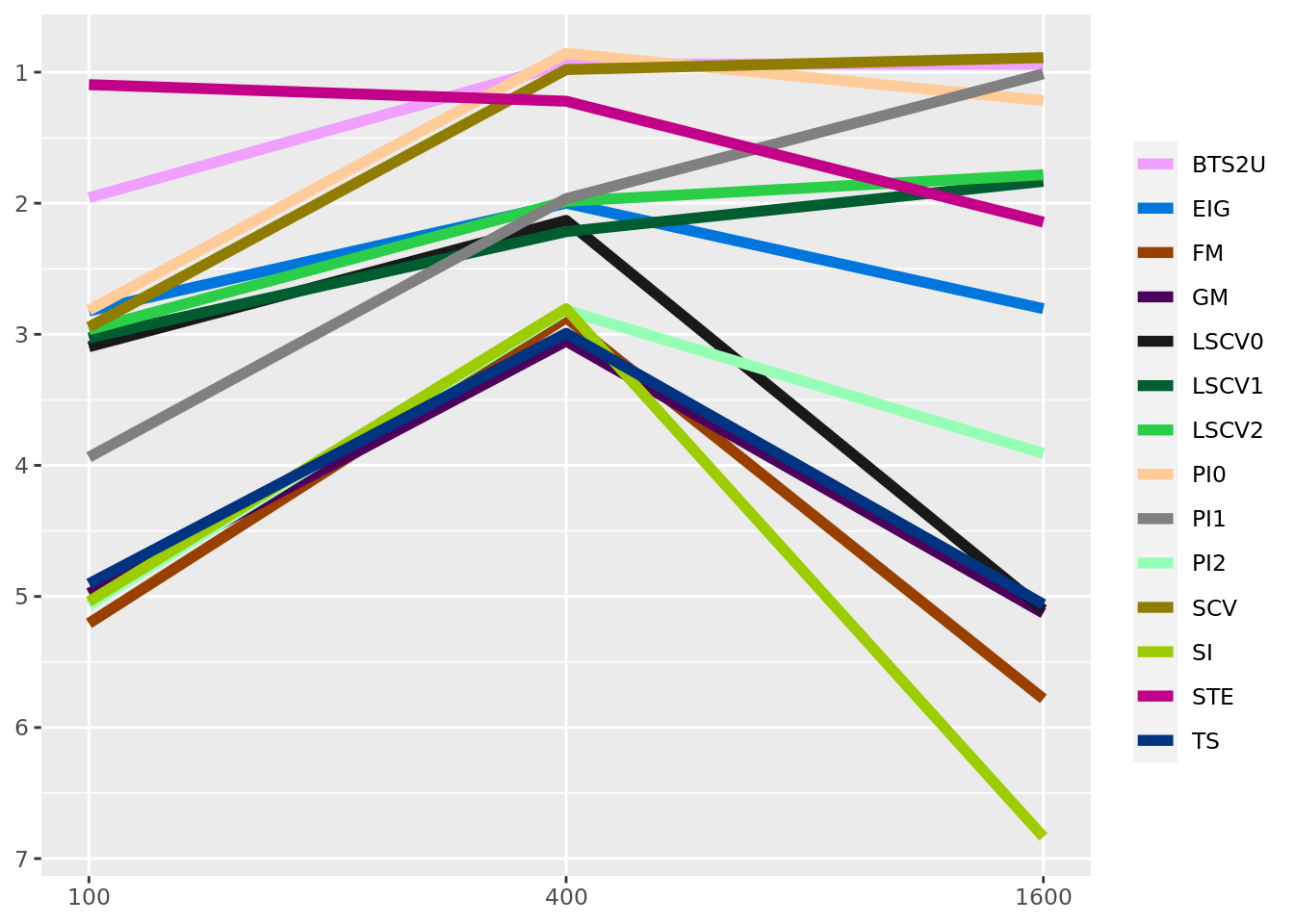}
        \caption{\fourthmodel{}}
        \label{fig:rankings-m24}
    \end{subfigure}
    \begin{subfigure}[b]{\figurewidth}
        \centering
        \includegraphics[width=\subfigurewidth]{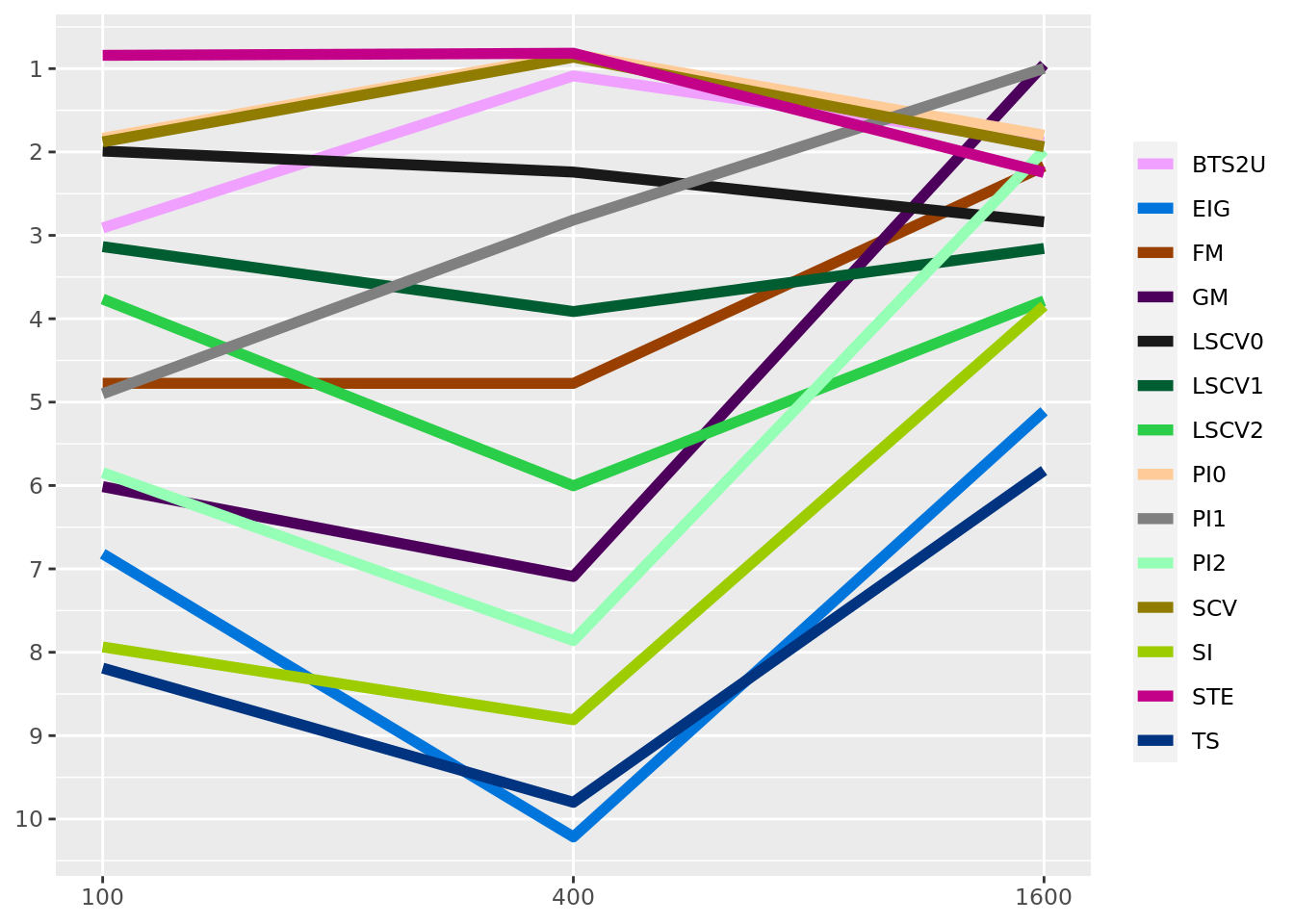}
        \caption{\fifthmodel{}}
        \label{fig:rankings-m25}
    \end{subfigure}
    \caption{%
        Intermediate rankings in \tablename~\ref{tab:intermediate-rankings} by test-bed.
        In every subfigure, the horizontal and vertical axes correspond to the sample size and the rank of the method, respectively.
        A small amount of jitter has been added to the ranks to appreciate overlapping trajectories better.
    }
    \label{fig:intermediate-rankings}
\end{figure}

\begin{figure}
    \centering
    \includegraphics[width=0.7\textwidth]{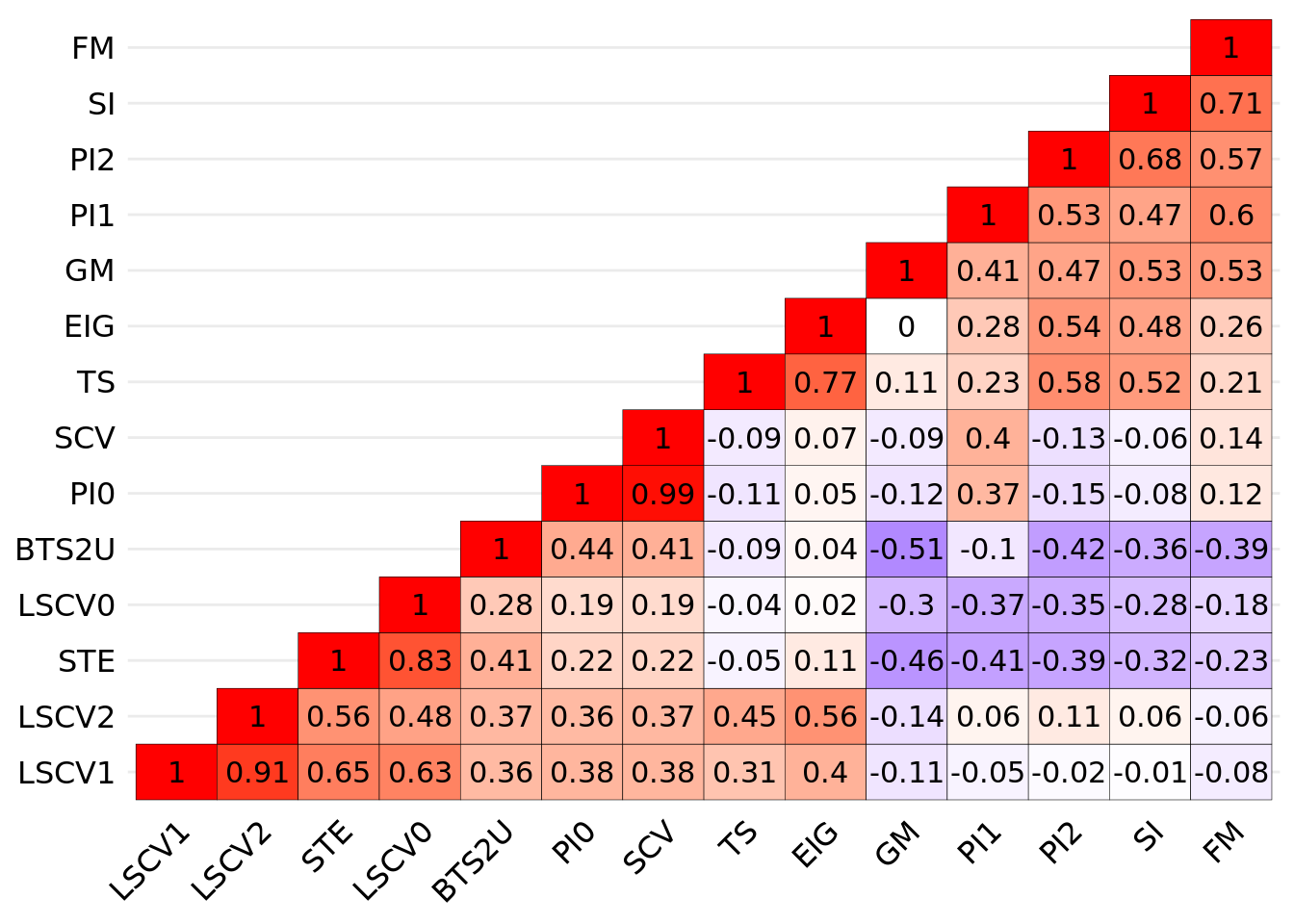}
    \caption{%
        Kendall correlations between the intermediate ranks in \tablename~\ref{tab:intermediate-rankings}.
    }
    \label{fig:rank-correlations}
\end{figure}

The two second-ranked methods apart from \kdepione{}, i.e., \kdelscvzero{} and \kdelscvone{}, also have some success.
Contrary to \kdepione{}, the \gls*{lscv} methods tend to under-smooth, helping to detect the short-lived mode in \figurename~\ref{fig:first-mixture-model}.
However, under-smoothing is counter-productive most of the time, leading to spurious modes as $\samplesize$ grows.
The last \gls*{lscv} member, the third-ranked \kdelscvtwo{}, underperforms across all settings, securing zero first ranks.

The fifth-ranked \kdepitwo{} goes deeper into regularisation than \kdepione{}, becoming too insensitive for most cases (see \figurename~\ref{fig:rankings-m21}, \figurename~\ref{fig:rankings-m24} and \figurename~\ref{fig:rankings-m25}).
There are two first ranks in \secondmodel{} and \thirdmodel{}, though, always with $\samplesize = 1600$.
Especially in \secondmodel{}, all the modes are long-lived and well-separated.
Hence, the risk of not detecting them is outweighed by that of making up spurious ones.
The fourth-placed method, \fishermarron{}, also shows top performance in \secondmodel{}-400 and \secondmodel{}-1600 for the same reason.

Among the sixth-ranked methods, we still see a top performance by \silverman{} in \secondmodel{}-1600 and \thirdmodel{}-1600.
Even though \silverman{} and \fishermarron{} appear highly correlated in \figurename~\ref{fig:rank-correlations}, the former is dominated by the latter across most configurations.
Nonetheless, both are \textit{niche} methods that excel in some straightforward cases, with large sample sizes and well-separated modes.
The same cannot be said about \tautstring{} and \telepathicbootstrap{}, which have no first positions and usually close the ranking, as depicted in \figurename~\ref{fig:rankings-m21}, \figurename~\ref{fig:rankings-m22} and \figurename~\ref{fig:rankings-m25}.

\tablename~\ref{tab:intermediate-rankings} also reveals noticeable differences among the first-ranked methods.
\gaussianmixture{} negatively correlates with the other four, which are positively correlated among themselves.
For instance, \gaussianmixture{} stands out with \thirdmodel{} while the rest struggle (see \figurename~\ref{fig:rankings-m23}) and the opposite happens with \fourthmodel{}.
On the other hand, \kdepizero{} and \kdelscv{} usually perform best with medium to large sample sizes (see \figurename~\ref{fig:rankings-m22}), whereas \kdeste{} operates better with small to medium ones (see \figurename~\ref{fig:rankings-m21}).
The three \glspl*{kde} are more or less even in the middle (\figurename~\ref{fig:rankings-m24} and \figurename~\ref{fig:rankings-m25}).
The performance of \btsselected{} is mainly correlated with that of \kdepizero{} and, to a lesser extent, \kdelscv{} and \kdeste{}.

Some curiosities are found in \tablename~\ref{tab:intermediate-rankings}.
\btsselected{} is one of only three methods, alongside \gaussianmixture{} and \kdeste{}, with an unmatched first rank, the \secondmodel{}-100 mentioned in the manuscript.
Additionally, \btsselected{} is the only method not named \gaussianmixture{} capable of sustaining the first rank at least once in each of the three sample sizes.
On the other hand, no method ranks first at least once in each of the five test-beds.
The most versatile methods in that sense are \gaussianmixture{}, \kdeste{} and \kdepione{}, with four.

\subsection{Distribution}

We will now look at the distribution of the predicted \gls*{nom} in each of the $\numtestbedstimessamplesizes = 15$ sampling configurations through \figurename~\ref{fig:distributions-m21}, \figurename~\ref{fig:distributions-m22}, \figurename~\ref{fig:distributions-m23}, \figurename~\ref{fig:distributions-m24}, and \figurename~\ref{fig:distributions-m25}.
Here, the reader will see the \textit{raw} accuracies and variabilities of each method in each scenario.
We shall concentrate our commentary on some selected cases as the conclusions for the rest are qualitatively very similar.

\figurename~\ref{fig:distributions-m22-100} shows the top performance mentioned above for \btsselected{}, with a small sample size.
\btsselected{} passes \kdeste{} based on the significance of the results.
The broad middle block of \secondmodel{} traps \kdeste{} and \kdelscvzero{} in over-predicting four modes.
Meanwhile, \gaussianmixture{} misses the mode to the right many times.
In turn, \kdepione{}, \fishermarron{} and \telepathicbootstrap{} behave even more conservatively, while \silverman{} and \kdepitwo{} get stuck at one and two modes, respectively.
Interestingly, \fishermarron{} extends beyond three modes more than \btsselected{}.

A more complicated scenario is in \figurename~\ref{fig:distributions-m24-1600} despite the large sample size.
The model \fourthmodel{} is like the classic \textit{claw} density used by~\citet{Davies2004} but with three short-lived \textit{fingers} instead of five.
\tautstring{} performs very well with the \textit{claw} but not with \fourthmodel{}, almost exclusively predicting one mode, similar to \fishermarron{}.
Overall, \fourthmodel{} produces high variability, and almost all the methods widely spread their predictions over the one to five range.
Having a lower accuracy, \btsselected{} ranks first, tied with three methods (\kdepizero{}, \kdepione{} and \kdelscv{}) based on statistical significance.
Very surprisingly, \telepathicbootstrap{} manages to rank third ahead of \gaussianmixture{} and \kdelscvzero{}.
The latter is \textit{off the chart}, as most of its predictions are above five.
Lastly, \silverman{} lies at the bottom, predicting one mode 100\% of the time.

The variability in \figurename~\ref{fig:distributions-m25-400} is the lowest of all the three cases considered.
On the one hand, \secondmodel{} is less complex than \fifthmodel{}, but data is more scarce in \figurename~\ref{fig:distributions-m22-100}.
On the other, \fourthmodel{} surpasses the complexity of \fifthmodel{}, but with larger samples in \figurename~\ref{fig:distributions-m24-1600}.
In this case, \btsselected{}, \kdepizero{}, \kdelscv{} and \kdeste{} rank at first, having very similar profiles between two and four modes.
Only the three \gls*{lscv} methods have predictions over four, while \silverman{}, \tautstring{} and \telepathicbootstrap{} are the only ones betting on one mode.
Finally, \fishermarron{} surprisingly outperforms \gaussianmixture{}, which is excessively conservative.

\ifnum\journalflag=0
    \newcommand{\subfigurewidthtwoandone}{0.7\textwidth}
\fi

\newcommand{\graphicswidthtwoandone}{\textwidth}

\begin{figure}
    \centering
    \begin{subfigure}[b]{\subfigurewidthtwoandone}
        \centering
        \includegraphics[width=\graphicswidthtwoandone]{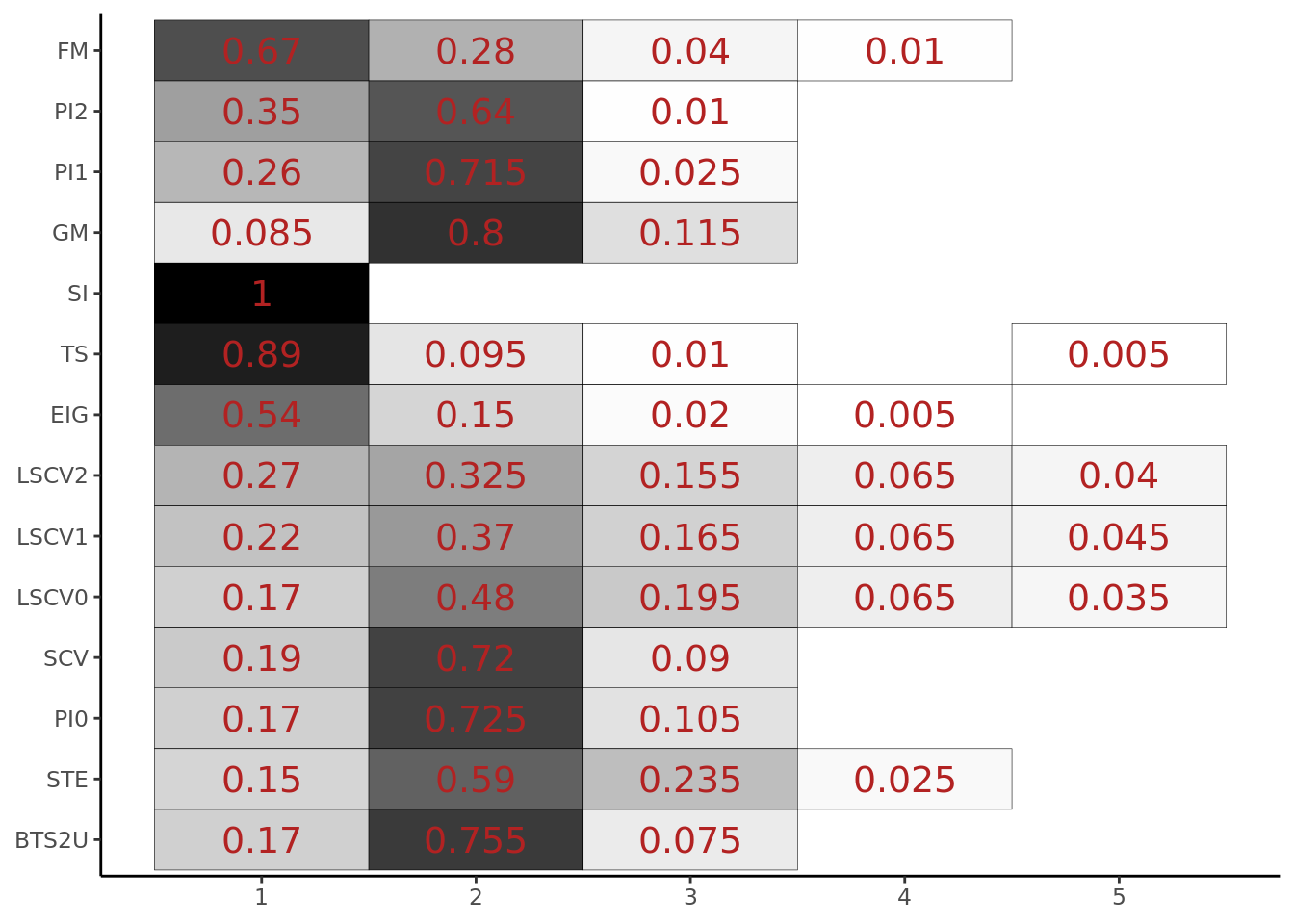}
        \caption{$\samplesize = 100$}
        \label{fig:distributions-m21-100}
    \end{subfigure}
    \begin{subfigure}[b]{\subfigurewidthtwoandone}
        \centering
        \includegraphics[width=\graphicswidthtwoandone]{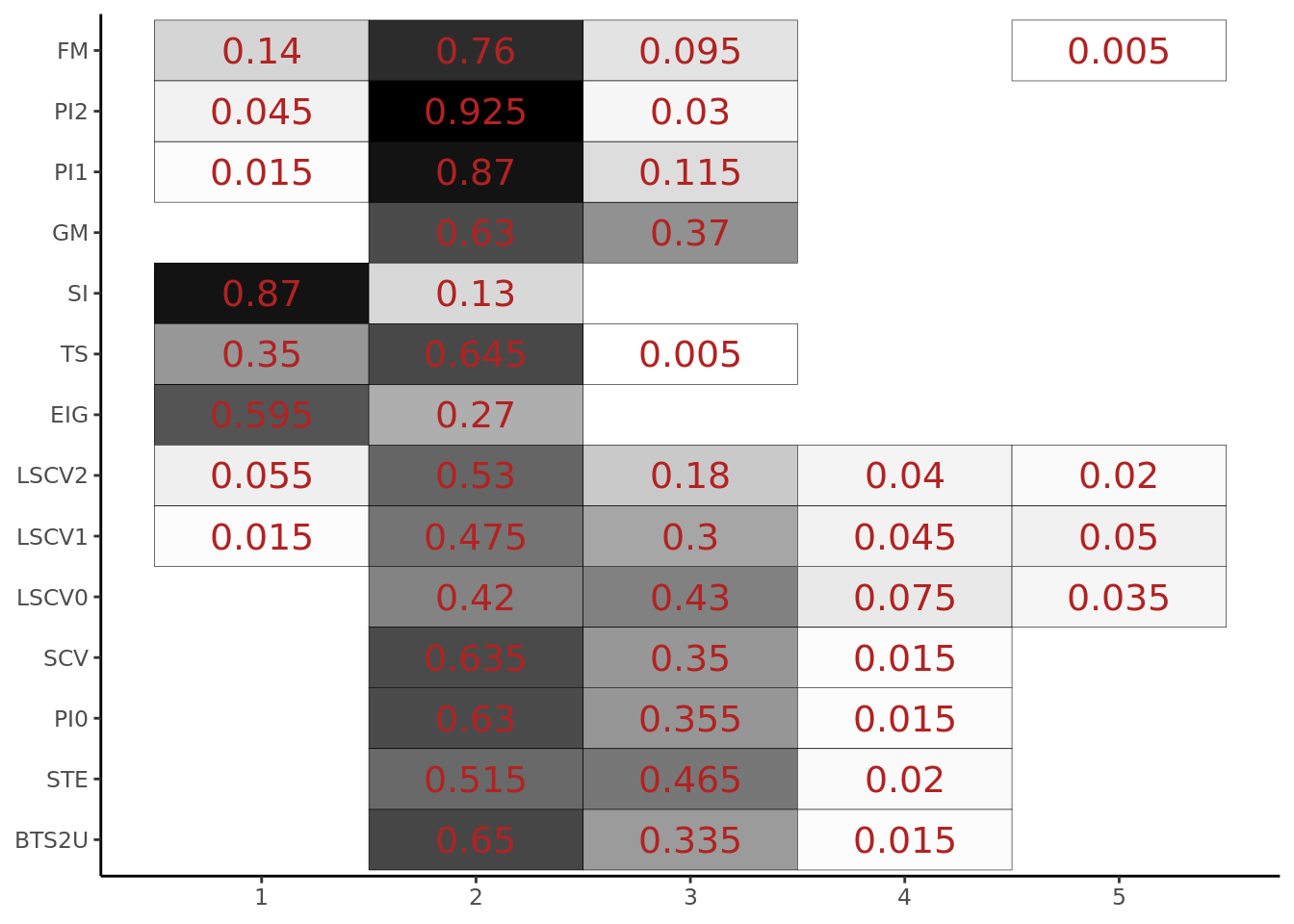}
        \caption{$\samplesize = 400$}
        \label{fig:distributions-m21-400}
    \end{subfigure}
    \begin{subfigure}[b]{\subfigurewidthtwoandone}
        \centering
        \includegraphics[width=\graphicswidthtwoandone]{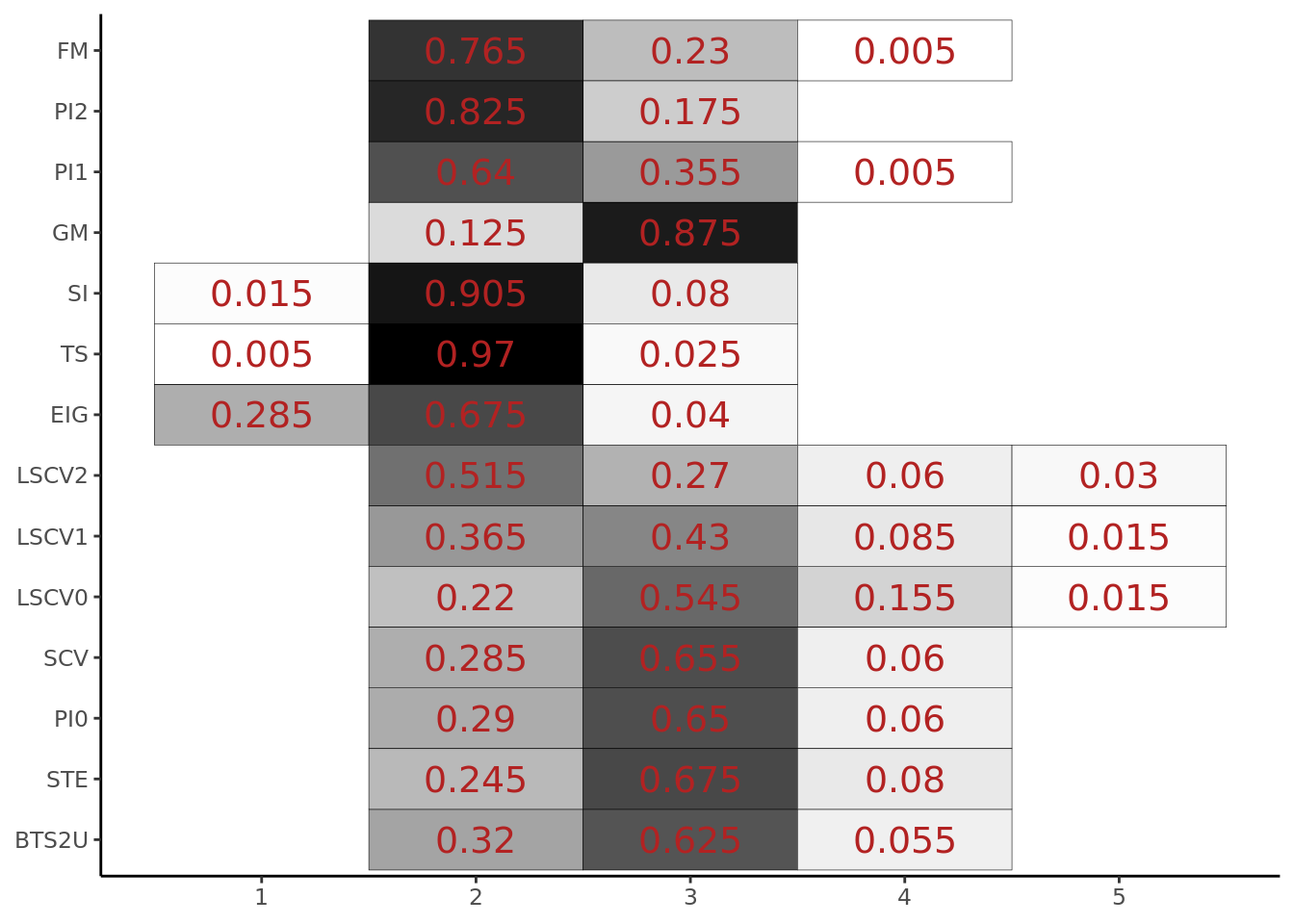}
        \caption{$\samplesize = 1600$}
        \label{fig:distributions-m21-1600}
    \end{subfigure}
    \caption{%
        Distribution of the predicted \gls*{nom} for the \firstmodel{} test-bed for several sample sizes $\samplesize$.
    }
    \label{fig:distributions-m21}
\end{figure}

\begin{figure}
    \centering
    \begin{subfigure}[b]{\subfigurewidthtwoandone}
        \centering
        \includegraphics[width=\graphicswidthtwoandone]{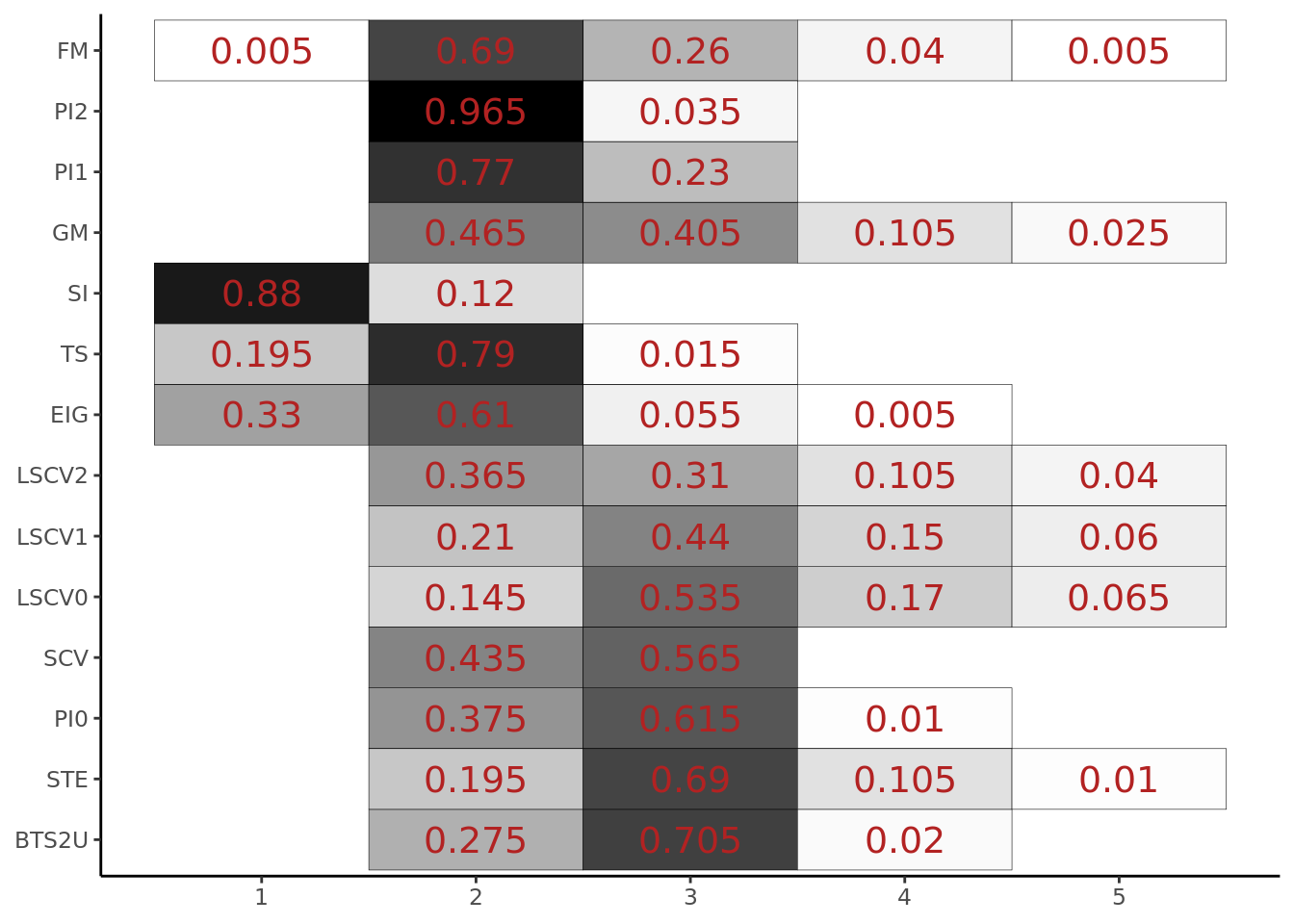}
        \caption{$\samplesize = 100$}
        \label{fig:distributions-m22-100}
    \end{subfigure}
    \begin{subfigure}[b]{\subfigurewidthtwoandone}
        \centering
        \includegraphics[width=\graphicswidthtwoandone]{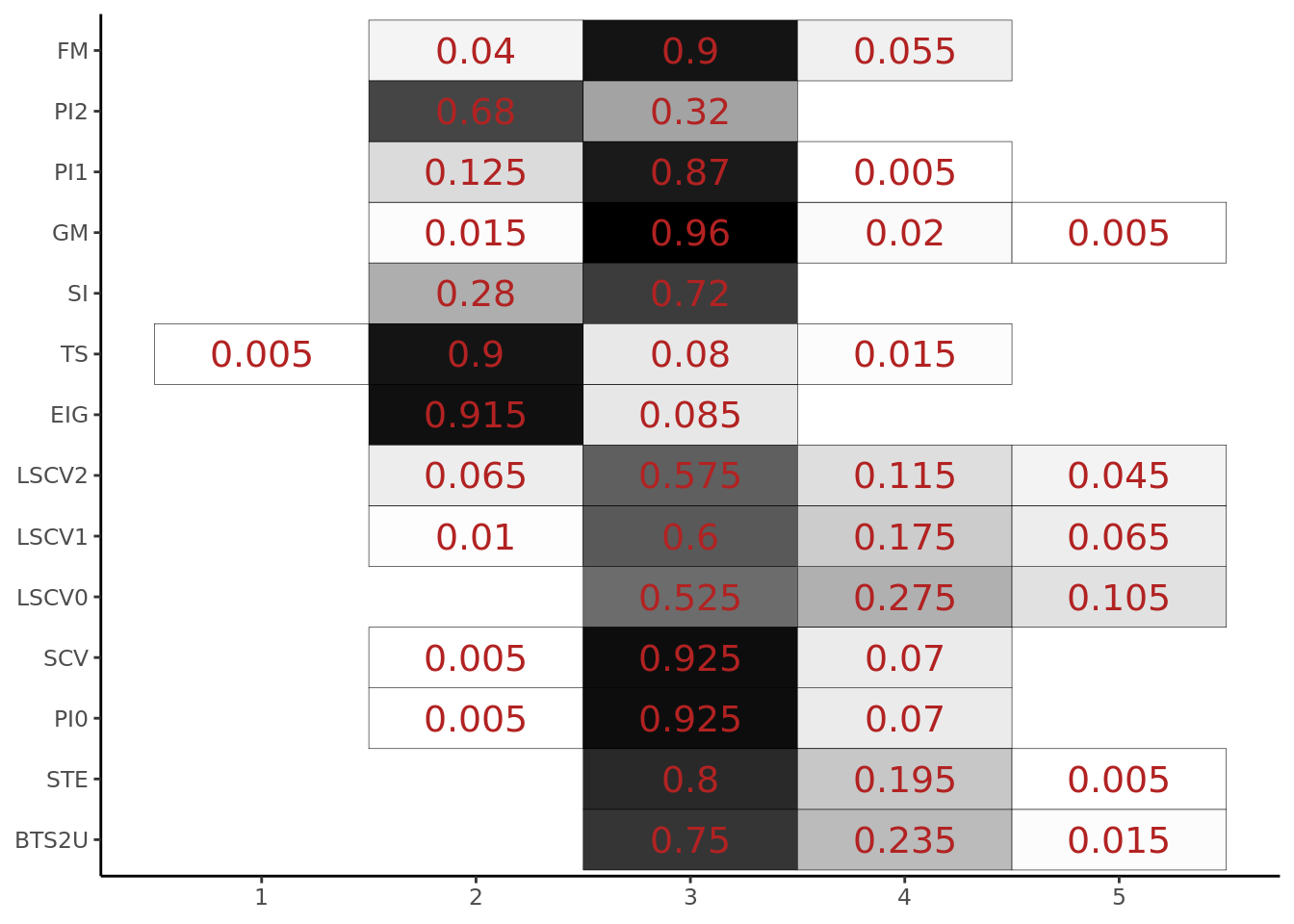}
        \caption{$\samplesize = 400$}
        \label{fig:distributions-m22-400}
    \end{subfigure}
    \begin{subfigure}[b]{\subfigurewidthtwoandone}
        \centering
        \includegraphics[width=\graphicswidthtwoandone]{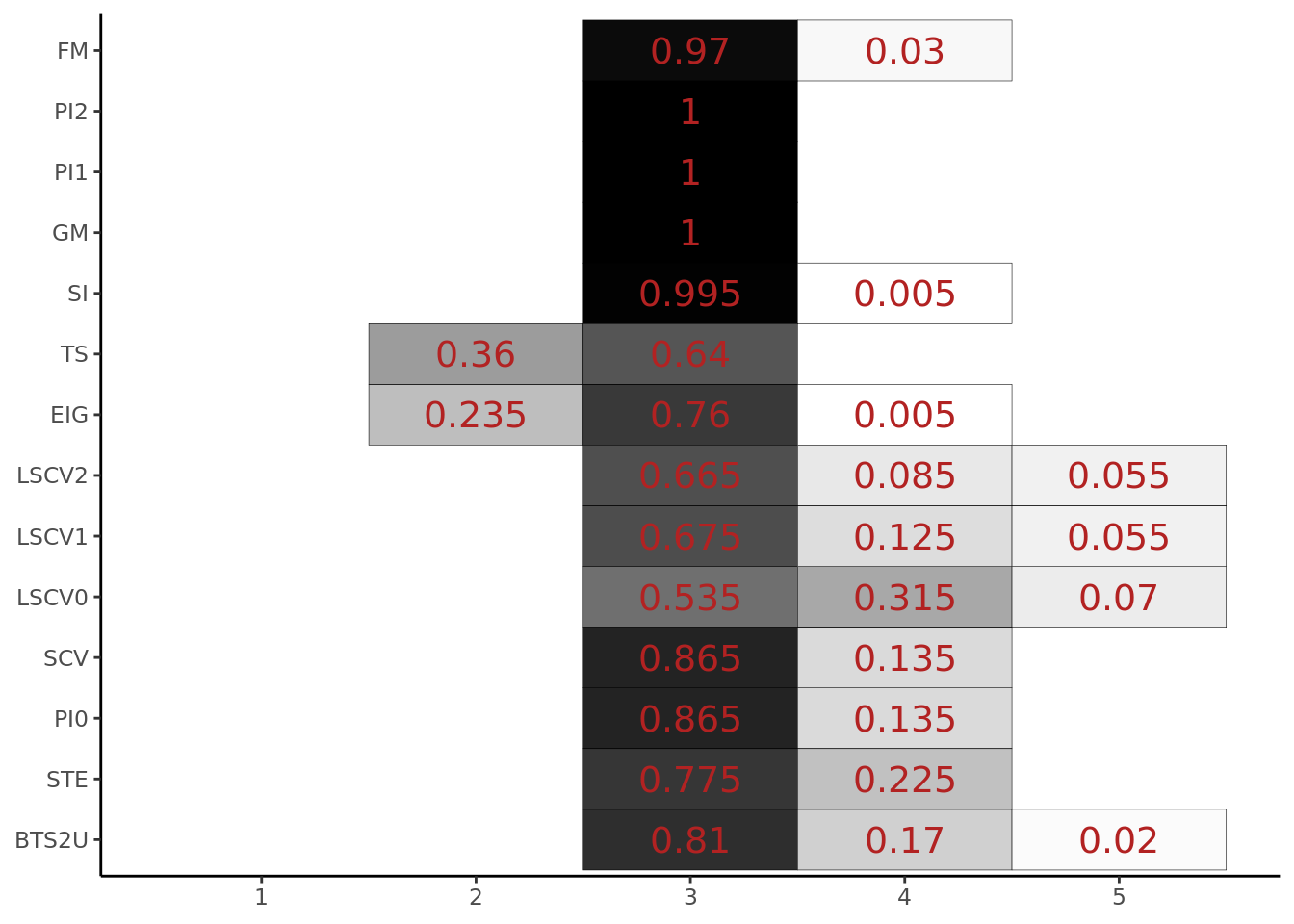}
        \caption{$\samplesize = 1600$}
        \label{fig:distributions-m22-1600}
    \end{subfigure}
    \caption{%
        Distribution of the predicted \gls*{nom} for the \secondmodel{} test-bed for several sample sizes $\samplesize$.
    }
    \label{fig:distributions-m22}
\end{figure}

\begin{figure}
    \centering
    \begin{subfigure}[b]{\subfigurewidthtwoandone}
        \centering
        \includegraphics[width=\graphicswidthtwoandone]{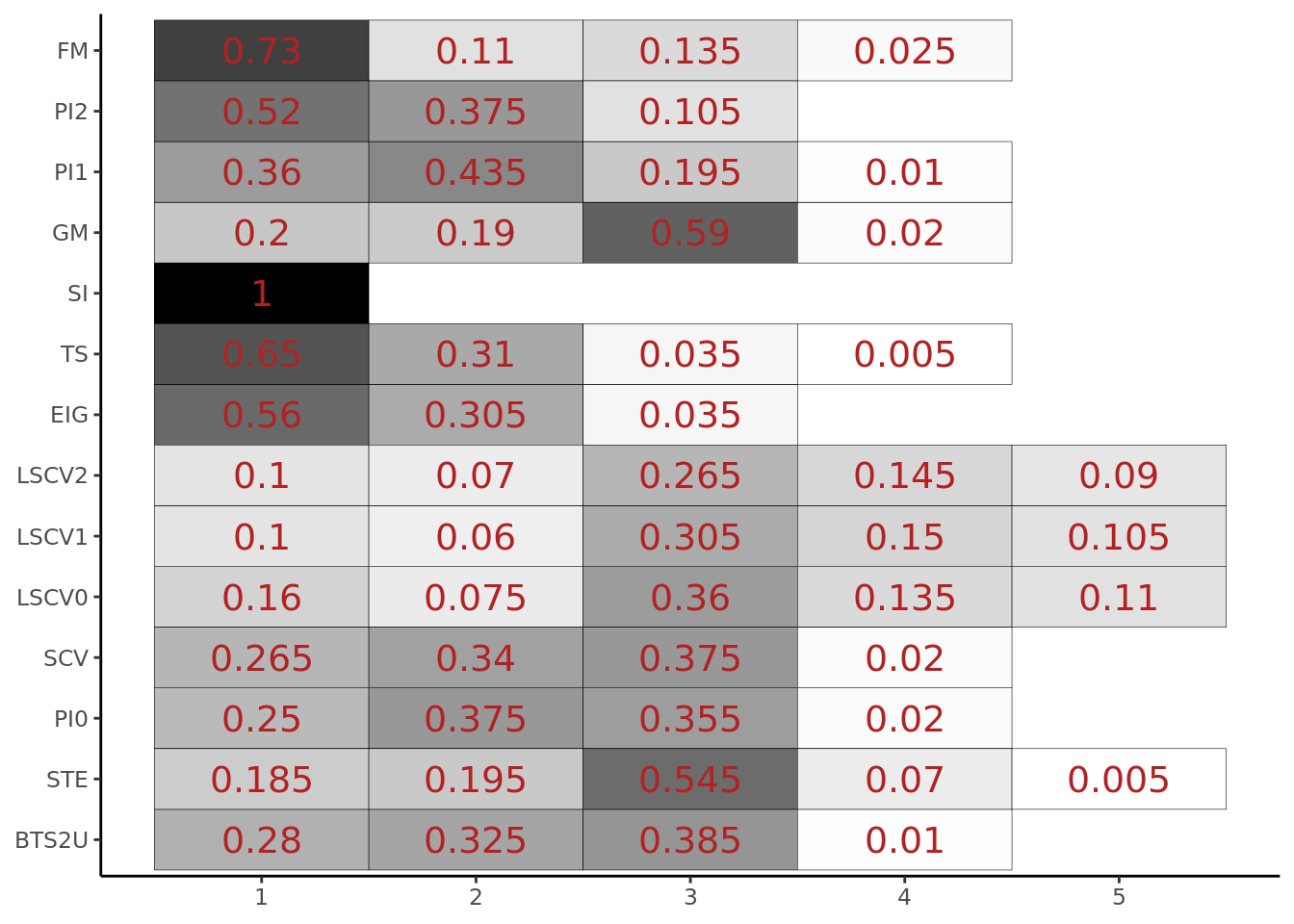}
        \caption{$\samplesize = 100$}
        \label{fig:distributions-m23-100}
    \end{subfigure}
    \begin{subfigure}[b]{\subfigurewidthtwoandone}
        \centering
        \includegraphics[width=\graphicswidthtwoandone]{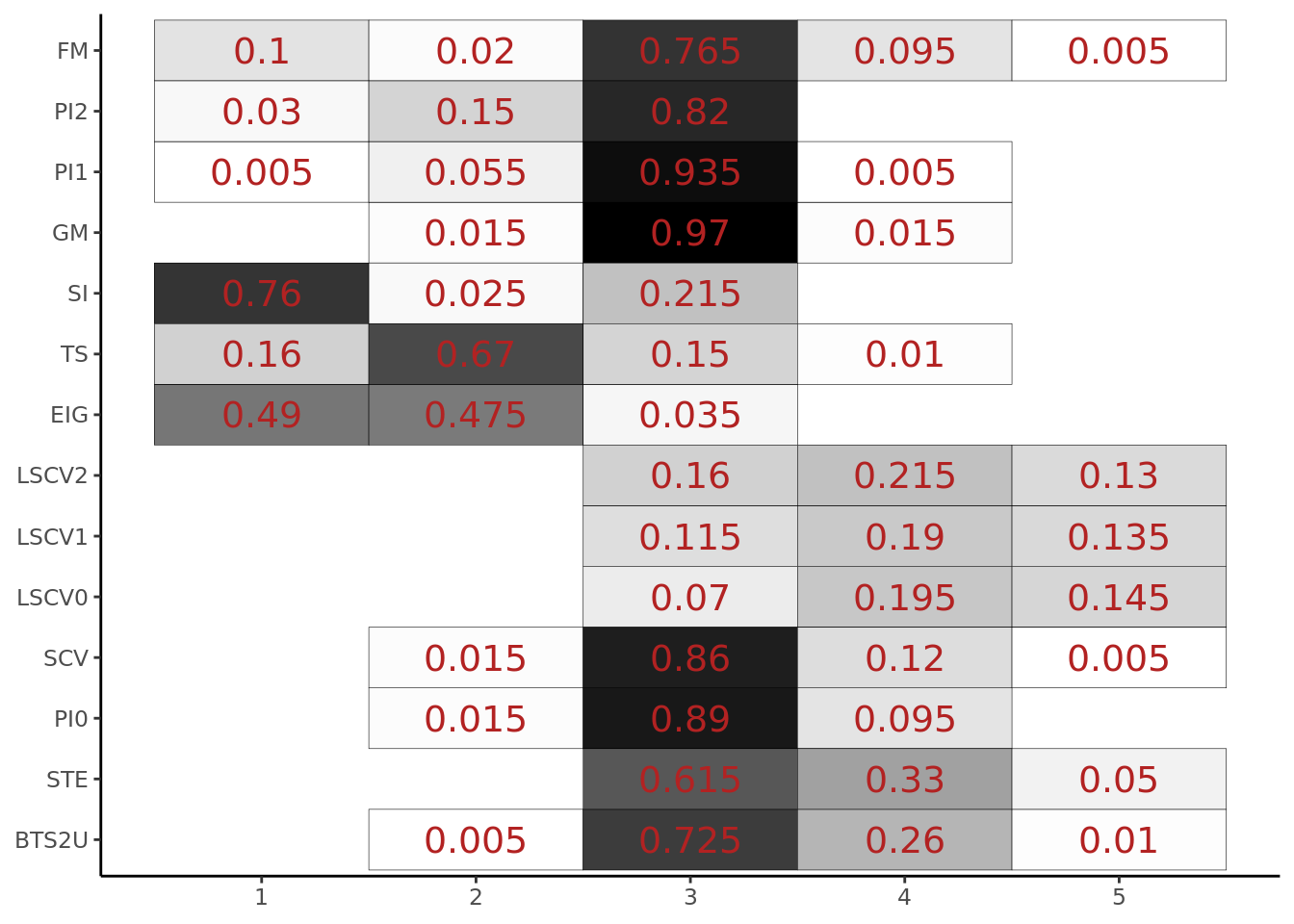}
        \caption{$\samplesize = 400$}
        \label{fig:distributions-m23-400}
    \end{subfigure}
    \begin{subfigure}[b]{\subfigurewidthtwoandone}
        \centering
        \includegraphics[width=\graphicswidthtwoandone]{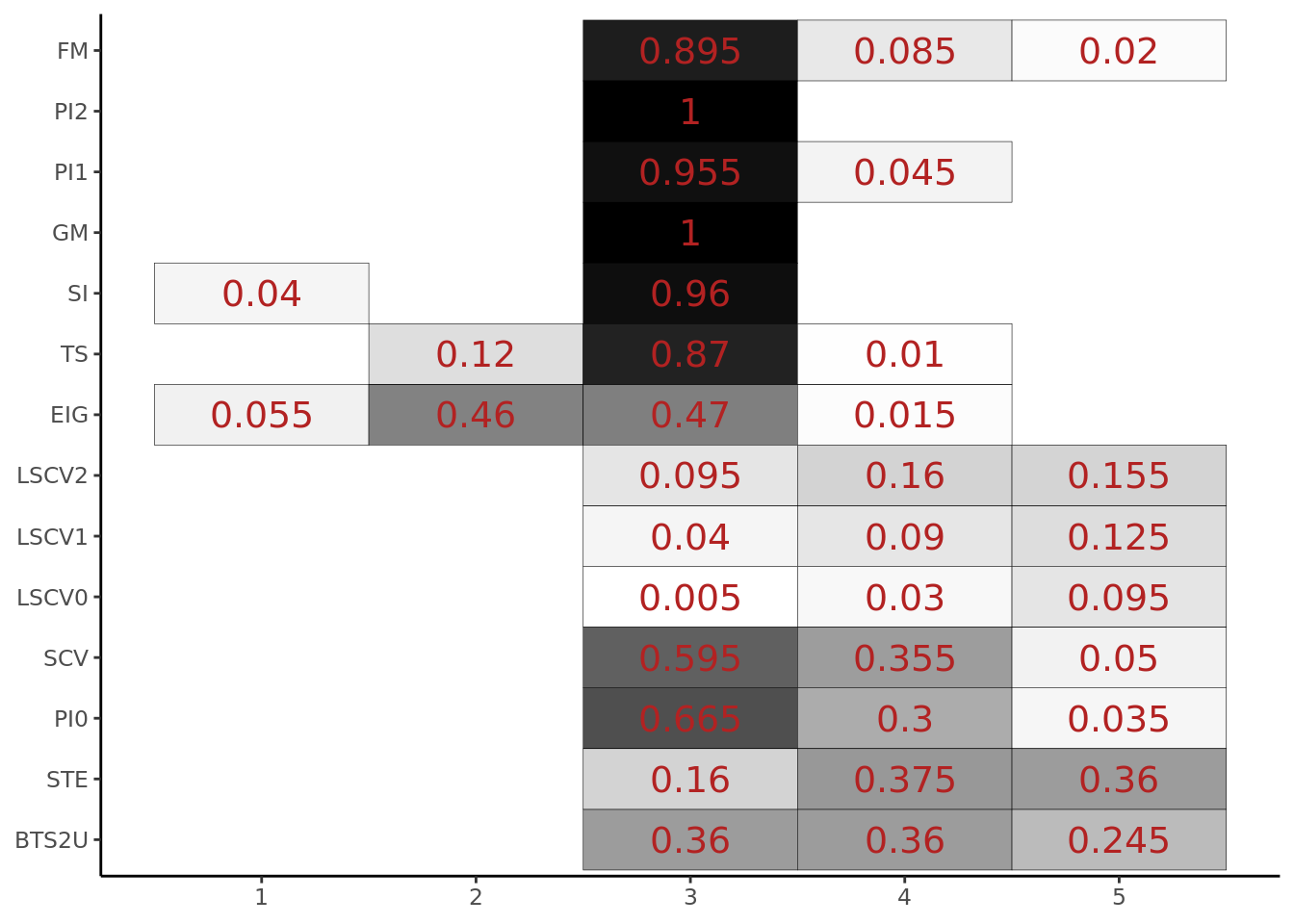}
        \caption{$\samplesize = 1600$}
        \label{fig:distributions-m23-1600}
    \end{subfigure}
    \caption{%
        Distribution of the predicted \gls*{nom} for the \thirdmodel{} test-bed for several sample sizes $\samplesize$.
    }
    \label{fig:distributions-m23}
\end{figure}

\begin{figure}
    \centering
    \begin{subfigure}[b]{\subfigurewidthtwoandone}
        \centering
        \includegraphics[width=\graphicswidthtwoandone]{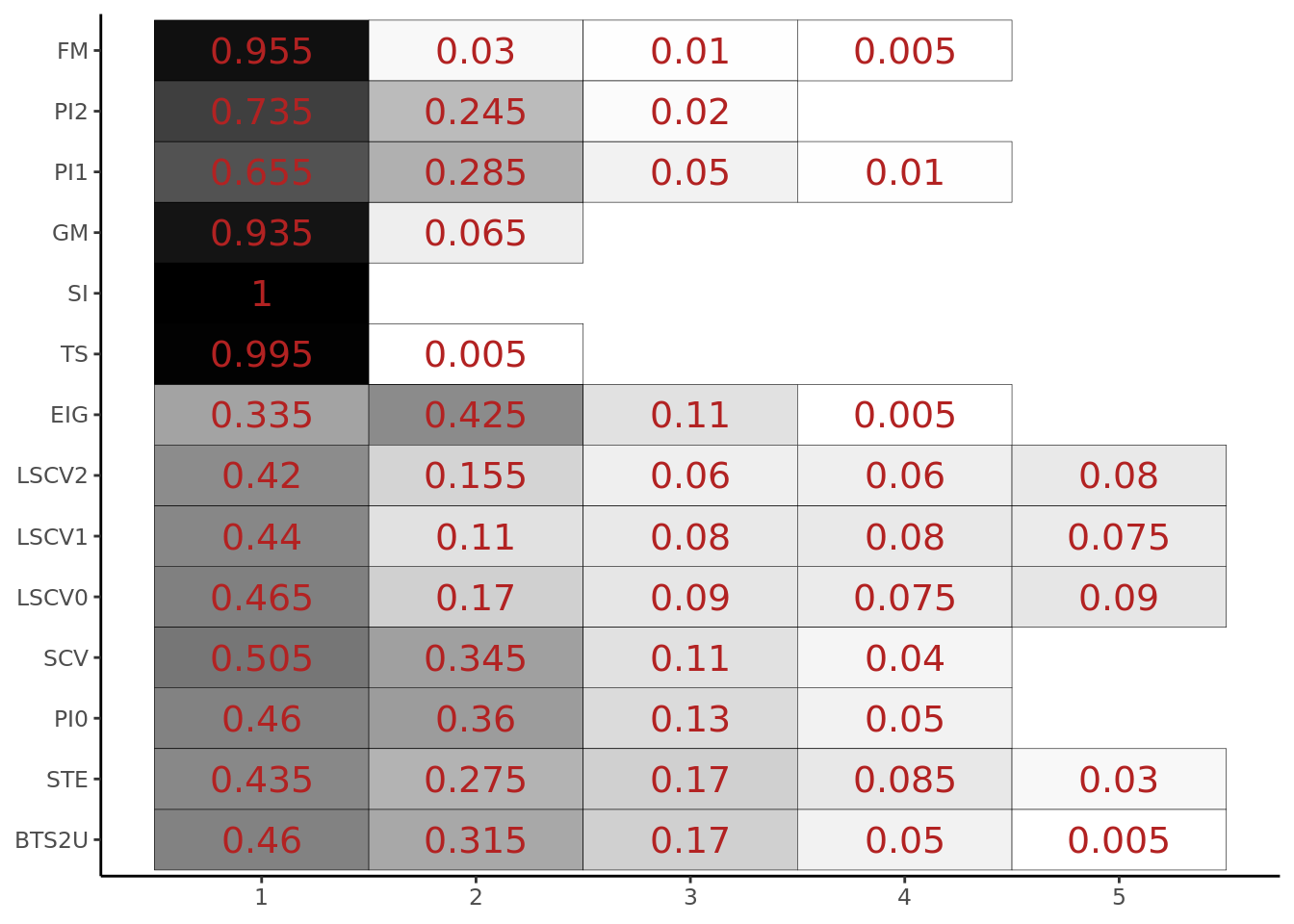}
        \caption{$\samplesize = 100$}
        \label{fig:distributions-m24-100}
    \end{subfigure}
    \begin{subfigure}[b]{\subfigurewidthtwoandone}
        \centering
        \includegraphics[width=\graphicswidthtwoandone]{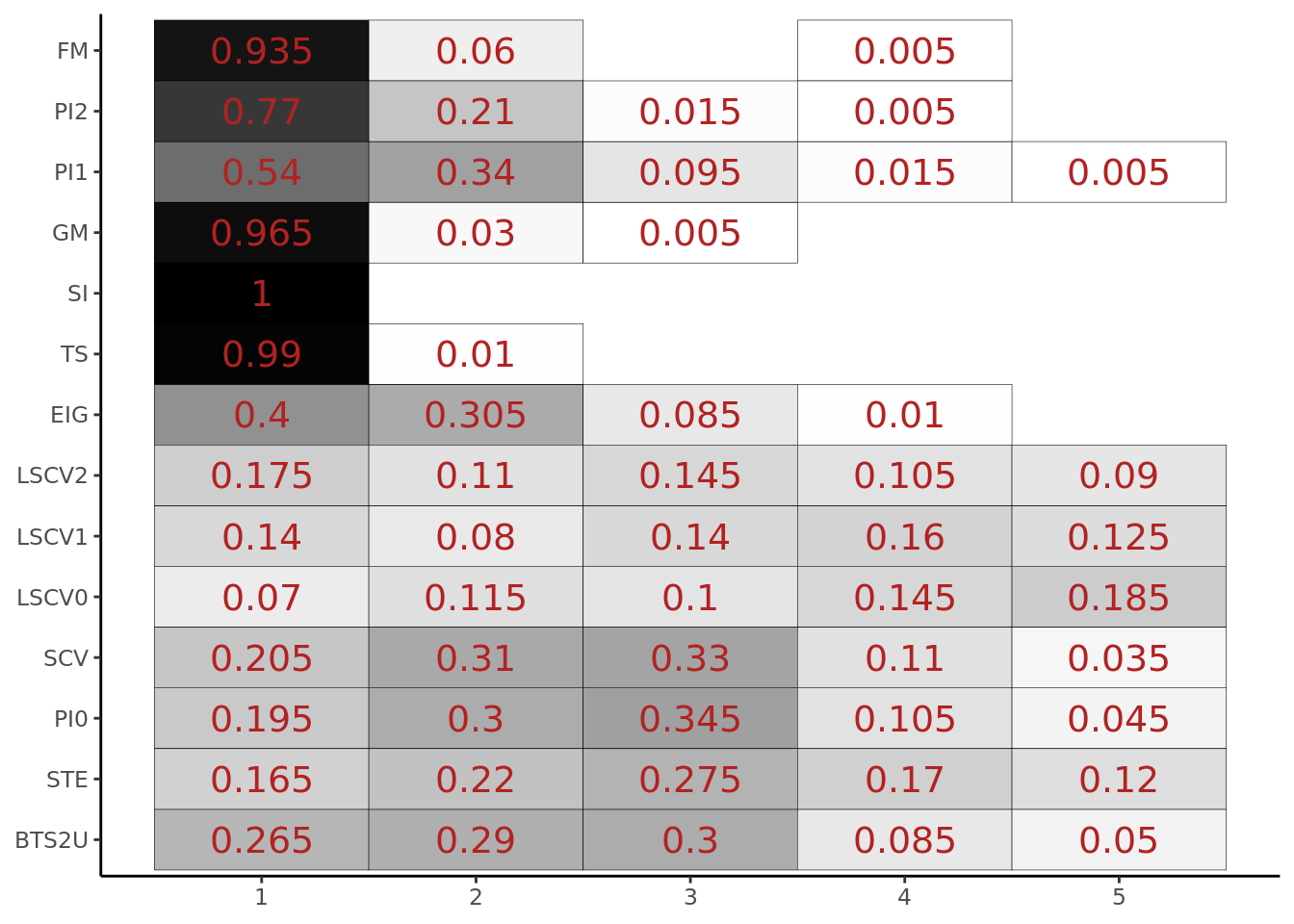}
        \caption{$\samplesize = 400$}
        \label{fig:distributions-m24-400}
    \end{subfigure}
    \begin{subfigure}[b]{\subfigurewidthtwoandone}
        \centering
        \includegraphics[width=\graphicswidthtwoandone]{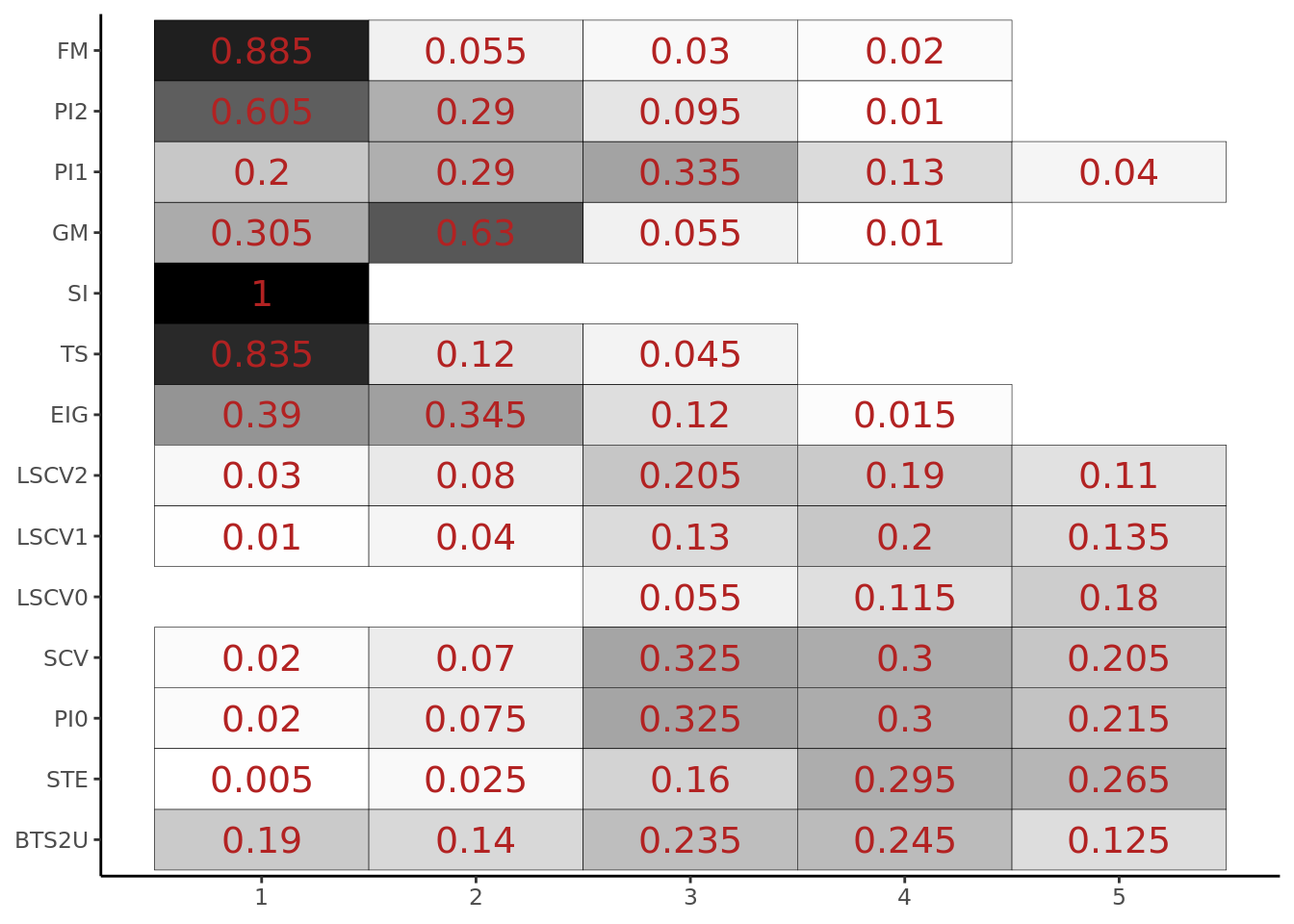}
        \caption{$\samplesize = 1600$}
        \label{fig:distributions-m24-1600}
    \end{subfigure}
    \caption{%
        Distribution of the predicted \gls*{nom} for the \fourthmodel{} test-bed for several sample sizes $\samplesize$.
    }
    \label{fig:distributions-m24}
\end{figure}

\begin{figure}
    \centering
    \begin{subfigure}[b]{\subfigurewidthtwoandone}
        \centering
        \includegraphics[width=\graphicswidthtwoandone]{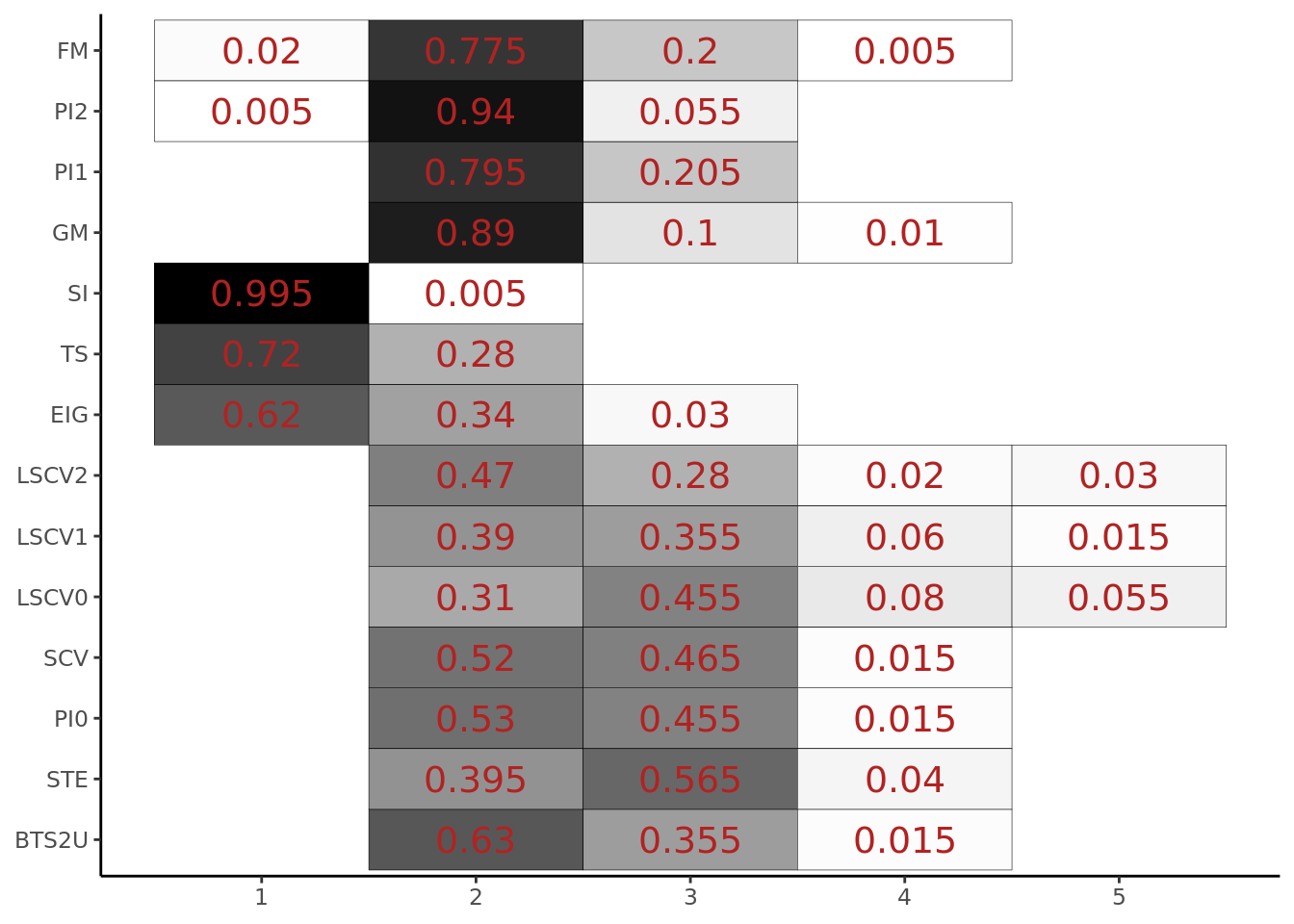}
        \caption{$\samplesize = 100$}
        \label{fig:distributions-m25-100}
    \end{subfigure}
    \begin{subfigure}[b]{\subfigurewidthtwoandone}
        \centering
        \includegraphics[width=\graphicswidthtwoandone]{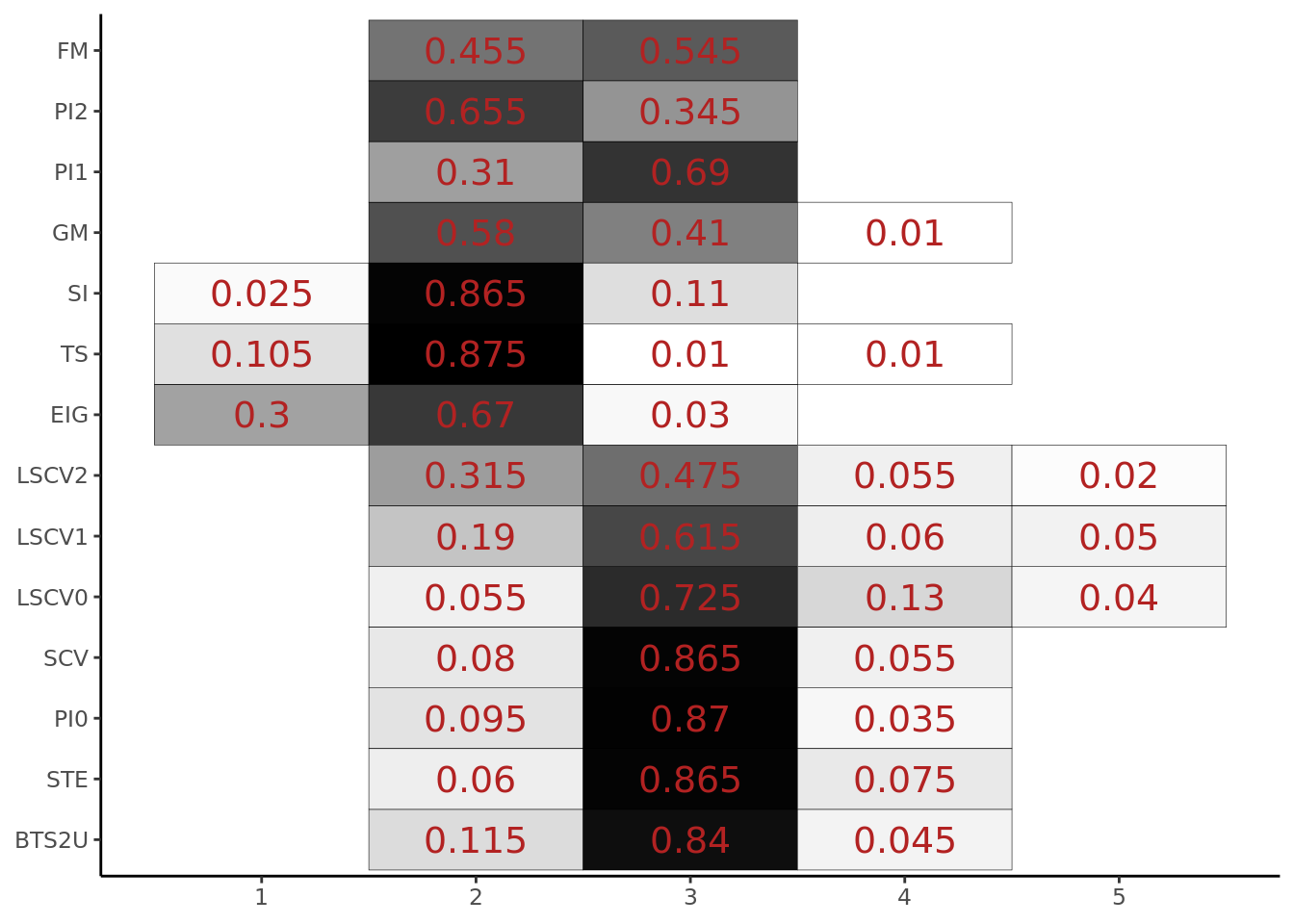}
        \caption{$\samplesize = 400$}
        \label{fig:distributions-m25-400}
    \end{subfigure}
    \begin{subfigure}[b]{\subfigurewidthtwoandone}
        \centering
        \includegraphics[width=\graphicswidthtwoandone]{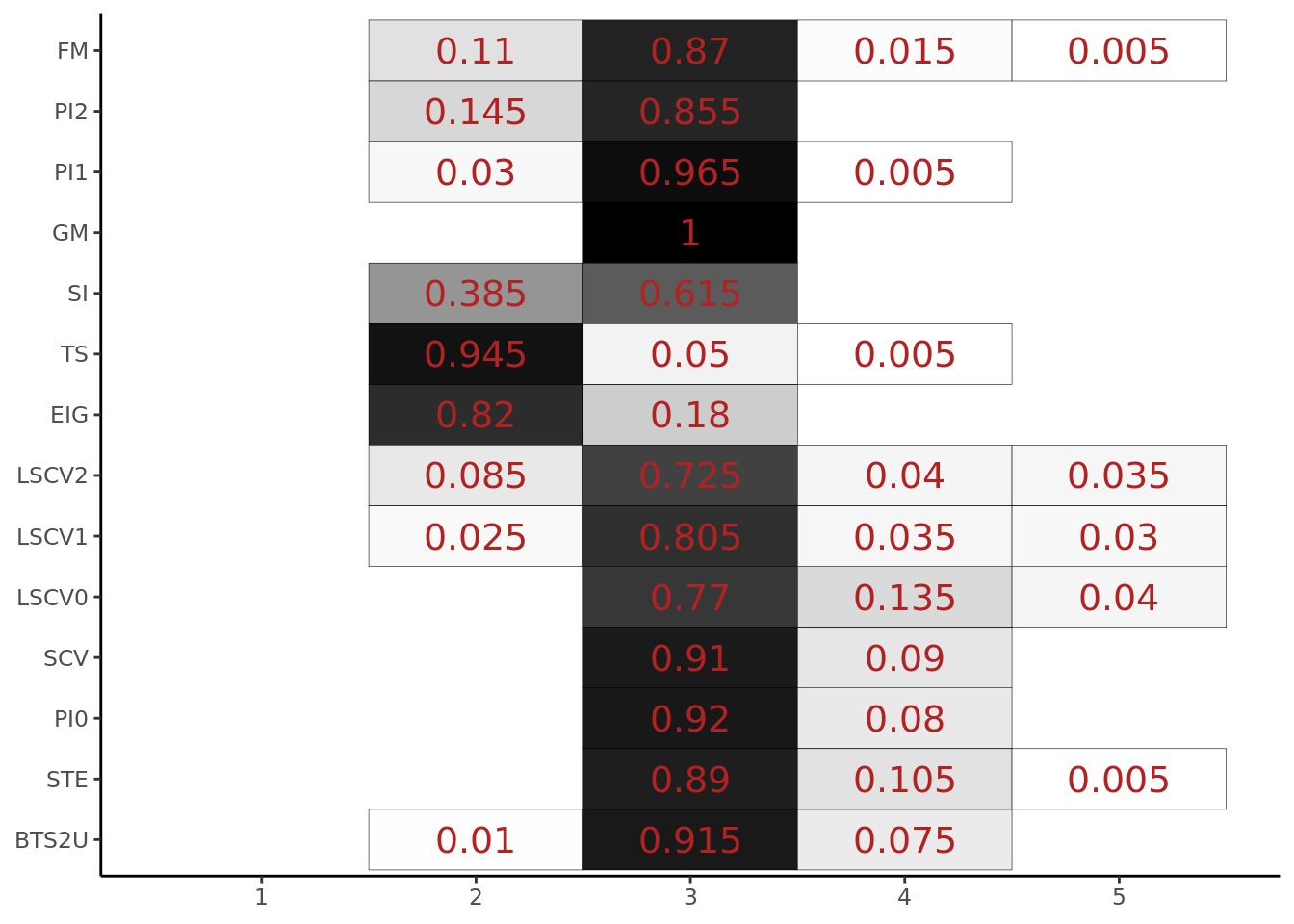}
        \caption{$\samplesize = 1600$}
        \label{fig:distributions-m25-1600}
    \end{subfigure}
    \caption{%
        Distribution of the predicted \gls*{nom} for the \fifthmodel{} test-bed for several sample sizes $\samplesize$.
    }
    \label{fig:distributions-m25}
\end{figure}

The previous figures suggest many relevant correlations between the contending methods, as depicted in \figurename~\ref{fig:estimation-correlations}.
We see several groups.
\btsselected{} is similar to \kdepizero{}, \kdelscv{} and \kdeste{}, while all the \gls*{lscv} variants are highly concordant.
A third group, loosely coupled, gathers \tautstring{}, \silverman{}, \gaussianmixture{}, \kdepione{}, \kdepitwo{} and \fishermarron{}, although \kdepione{} also correlates with \kdepizero{}.
The fourth and final group comprises the outlying \telepathicbootstrap{}, the only method with all its correlations below 0.5.

\begin{figure}
    \centering
    \includegraphics[width=0.7\textwidth]{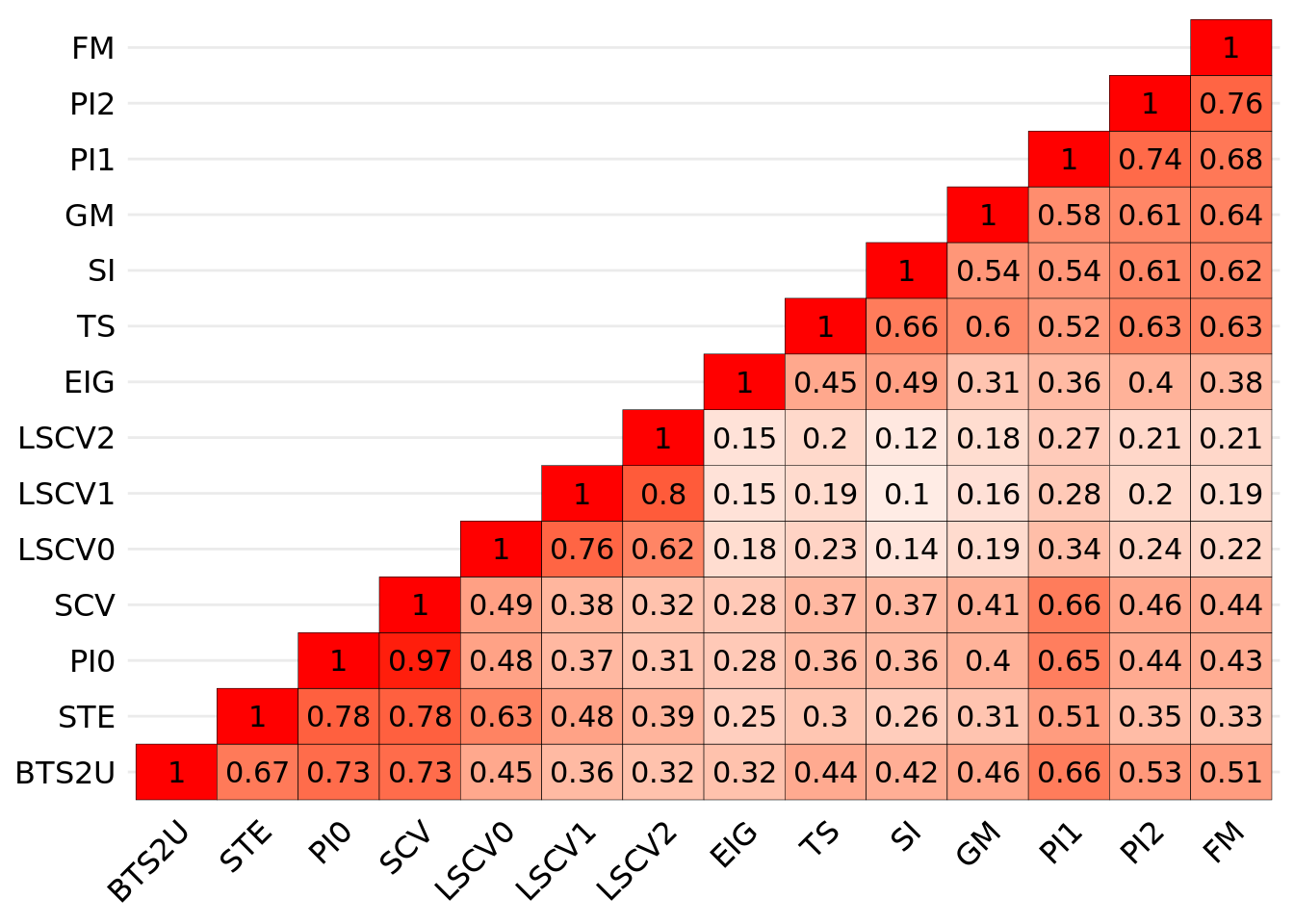}
    \caption{%
        Kendall correlations between predictions along the $\numtestbedstimessamplesizes \times \numexperimentreplicationsparam = 3000$ samples.
    }
    \label{fig:estimation-correlations}
\end{figure}

\subsection{BTS variants}

In Section~\ref{sec:setup}, we chose \btsselected{} as our reference \gls*{bts} variant to compare with the alternative traditional methods.
Here, we will compare all the \gls*{bts} variants in a similar exercise.

\tablename~\ref{tab:final-ranking-bts} shows the global ranking for the \gls*{bts} variants.
\btszero{} ranks behind the rest, including \btsselected{}.
The intermediate rankings are reported in \tablename~\ref{tab:intermediate-rankings-bts} and depicted in \figurename~\ref{fig:intermediate-rankings-bts}.
The rank correlations appear in \figurename~\ref{fig:rank-correlations-bts}.

The problem with \btszero{} is the sample size: with large samples, it improves considerably, just like \kdepizero{} and \kdepione{}.
On the other hand, the refined versions \btstwosample{}, \btstwojeffreys{} and \btsselected{} are not significantly better than their respective predecessors, \btsonesample{}, \btsonejeffreys{} and \btsoneuniform{}.

\begin{table}
    \centering
    \footnotesize
    \begin{tabular}{ccccccc}
    \toprule
    \texttt{BTS0} & \texttt{BTS1S} & \texttt{BTS1J} & \texttt{BTS1U} & \texttt{BTS2S} & \texttt{BTS2J} & \texttt{BTS2U} \\
    \midrule
    2             & \textbf{1}     & \textbf{1}     & \textbf{1}     & \textbf{1}     & \textbf{1}     & \textbf{1}     \\
    \bottomrule
\end{tabular}

    \captionsetup{font=footnotesize}
    \caption{%
        Global ranking among \gls*{bts} variants.
    }
    \label{tab:final-ranking-bts}
\end{table}

\begin{table}
    \centering
    \footnotesize
    \begin{tabular}{lrccccccc}
    \toprule
                 &      & \texttt{BTS0} & \texttt{BTS1S} & \texttt{BTS1J} & \texttt{BTS1U} & \texttt{BTS2S} & \texttt{BTS2J} & \texttt{BTS2U} \\
    \midrule
    \texttt{M21} & 100  & 2             & 2              & \textbf{1}     & \textbf{1}     & 2              & \textbf{1}     & \textbf{1}     \\
    \texttt{M21} & 400  & 3             & 2              & 2              & \textbf{1}     & 2              & 2              & \textbf{1}     \\
    \texttt{M21} & 1600 & 3             & 2              & 2              & \textbf{1}     & 2              & 2              & \textbf{1}     \\
    \addlinespace
    \texttt{M22} & 100  & 3             & 2              & 2              & \textbf{1}     & 2              & 2              & \textbf{1}     \\
    \texttt{M22} & 400  & \textbf{1}    & 2              & 2              & 3              & 2              & 2              & 3              \\
    \texttt{M22} & 1600 & \textbf{1}    & 2              & 2              & 3              & 2              & 2              & 3              \\
    \addlinespace
    \texttt{M23} & 100  & 4             & 3              & 2              & \textbf{1}     & 3              & 2              & \textbf{1}     \\
    \texttt{M23} & 400  & \textbf{1}    & 2              & 2              & 3              & 2              & 2              & 3              \\
    \texttt{M23} & 1600 & \textbf{1}    & 2              & 2              & 2              & 2              & 2              & 2              \\
    \addlinespace
    \texttt{M24} & 100  & 2             & 2              & \textbf{1}     & \textbf{1}     & 2              & \textbf{1}     & \textbf{1}     \\
    \texttt{M24} & 400  & 4             & 3              & 2              & \textbf{1}     & 3              & 2              & \textbf{1}     \\
    \texttt{M24} & 1600 & \textbf{1}    & \textbf{1}     & \textbf{1}     & \textbf{1}     & \textbf{1}     & \textbf{1}     & \textbf{1}     \\
    \addlinespace
    \texttt{M25} & 100  & 4             & 3              & 2              & \textbf{1}     & 3              & 2              & \textbf{1}     \\
    \texttt{M25} & 400  & 2             & \textbf{1}     & \textbf{1}     & \textbf{1}     & \textbf{1}     & \textbf{1}     & \textbf{1}     \\
    \texttt{M25} & 1600 & \textbf{1}    & 2              & 2              & 2              & 2              & 2              & 2              \\
    \bottomrule
\end{tabular}

    \captionsetup{font=footnotesize}
    \caption{%
        Intermediate rankings by test-bed and sample size configuration among \gls*{bts} variants.
    }
    \label{tab:intermediate-rankings-bts}
\end{table}

\begin{figure}
    \centering
    \begin{subfigure}[b]{\figurewidth}
        \centering
        \includegraphics[width=\subfigurewidth]{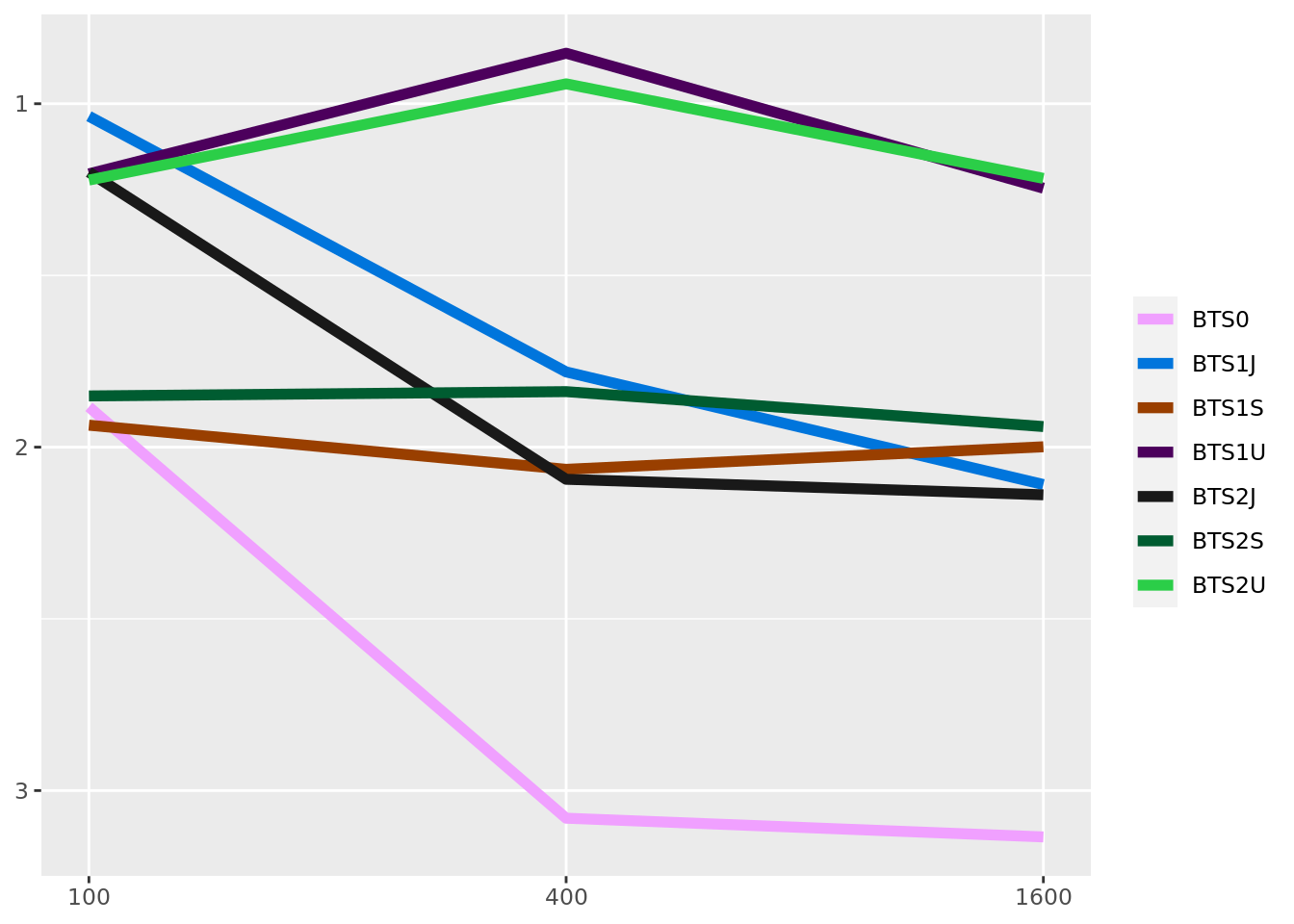}
        \caption{\firstmodel{}}
    \end{subfigure}
    \begin{subfigure}[b]{\figurewidth}
        \centering
        \includegraphics[width=\subfigurewidth]{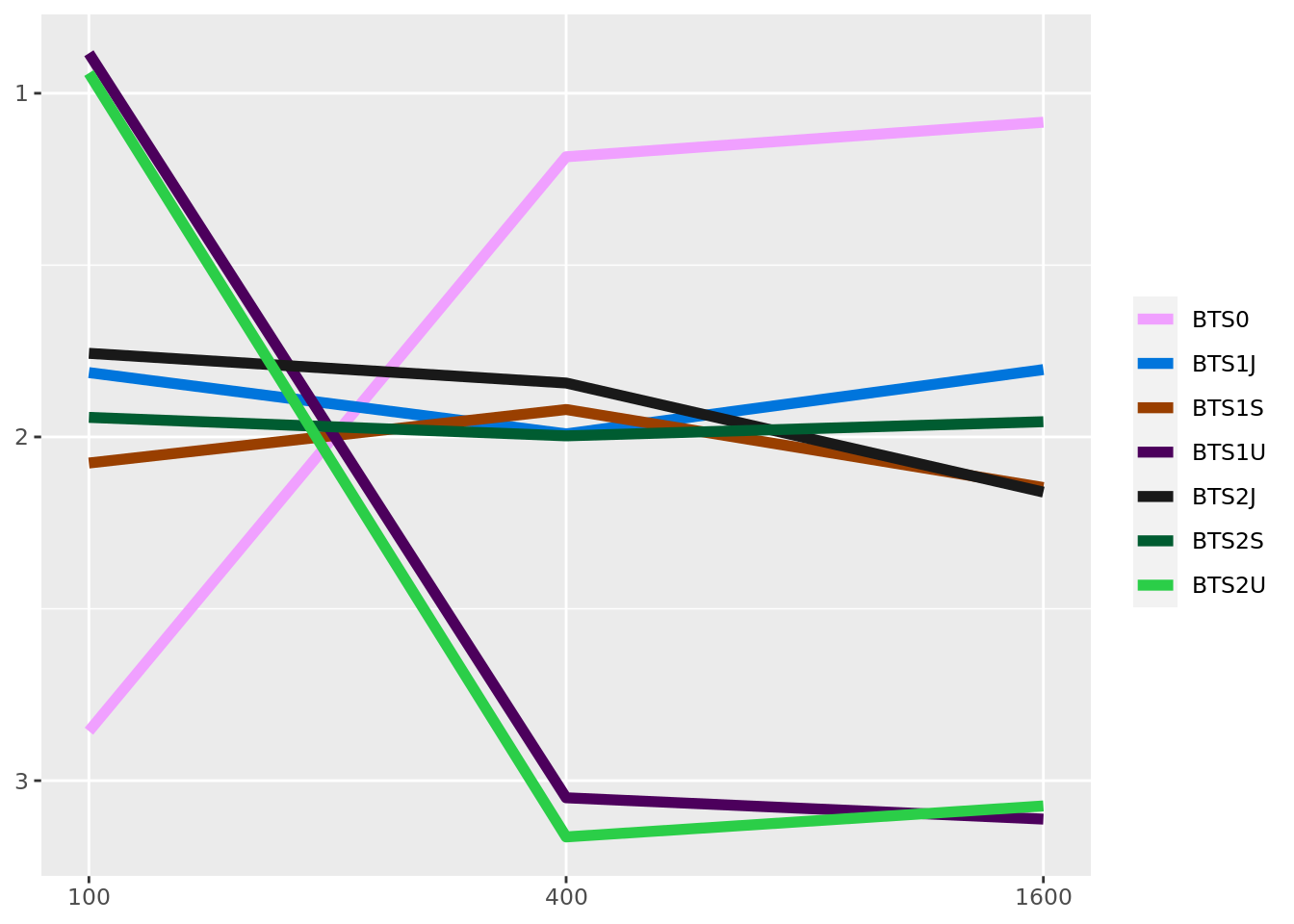}
        \caption{\secondmodel{}}
    \end{subfigure}
    \begin{subfigure}[b]{\figurewidth}
        \centering
        \includegraphics[width=\subfigurewidth]{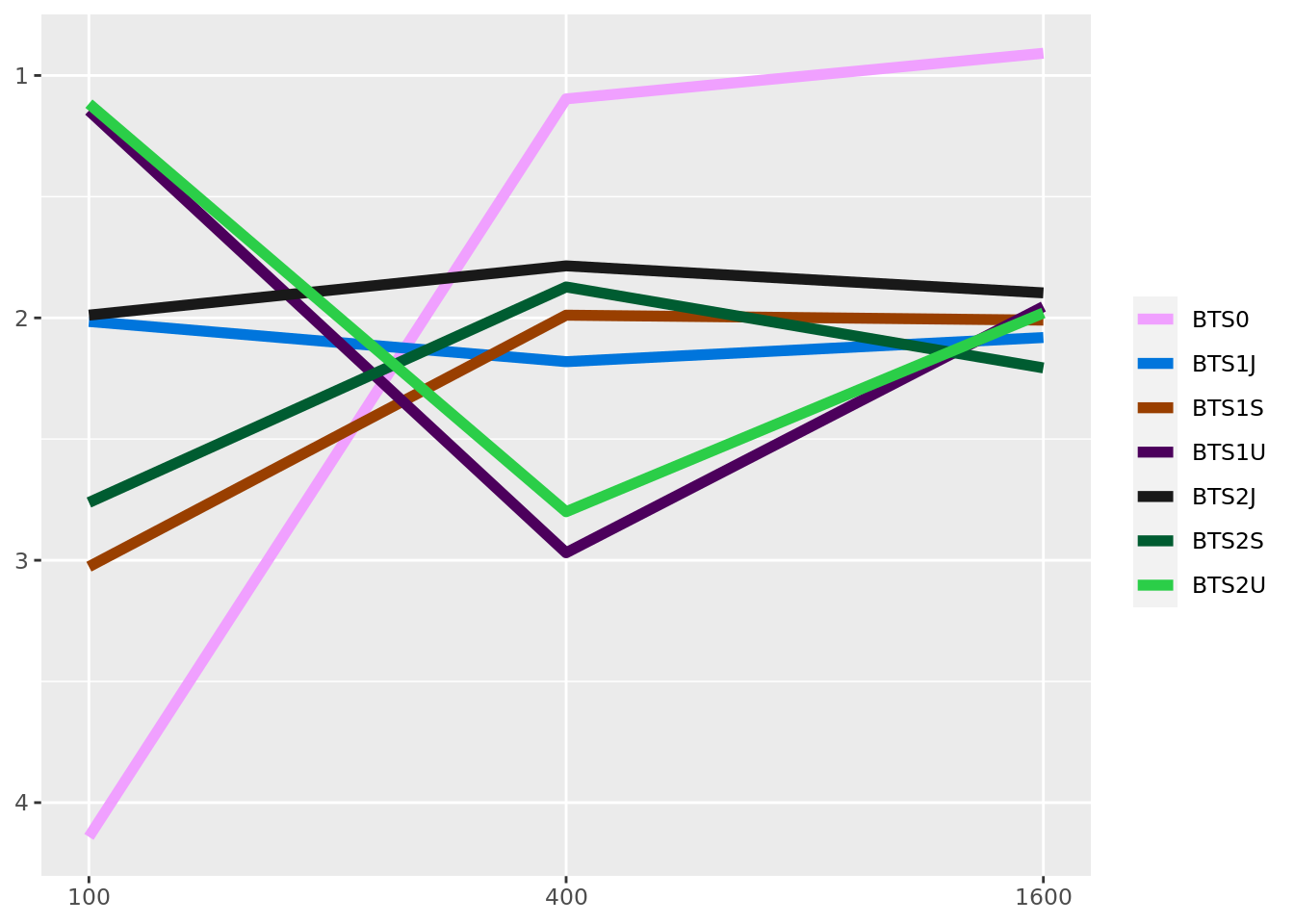}
        \caption{\thirdmodel{}}
        \label{fig:rankings-m23-bts}
    \end{subfigure}
    \begin{subfigure}[b]{\figurewidth}
        \centering
        \includegraphics[width=\subfigurewidth]{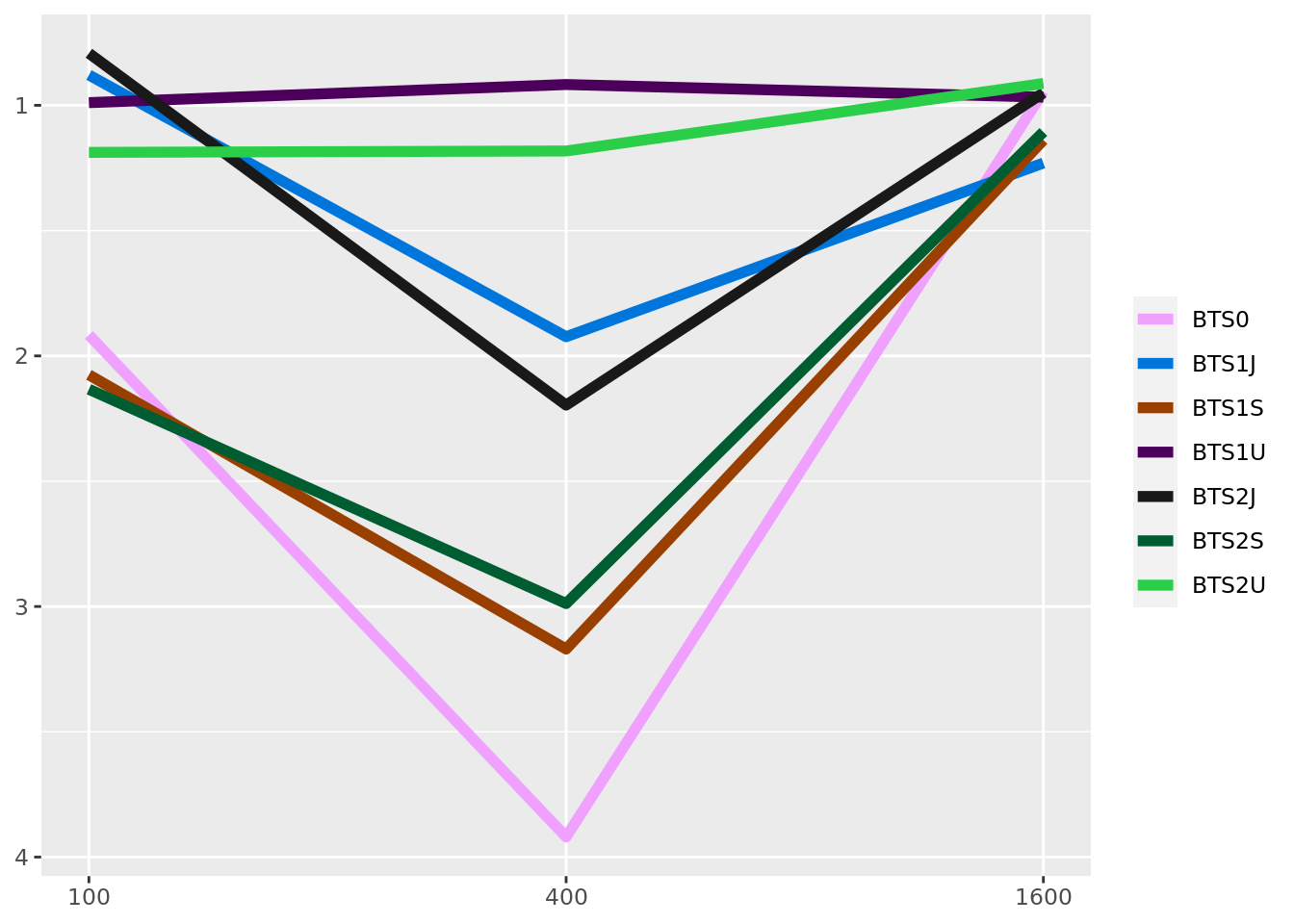}
        \caption{\fourthmodel{}}
    \end{subfigure}
    \begin{subfigure}[b]{\figurewidth}
        \centering
        \includegraphics[width=\subfigurewidth]{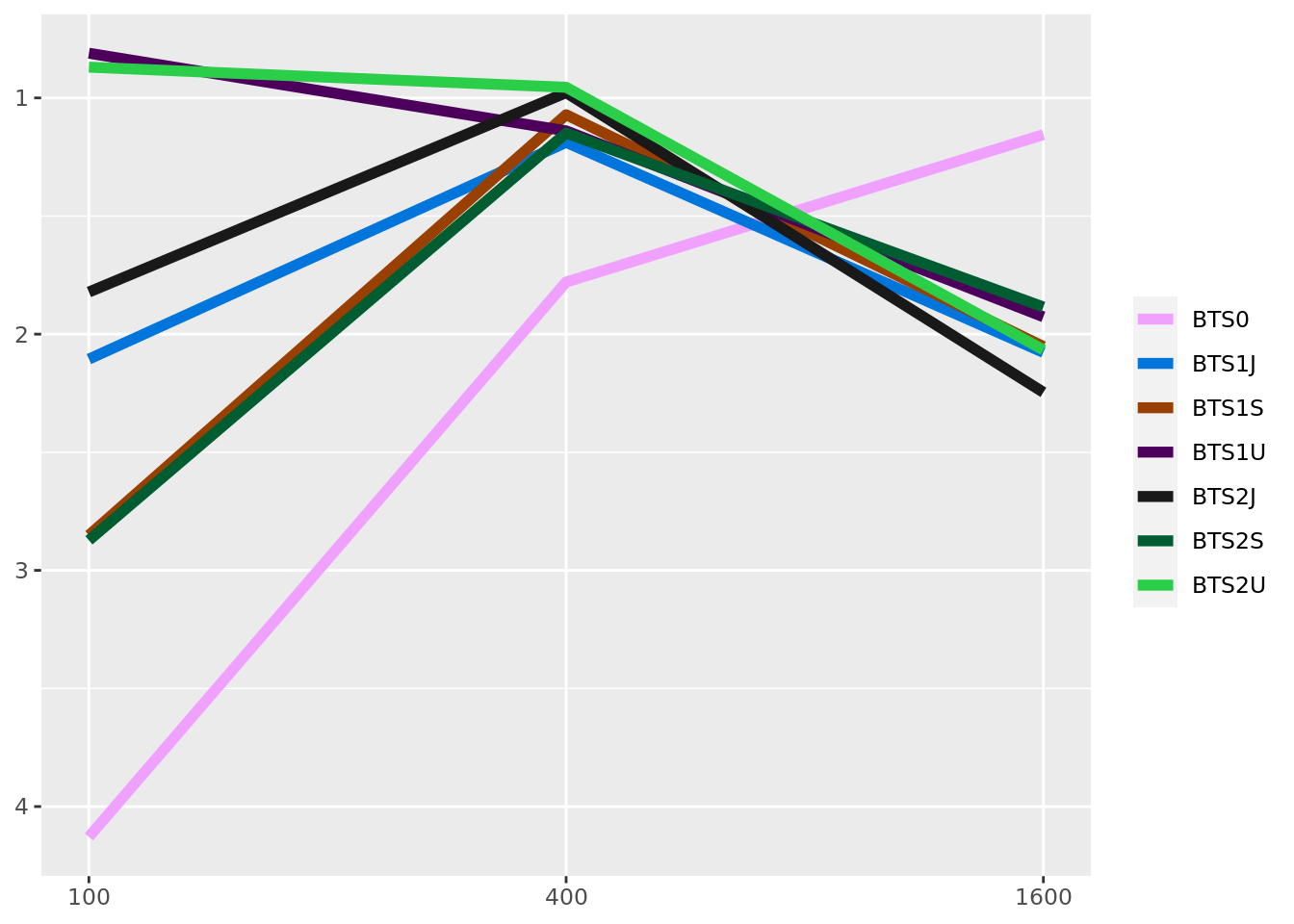}
        \caption{\fifthmodel{}}
    \end{subfigure}
    \caption{%
        Intermediate rankings by test-bed among \gls*{bts} variants.
    }
    \label{fig:intermediate-rankings-bts}
\end{figure}

\begin{figure}
    \centering
    \includegraphics[width=0.7\textwidth]{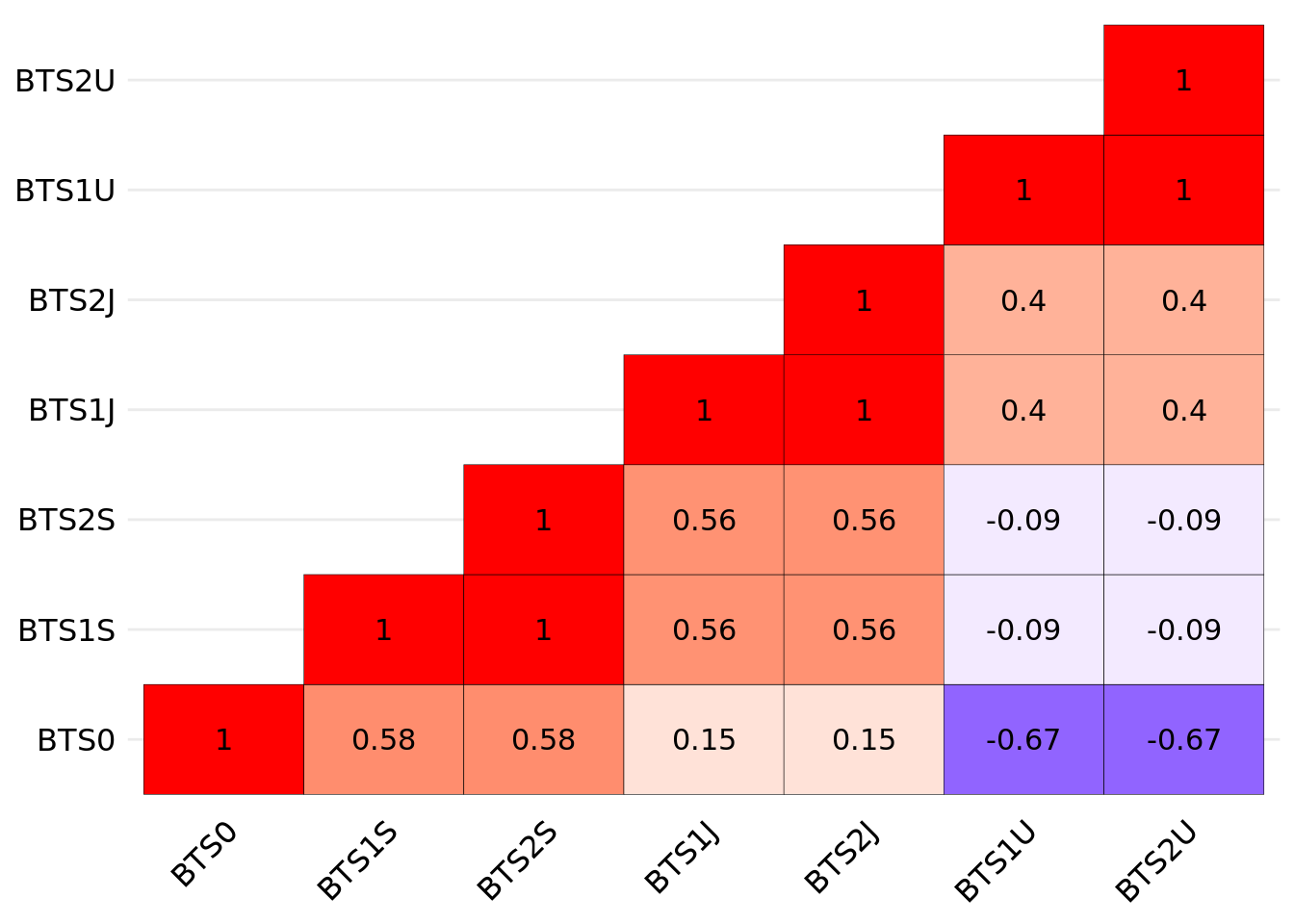}
    \caption{%
        Kendall correlations between the intermediate ranks in \tablename~\ref{tab:intermediate-rankings-bts}.
    }
    \label{fig:rank-correlations-bts}
\end{figure}

\figurename~\ref{fig:estimation-correlations-bts} shows the correlations between the predictions of the methods.
The refined versions are highly correlated with their respective base methods.
In turn, we observe that the more biased the prior probabilities $\priorprobnummodesparam$ are, the higher the correlation with \btszero{}.

\begin{figure}
    \centering
    \includegraphics[width=0.7\textwidth]{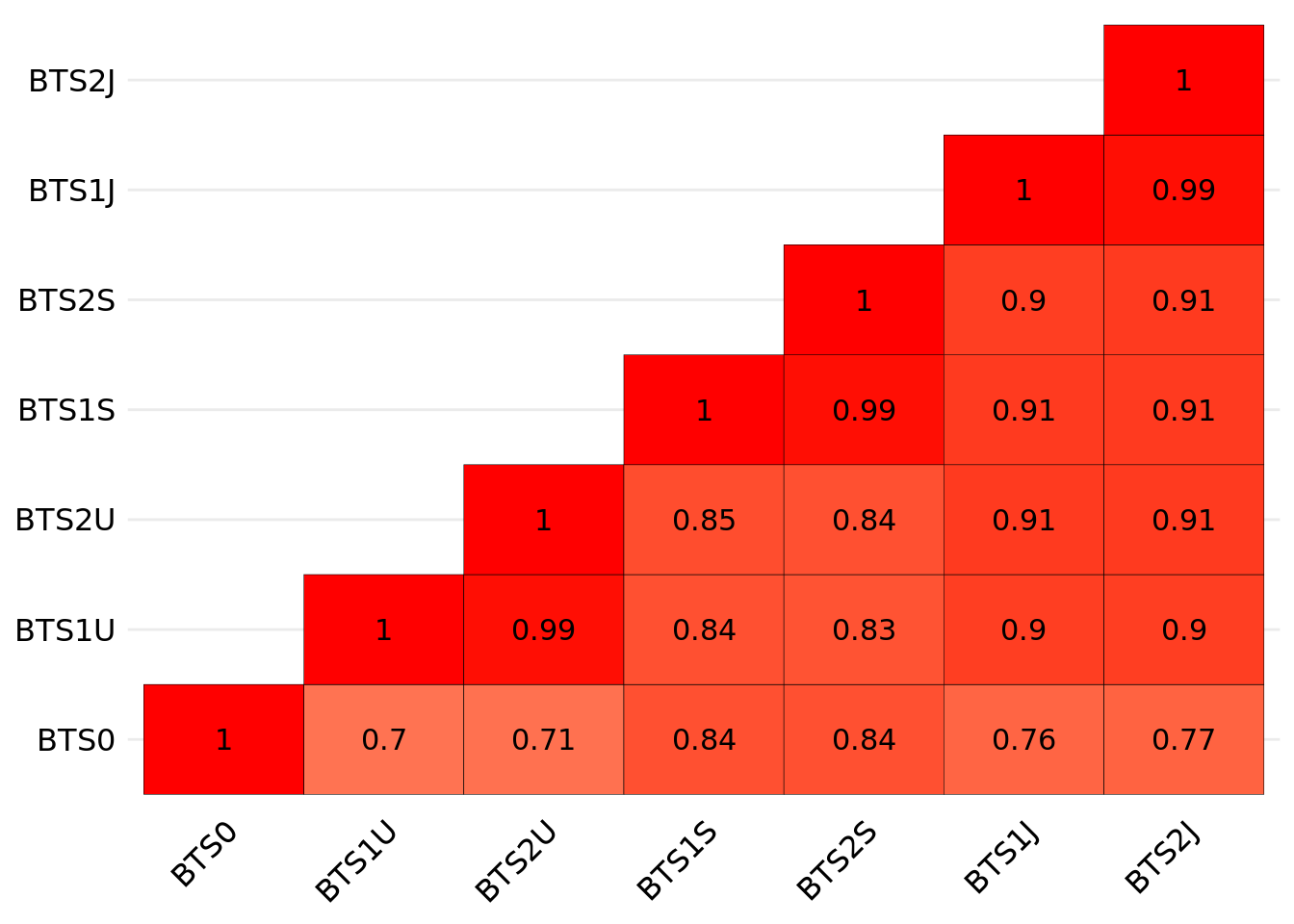}
    \caption{%
        Kendall correlations between predictions for the \gls*{nom} along the $\numtestbedstimessamplesizes \times \numexperimentreplicationsparam = 3000$ samples for the \gls*{bts} variants.
    }
    \label{fig:estimation-correlations-bts}
\end{figure}

        \section{Dimensionality reduction}

This appendix expands on the dimensionality reduction in the analysis phase of \gls*{bts}.
First, we outline the technicalities of \gls*{sfpca}.
Then, we address the requests of two reviewers for further justification on picking just the first \gls*{pc}.

\label{sec:dim-reduction}

\subsection{SFPCA}

\label{sec:sfpca}

The following lines discuss \gls*{sfpca} in our context for $\zbsplinepdfspace$.
We translate the original theory in~\citet{Hron2016} to our notation so the reader can more easily identify the elements involved.

Let us express the \gls*{clr} of each of the $\zbsplinedim$ \glspl*{pc} in terms of ZB-spline basis functions as $\clr{\sfpcapceigenfun{\genindexi}} = \sum_{\genindexj = 1}^{\zbsplinedim} \sfpcacoordinate{\genindexi}{\genindexj} \zbasisspline_{\genindexj}$.
That is, each $\sfpcapceigenfun{\genindexi}$ corresponds to the column vector $\sfpcacoordinatesvec{\genindexi} = (\sfpcacoordinate{\genindexi}{1}, \dots, \sfpcacoordinate{\genindexi}{\zbsplinedim}) \in \realszbsplinedim$.
Similarly, let us also expand the centred functional data as
$
    \clr{\sfpcacentreddatum}
    =
    \sum_{\genindexj = 1}^{\zbsplinedim} \sfpcasamplecoordinates{\genindexi}{\genindexj} \zbasisspline_{\genindexj}
$,
i.e., the coordinates forming the rows of the matrix $\sfpcasamplemat \in \reals^{\smoothingsamplesize \times \zbsplinedim}$.

To obtain $\sfpcacoordinatesvec{\genindexi}$, we \textit{first} have to solve the $\genindexi$-th largest eigenvalue $\sfpcaeigenvalue{\genindexi}$ problem
\begin{equation}
    \label{eq:sfpca-eigenvalue-problem}
    \frac{1}{\smoothingsamplesize}
    \sqrtbayesinnerproductmatrix
    \transpose{\sfpcasamplemat}
    \sfpcasamplemat
    \sqrtbayesinnerproductmatrix
    \sfpcauniteigenvector{\genindexi}
    =
    \sfpcaeigenvalue{\genindexi}
    \sfpcauniteigenvector{\genindexi}
    \,,
\end{equation}
where $\sfpcauniteigenvector{\genindexi} \in \realszbsplinedim$ has Euclidean norm one, i.e., $\transpose{\sfpcauniteigenvector{\genindexi}} \sfpcauniteigenvector{\genindexi} = 1$, and $\sqrtbayesinnerproductmatrix$ is the square root of the ZB-spline inner product matrix $\bayesinnerproductmatrix$ defined in Section~\ref{sec:preliminaries}.
Problem~\eqref{eq:sfpca-eigenvalue-problem} corresponds to the usual principal component analysis in $\realszbsplinedim$ for the transformed data matrix $\sfpcasamplemat \sqrtbayesinnerproductmatrix$~\citep[p. 5]{Hron2016} and can be solved directly using the routine \texttt{eigen} in \rlanguage{}.
\textit{Then}, finally, one takes $\sfpcacoordinatesvec{\genindexi}$ satisfying $\sqrtbayesinnerproductmatrix \sfpcacoordinatesvec{\genindexi} = \sfpcauniteigenvector{\genindexi}$, while the corresponding eigenvalue is $\sfpcaeigenvalue{\genindexi}$.

Considering the solutions of all the $\sfpcaeigenvalue{\genindexi}$-problems~\eqref{eq:sfpca-eigenvalue-problem}, it can be easily checked that
$
    \bayesinnerproduct{\sfpcapceigenfun{\genindexi}}{\sfpcapceigenfun{\genindexj}}
    =
    \transpose{\sfpcacoordinatesvec{\genindexi}}
    \bayesinnerproductmatrix
    \sfpcacoordinatesvec{\genindexj}
    =
    \transpose{\sfpcauniteigenvector{\genindexi}} \sfpcauniteigenvector{\genindexj}
    =
    \kroneckerdd{\genindexi}{\genindexj}
$,
the Kronecker delta, meaning the \glspl*{pc} form an orthonormal basis.
To grasp the true meaning of \glspl*{pc}, specifically their variability-maximising property, we refer the reader to~\citet{Hron2016}.
In particular, the sum of squared scores $\sum_{\genindexi = 1}^{\smoothingsamplesize} \sfpcascore_{\genindexi}^2$ (see Section~\ref{sec:analysis-stage}) is maximal when a unitary $\sfpcapceigenfun{1}$ is the first \gls*{pc}~\citep[Equation 7]{Hron2016}.

\subsection{Dimension justification}

In Section~\ref{sec:analysis-stage} of the manuscript, we sketched some convenience and simplicity reasons behind keeping just one dimension coincidental with the goals of the analysis phase.
Namely, we alluded to making robust inferences and improving interpretability.
We also claimed that one \gls*{pc} provided enough power in the \gls*{bts} context but gave no justification beyond the scree plot in \figurename~\ref{fig:sfpca-scree-plot}, complemented in this \gls*{sm} with \figurename~\ref{fig:ohtani-sfpca-scree-plot}.
Let us now imagine the consequences of including several \glspl*{pc}.

First, the multiparameter generalisation of~\eqref{eq:sfpca-model} is
\begin{equation}
    \label{eq:sfpca-model-multiparameter}
    \condprob{\funcarg}{\sfpcaparamnum{1}, \dots, \sfpcaparamnum{\sfpcanumpcs}}
    =
    \sfpcamean
    \bayesperturbation
    \bigbayesperturbation_{\genindexi = 1}^{\sfpcanumpcs}
    (\sfpcaparamnum{\genindexi}
    \bayespowering
    \sfpcastdevnum{\genindexi})
    \,,
\end{equation}
where $\upmu$ stays the same as in~\eqref{eq:sfpca-model}, $\sfpcastdevnum{\genindexi} = \sqrt{\sfpcaeigenvalue{\genindexi}} \bayespowering \sfpcapceigenfun{\genindexi}$, and typically $\sfpcanumpcs \ll \zbsplinedim$.
A similar procedure based on centred projections allows prescribing a rectangular support for the parameter vector $(\sfpcaparamnum{1}, \dots, \sfpcaparamnum{\sfpcanumpcs})$.

Using~\eqref{eq:sfpca-model-multiparameter}, new information would enter into our analysis.
Though valuable for other purposes, such information would be noisy and redundant for estimating the \gls*{nom}.
By \textit{debugging} samples like that in \figurename~\ref{fig:posterior-h-alpha}, we observed that many different but similar \glspl*{pdf} lead to the same \gls*{nom}.
The second and subsequent \glspl*{pc} generally add variability regarding other \gls*{pdf} properties.
For instance, in some scenarios, mode A is higher than mode B, while in others, it is the other way around.
Similarly, modes A and B appear in slightly different positions in some scenarios.
Taking all into account, we would end up with a complex model like~\eqref{eq:sfpca-model-multiparameter}, where several $\sfpcaparamnum{\genindexi}$ have to be tuned, increasing the computational cost and reducing the robustness of the inference process.
Moreover, making sense of such an analysis would be challenging compared to the simplicity of \figurename~\ref{fig:sfpca-mode-tree}.
In particular, note that the mode tree in \figurename~\ref{fig:sfpca-mode-tree} does not say anything about the height of a mode.

The above explanation about redundancy and noise will be more apparent after examining the modes of variation for the subsequent \glspl*{pc} in \figurename~\ref{fig:sfpca-subsequent-modes-of-variation}.
First, note that the amplitude of the \textit{oscillations} around the mean gets increasingly weak from \figurename~\ref{fig:sfpca-splines} to \figurename~\ref{fig:sfpca-splines-2} and then \figurename~\ref{fig:sfpca-splines-3}.
In fact, the \textit{oscillation} for $\sfpcapceigenfun{3}$ is already virtually imperceptible in \figurename~\ref{fig:sfpca-splines-3}, being $\sfpcapceigenfun{3}$ arguably noisy.
The case of \figurename~\ref{fig:sfpca-splines-2} is far more interesting, however.
There, we see that the \gls*{pdf} in which the two left-most modes are most pronounced (the blue curve) is also the one in which the right-most mode has the lowest height.
That is precisely the opposite case of \figurename~\ref{fig:sfpca-splines}, where the three modes grow simultaneously.
This means that $\sfpcapceigenfun{2}$ is associated with calibrating the relative height of the modes, which is redundant for estimating the \gls*{nom}.

\begin{figure}
    \centering
    \begin{subfigure}[b]{\figurewidth}
        \centering
        \includegraphics[width=\subfigurewidth]{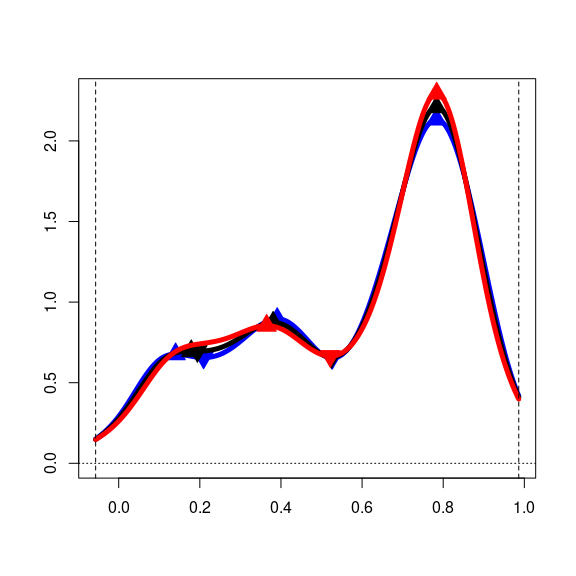}
        \caption{Mode of variation of $\sfpcapceigenfun{2}$}
        \label{fig:sfpca-splines-2}
    \end{subfigure}
    \begin{subfigure}[b]{\figurewidth}
        \centering
        \includegraphics[width=\subfigurewidth]{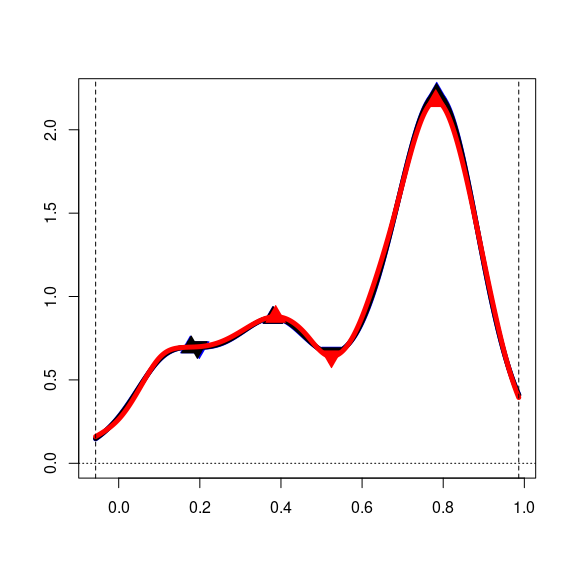}
        \caption{Mode of variation of $\sfpcapceigenfun{3}$}
        \label{fig:sfpca-splines-3}
    \end{subfigure}
    \caption{
        Modes of variation for subsequent \glspl*{pc} following the example in \figurename~\ref{fig:sfpca-results}.
        Both subfigures have the same structure and colouring conventions as \figurename~\ref{fig:sfpca-splines}, but replacing $\sfpcapceigenfun{1}$ with $\sfpcapceigenfun{2}$ and $\sfpcapceigenfun{3}$, respectively.
        Also, the black curve in both cases is the same as in \figurename~\ref{fig:sfpca-splines}, corresponding to the mean $\sfpcamean$.
        The support bounds $\sfpcaparammin$ and $\sfpcaparammax$ are calculated analogously as for~\eqref{eq:jeffreys-prior}, considering $\sfpcapceigenfun{2}$ or $\sfpcapceigenfun{3}$ instead of $\sfpcapceigenfun{1}$.
        On the left, the lower bound (blue) has three modes, while the upper bound (red) has two.
        By contrast, the situation in the right-hand side subfigure is unclear, given the small amplitude of the \textit{oscillation} around the mean.
    }
    \label{fig:sfpca-subsequent-modes-of-variation}
\end{figure}

There is a natural reason why one \gls*{pc} works so well.
One has to consider that the set of functions $\compositionalsplinesample$ we aim to summarise with \gls*{sfpca} has a relatively low complexity compared to arbitrary datasets in functional data analysis.
The random mechanism generating them is comparatively straightforward: a joint use of \gls*{kde} and compositional splines.
The complementary interaction between $\bandwidth$ and $\oneminuscurvaturepenalty$ in \figurename~\ref{fig:posterior-h-alpha} should be relatively simple to capture with a single \gls*{pc}.
In this respect, we note that an oblique straight boundary, like in \figurename~\ref{fig:posterior-h-alpha}, is standard but not universal, as shown in \figurename~\ref{fig:ohtani-posterior-h-alpha}.
In the latter case, the boundary is vertical, indicating an even simpler scenario dominated by $\bandwidth$.

Finally, adding more \glspl*{pc} would also be inconvenient for two reasons.
First, we would no longer be able to pick a median spline representative $\selectedmodelnummodesparam$ of each $\nummodesparam$-modality hypothesis for the testing phase, at least not directly and efficiently, as in the one-dimensional case.
Also, the Jeffreys prior in~\eqref{eq:jeffreys-prior} in the one-dimensional case has the desirable property of being a \textit{reference prior}, which does not happen in the multiparameter setting~\citep{Bernardo1994}.

        \section{Computing}

\label{sec:computing}

This section comments on the computational environment used throughout our investigation.
Emerging technologies such as \textit{containerisation} and \textit{cloud computing} are increasingly drawing the attention of the research community as a means to enhance reproducibility.

\paragraph{Containerisation}

All our research outputs have been produced within a \textit{Docker} container~\citep{Merkel2014}.
The figures and tables were generated upon building a Docker image as part of the installed vignettes of an \rlanguage{} package.
The \texttt{Dockerfile} with the specification of that image will be distributed along the \gls*{bts} package for \rlanguage{}.
Next, the built image was made available for the simulation study in a private \textit{cloud} environment through the \textit{Docker Hub} container registry.

\paragraph{Cloud computing}

The simulation study was carried out with the help of Microsoft's \textit{Azure Databricks} service~\citep{Etaati2019}.
The whole experiment was scheduled as a Databricks \textit{workflow}, where each of the $\numtestbedstimessamplesizes = 15$ sampling configurations was a Databricks \textit{task} executing a Databricks \rlanguage{} \textit{notebook} with distinct parameters.
Then, the $\numexperimentreplicationsparam = \numexperimentreplications$ replications took the form of parallelisable Spark \textit{tasks}~\citep{Karau2017}.
The required \rlanguage{} files and a JSON file specifying the workflow will be distributed along the \gls*{bts} package.
Finally, the workflow job was assigned to a Databricks cluster with 25 type \texttt{Standard\_DS3\_v2} workers, each counting on four cores.
Therefore, 100 threads ran in parallel at any given time for a total execution time of approximately 20 hours.

    \end{refsection}

    \renewcommand*{\mkbibcompletename}[1]{\textsc{#1}}
    \printbibliography[section=2, filter=references, title={References}, sorting=nyt]
    \printbibliography[section=2, filter=resources, title={Resources}, sorting=none]
    \renewcommand*{\mkbibcompletename}[1]{#1}
\fi

\end{document}